\documentclass[fleqn,usenatbib]{mnras}

\usepackage{newtxtext,newtxmath}
\usepackage{tablefootnote}

\usepackage[T1]{fontenc}
\usepackage[table,x11names]{xcolor}
\usepackage{pdflscape}
\DeclareRobustCommand{\VAN}[3]{#2}
\let\VANthebibliography\thebibliography
\def\thebibliography{\DeclareRobustCommand{\VAN}[3]{##3}\VANthebibliography}

\usepackage{graphicx}	
\usepackage{amsmath}	

\title[C-MetaLL Survey: I]{Cepheid Metallicity in the Leavitt Law (C- MetaLL) survey: I. HARPS-N@TNG spectroscopy of 47 Classical Cepheid and 1 BL Her variables\thanks{Based on observations made with the Italian Telescopio Nazionale Galileo (TNG) operated by the Fundación Galileo Galilei (FGG) of the Istituto Nazionale di Astrofisica (INAF) at the Observatorio del Roque de los Muchachos (La Palma, Canary Islands, Spain).}}

\author[V. Ripepi et al.]{
V. Ripepi,$^{1}$\thanks{E-mail: vincenzo.ripepi@inaf.it}
G. Catanzaro,$^{2}$
R. Molinaro,$^{1}$
M. Gatto,$^{1,3}$
G. De Somma,$^{1,3,4}$
M. Marconi,$^{1}$
M. Romaniello,$^{5}$
\newauthor
S. Leccia,$^{1}$
I. Musella,$^{1}$
E. Trentin,$^{1,6}$
G. Clementini,$^{7}$
V. Testa,$^{8}$ 
F. Cusano,$^{7}$
J. Storm,$^{6}$
\\
$^{1}$ INAF-Osservatorio Astronomico di Capodimonte, Salita Moiariello 16, 80131, Naples, Italy\\
$^{2}$ INAF-Osservatorio Astrofisico di Catania, Via S.Sofia 78, 95123, Catania, Italy\\
$^{3}$ Dipartimento di Fisica "E. Pancini", Università di Napoli "Federico II", Via Cinthia, 80126 Napoli, Italy\\  
$^{4}$ Istituto Nazionale di Fisica Nucleare (INFN)-Sez. di Napoli, Via Cinthia, 80126 Napoli, Italy\\ 
$^{5}$ European Southern Observatory (ESO), Karl-Schwarzschild-Str., 85748 Garching, Germany \\
$^{6}$ Leibniz-Institut f\"ur Astrophysik Potsdam (AIP), An der Sternwarte 16, D-14482 Potsdam, Germany \\
$^{7}$ INAF-Osservatorio di Astrofisica e Scienza dello Spazio, Via Gobetti 93/3, I-40129 Bologna, Italy\\ 
$^{8}$ INAF – Osservatorio Astronomico di Roma, via Frascati 33, I-00078 Monte Porzio Catone, Italy 
}

\date{Accepted XXX. Received YYY; in original form ZZZ}

\pubyear{2015}

\begin{document}
\label{firstpage}
\pagerange{\pageref{firstpage}--\pageref{lastpage}}
\maketitle

\begin{abstract}
Classical Cepheids (DCEPs) are the most important primary indicators of the extragalactic distance scale. Establishing the dependence on metallicity of their period--luminosity and period--Wesenheit ($PLZ$/$PWZ$) relations has deep consequences on the calibration of secondary distance indicators that lead to the final estimate of the Hubble constant (H$_0$). 
We collected high-resolution spectroscopy for 47 DCEPs plus 1 BL Her variables with HARPS-N@TNG and derived accurate atmospheric parameters, radial velocities and metal abundances. We measured spectral lines for 29 species and characterized their chemical abundances, finding very good agreement with previous results.
We re-determined the ephemerides for the program stars and measured their intensity-averaged magnitudes in the $V,I,J,H,K_s$ bands.  
 We complemented our sample with literature data and used the Gaia Early Data Release 3 (EDR3) to investigate the $PLZ$/$PWZ$ relations for Galactic DCEPs in a variety of filter combinations. We find that the solution without any metallicity term is ruled out at more than the 5 $\sigma$ level. 
Our best estimate for the metallicity dependence of the intercept of the $PLK_s$, $PWJK_s$, $PWVK_s$ and $PWHVI$ relations with three parameters, is $-0.456\pm$0.099, $-0.465\pm$0.071, $-0.459\pm$0.107 and $-0.366\pm$0.089 mag/dex, respectively. These values are significantly larger than the recent literature. The present data are still inconclusive to establish whether or not also the slope of the relevant relationships depends on metallicity. Applying a correction to the standard zero point offset of the Gaia parallaxes has the same effect of reducing by $\sim$22\% the size of the metallicity dependence on the intercept of the PLZ/PWZ relations. 
\end{abstract}




\begin{table*}
      \caption{Properties of the program stars. The different columns provide: (1) name of the targets according to the SIMBAD database \citep{Wenger2000}; (2) Right Ascension; (3) Declination; (4) exposure time (in seconds) of the acquired spectra.}
         \label{Tab:log}
\begin{tabular}{lccc}
\hline \hline
            \noalign{\smallskip}
  \multicolumn{1}{c}{ID} &
  \multicolumn{1}{c}{RA (J2000)} &
  \multicolumn{1}{c}{Dec (J2000)} &
  \multicolumn{1}{c}{T$_{\rm exp}$ (s)} \\
            \noalign{\smallskip}
            
  \multicolumn{1}{c}{(1)} &
  \multicolumn{1}{c}{(2)} &
  \multicolumn{1}{c}{(3)} &
  \multicolumn{1}{c}{(4)} \\
            \noalign{\smallskip}
\hline
            \noalign{\smallskip}
ASASSN\_J180946.70$-$182238.2 & 18:09:46.68 & $-$18:22:38.2 & 3$\times$1800\\
  ASAS\_J052610+1151.3 & 05:26:09.64 & +11:51:13.2 & 3$\times$1800\\
  ASAS\_J061022+1438.6 & 06:10:22.23 & +14:38:40.2 & 3$\times$1200\\
  ASAS\_J063519+2117.8 & 06:35:19.18 & +21:17:48.1 & 1$\times$3000+2$\times$3600\\
  ASAS\_J065413+0756.5 & 06:54:13.53 & +07:56:29.1 & 1$\times$3000+2$\times$3600\\
  ASAS\_J070832$-$1454.5 & 07:08:31.69 & $-$14:54:26.8 & 3$\times$1200\\
  ASAS\_J070911$-$1217.2 & 07:09:10.80 & $-$12:17:09.2 & 3$\times$1800\\
  ASAS\_J072424$-$0751.3 & 07:24:24.52 & $-$07:51:19.6 & 3$\times$1800\\
  ASAS\_J074401$-$1707.8 & 07:44:00.61 & $-$17:07:44.0 & 1$\times$3000+2$\times$3600\\
  ASAS\_J074412$-$1704.9 & 07:44:11.92 & $-$17:04:51.6 & 3$\times$1800\\
  ASAS\_J162326$-$0941.0 & 16:23:26.38 & $-$09:40:59.3 & 3$\times$1200\\
  ASAS\_J180342$-$2211.0 & 18:03:41.74 & $-$22:10:58.7 & 3$\times$1500\\
  ASAS\_J182714$-$1507.1 & 18:27:13.45 & $-$15:07:04.7 & 3$\times$1800\\
  ASAS\_J183347$-$0448.6 & 18:33:46.84 & $-$04:48:32.9 & 3$\times$1200\\
  ASAS\_J183652$-$0907.1 & 18:36:52.35 & $-$09:07:04.8 & 1$\times$1800+2$\times$2700\\
  ASAS\_J183904$-$1049.3 & 18:39:04.03 & $-$10:49:21.2 & 1$\times$1200+2$\times$1800\\
  ASAS\_J192007+1247.7 & 19:20:06.96 & +12:47:43.0 & 3$\times$1500\\
  ASAS\_J192310+1351.4 & 19:23:10.24 & +13:51:24.0 & 3$\times$2700\\
  BD+59\_12 & 00:12:40.0 & +60:13:35.1 & 3$\times$1800\\
  CF\_Cam & 03:35:12.01 & +58:17:40.9 & 3$\times$3600\\
  DR2\_468646563398354176 & 04:06:26.84 & +56:22:57.4 & 3$\times$2400\\
  DR2\_514736269771300224 & 02:15:31.46 & +63:31:04.0 & 3$\times$2400\\
  HD\_160473 & 17:41:08.26 & $-$23:28:27.5 & 3$\times$1500\\
  HD\_344787 & 19:43:29.39 & +23:10:40.6 & 3$\times$1500\\
  HO\_Vul & 19:58:53.51 & +24:23:42.2 & 3$\times$3000\\
  OGLE-GD-CEP-0066 & 06:47:26.46 & +04:11:27.3 & 1$\times$3000+2$\times$3600\\
  OGLE-GD-CEP-0104 & 07:25:41.60 & $-$19:53:48.0 & 3$\times$3600\\
  OO\_Pup & 07:33:38.80 & $-$16:19:01.6 & 3$\times$2700\\
  OP\_Pup & 07:39:21.49 & $-$17:20:49.1 & 3$\times$2700\\
  OR\_Cam & 03:34:23.74 & +58:24:50.0 & 3$\times$2100\\
  TX\_Sct & 18:28:14.50 & $-$11:15:08.8 & 3$\times$1800\\
  V1495\_Aql & 18:53:17.88 & $-$00:06:27.3 & 3$\times$2400\\
  V1496\_Aql & 18:54:59.53 & $-$00:04:36.4 & 3$\times$1200\\
  V1788\_Cyg & 20:42:37.18 & +38:27:25.4 & 3$\times$3600\\
  V2475\_Cyg & 20:13:56.21 & +35:19:41.4 & 3$\times$3600\\
  V355\_Sge & 19:35:44.75 & +18:56:42.6 & 3$\times$1800\\
  V363\_Cas & 00:15:14.33 & +60:20:25.7 & 3$\times$1800\\
  V371\_Gem & 06:10:19.36 & +24:01:15.1 & 3$\times$1800\\
  V383\_Cyg & 20:28:58.16 & +34:08:06.4 & 3$\times$2700\\
  V389\_Sct & 18:27:35.88 & $-$11:16:42.9 & 3$\times$2100\\
  V536\_Ser & 18:11:30.59 & $-$15:55:33.5 & 3$\times$1800\\
  V5567\_Sgr & 18:21:05.53 & $-$18:27:19.7 & 3$\times$1200\\
  V598\_Per & 03:11:07.34 & +55:30:29.7 & 3$\times$3600\\
  V824\_Cas & 00:22:31.30 & +63:01:59.1 & 3$\times$2400\\
  V912\_Aql & 18:48:16.32 & +00:49:03.6 & 3$\times$1800\\
  V914\_Mon & 06:45:53.32 & +10:03:41.3 & 3$\times$1800\\
  V946\_Cas & 02:44:19.39 & +64:45:57.5 & 3$\times$3000\\
  V966\_Mon & 06:27:34.78 & +09:49:49.7 & 3$\times$3600\\
  X\_Sct & 18:31:19.74 & $-$13:06:29.4 & 3$\times$1200\\
  ZTF\_J000234.99+650517.9 & 00:02:34.99 & +65:05:17.9 & 3$\times$2400\\
            \noalign{\smallskip}
\hline\end{tabular}
\end{table*}

\section{Introduction}

Classical Cepheids (DCEPs) represent a fundamental step of the extra-galactic distance ladder thanks to the Period--Luminosity (PL) and Period--Wesenheit (PW) relations that holds for these objects \citep[see e.g.][]{Leavitt1912,Madore1982,Caputo2000,Riess2016}. 
These relations are usually calibrated by means of geometric methods such as trigonometric parallaxes, eclipsing binaries and water masers, and, in turn, used to calibrate secondary distance indicators, including the Type Ia Supernovae (SNIa), which are sufficiently intrinsically bright to allow measuring the distances of far galaxies in the unperturbed Hubble flow. 
The Hubble constant (H$_0$) is then estimated by measuring the slope of the relation between the distance to galaxies in the Hubble flow and their recession velocities. This three-step path has long be known as the distance ladder, and used in the last decades to estimate the H$_0$ value \citep[e.g.][]{Sandage2006,Freedman2012,Riess2016}. 

The actual value of H$_0$ has become a matter of hot debate in the last years, as its value estimated by means of the distance ladder \citep[see e.g.][and references therein]{Riess2020} is significantly  discrepant with respect to the H$_0$ value estimated by the Planck Cosmic Microwave Background (CMB) project adopting the flat $\Lambda$ Cold Dark Matter ($\Lambda$CDM) model \citep[e.g.][and references therein]{Freedman2017, Verde2019}. This occurrence did not change even with the recent Early Data Release 3 of the Gaia mission \citep[EDR3][]{Gaia2016,Gaia2021}. Indeed, the latest estimate by the group led by A. Riess is H$_0$=73.2$\pm$1.3 km s$^{-1}$ Mpc$^{-1}$ \citep[][]{Riess2021}, in tension by  4.2$\sigma$   with the Planck flat $\Lambda$CMB value of H$_0$=67.4$\pm$0.5 km s$^{-1}$ Mpc$^{-1}$ \citep{Planck2018}. 

To better quantify and characterize this tension it is mandatory to investigate the remaining systematics concerning the various paths to estimate H$_0$. In particular, it is critical to reduce both the random and the systematic errors, in each rung of the distance ladder calibration. In this way it will be possible to enhance the accuracy of the estimated H$_0$ value and, in turn, to put rigorous constraints on the different cosmological models. 

In this context, an enduring source of uncertainty in the cosmic distance scale is represented by the  dependence on the metallicity of the Cepheid PL/PW relations which are the standard candles used to calibrate the secondary distance indicators such as the SNIa. 
Indeed, the metal content of DCEPs is thought to affect both the slope and intercept of their $PL/PW$ relations, especially in the optical filters, whereas this dependence is expected to be mild for various band combinations involving near-infrared (NIR) filters \citep[see e.g.][and references therein]{Fiorentino2007,Ngeow2012b,Dicriscienzo2013,Fiorentino2013,Gieren2018}.  However, the metallicity dependence should be considered to mitigate the systematic effects in the definition of the cosmic distance ladder \citep[][and references therein]{Romaniello2008,Bono2010}. 

Previous estimates for the metallicity dependence of the PL/PW relations published in the literature differ significantly from each other  \citep[e.g.][]{Macri2006,Romaniello2008,Bono2010,Freedman2011,Shappee2011,Pejcha2012,Groenewegen2013,Kodric2013,Fausnaugh2015,Riess2016}. These diverse results are mainly caused by the significant uncertainties still affecting metal abundance measures for DCEPs beyond the Milky Way (MW). The situation has completely changed with the advent of the Gaia mission, whose astrometric measures allow us to measure locally, in the MW the metallicity dependence of the $PL/PW$ relations.

The first such an attempt was carried out by \citet{Groenewegen2018} (G18 hereafter) who presented an analysis of the metallicity dependence of Galactic DCEPs PL/PW relations in the NIR based on {\it Gaia} data release 2 \citep[DR2][]{Gaia2016,Gaia2018} parallaxes. This relatively large sample of Galactic DCEPs (more than 400) also has NIR photometry and individual [Fe/H] measurements obtained from  high-resolution spectroscopy. As result, G18 did not find a metallicity dependence of the intercept of the NIR PL/PW better than 1$\sigma$ significance, mainly due to the still too imprecise DR2 parallaxes and uncertain Gaia zero point offset (ZPO) of the parallaxes themselves. A similar result was obtained by \citet{Ripepi2019} in the {\it Gaia} band-passes. More recently, \citet{Ripepi2020a} analysed the metallicity dependence of the NIR PL/PW relations for Galactic DCEPs using a sample similar to that by G18. They found indications of a dependence of not only the intercept of the NIR PL/PW relations, but, for the first time, also of the slopes of these relations. However, the statistical significance of these findings was inconclusive. 

More recently, the SH0ES team \citep{Riess2016,Riess2021} used a sample of 75 DCEPs with accurate Hubble Space Telescope (HST) photometry \citep{Riess2018} and Gaia EDR3 parallaxes, to derive a $PW$ relation in the $(H,V,I)$ bands taking into account a metallicity term on the intercept, equal to -0.18$\pm$0.13 or -0.20$\pm$0.13 mag/dex (hence slightly more significant than 1$\sigma$) and in agreement with the value used by the previous works of the group, i.e. -0.17 mag/dex. The large error is likely due to the small sample adopted for that study and the modest metallicity interval spanned by the adopted DCEP sample.
A similar result for the intercept metallicity dependence was obtained by \citet[][]{Breuval2021} which combined three samples of DCEPs: MW, Large Magellanic Cloud (LMC) and Small Magellanic Cloud (SMC) to estimate the metallicity term of the $PL$ relation. To this aim they used the accurate geometric LMC and SMC distances based on late-type detached eclipsing binary systems (DEBs) recently published by \citet{Pietrzynski2019} and \citet{Graczyk2020}, respectively. In the MW, they adopted DCEPs  parallaxes from the EDR3. \citet[][]{Breuval2021} reported a metallicity effect of $-$0.221$\pm$0.051 mag/dex on the intercept of the $PL$ in the $K_s$ band. This metallicity effect estimate is not entirely based on the DCEP properties but mixes DCEPs and DEBs as standard candles.
Both the above-quoted works assumed that the metallicity affects only the intercept of the DCEPs $PL$/$PW$ relations, but this is  a hypothesis not corroborated by the theoretical results as mentioned above.   

In this context, we are carrying out a long term program dubbed “Cepheid–Metallicity in the Leavitt Law” (C-MetaLL) devoted to the direct measure of the metallicity dependence of the DCEPs $PL$/$PW$ relations using only MW pulsators with accurate metallicities from high-resolution spectroscopy, optical and NIR photometry, and Gaia parallaxes. 
The main scope of this project is to enlarge by more than 50\% the number of DCEPs with metallicities from high-resolution spectroscopy and NIR photometry than currently available in the literature. To reach our goal, we are gathering  high-resolution spectra of Galactic DCEPs with HARPS-N@TNG\footnote{High Accuracy Radial velocity Planet Searcher - North at Telescopio Nazionale Galileo: http://www.tng.iac.es/instruments/harps/}, UVES@VLT\footnote{Ultraviolet and Visual Echelle Spectrograph at Very Large Telescope: https://www.eso.org/sci/facilities/paranal/instruments/uves.html} and PEPSI@LBT\footnote{Potsdam Echelle Polarimetric and Spectroscopic Instrument at Large Binocular Telescope: https://pepsi.aip.de/}, as well as  NIR photometry with the REM telescope\footnote{Rapid Eye Mount telescope: https://www.eso.org/public/italy/teles-instr/lasilla/rem/}. In particular, we aim at increasing the metallicity range of observed DCEPs towards the metal-poor ([Fe/H]$<-0.3-0.4$ dex) regime where only a dozen DCEPs exist with literature measures (see Sect. 5). 

In this paper we present the first results of the C-MetaLL project, presenting high-resolution spectroscopy for a sample of 47 DCEPs and 1 BL Her (BLHER in the following)\footnote{As we show in the following, star ASAS\_J162326-0941.0, formerly classified as DCEP in the literature, turned out to be a BLHER variable.} observed with HARPS-N@TNG. This sample is complemented by literature data to estimate new $PL$-Metallicity ($PLZ$) and $PW$-Metallicity ($PWZ$) relations. 

The paper is organized as follows: 
in Sect. 2 we describe  observations and data analysis; in Sect. 3 we derive atmospheric parameters for the program stars; in Sect. 4 we characterise the program stars from the photometric point of view while in Sect. 5 we merge our sample with literature data; in Sect.6 we discuss the astrometric properties of our and the literature samples; in Sect. 7 we derive the $PLZ$/$PWZ$ relations; in Sect. 8 we discuss our results. Finally, in Sect. 9 we summarize the main findings of this work.

%


\begin{table*}
\caption{Atmospheric parameters. For each spectrum of our targets we report: identification (column 1); heliocentric Julian date at mid exposure (column 2), pulsation phase (column 3), 
effective temperature (column 4), gravity (column 5), microturbulent, broadening parameter and radial velocity (columns 6, 7 and 8).}
\label{table_sum_spectro}
\begin{tabular}{lcccccrr}
\hline 
\hline
~~~~~~ID                        &      HJD   & $\phi$ &T$_{\rm eff}$ & $\log g$ &    $\xi$      & v$_{br}$~~~~      & v$_{\rm rad}$~~~~~~ \\
                                & 2400000.+  &        & (K)          &          & (km s$^{-1}$) & (km s$^{-1}$) & (km s$^{-1}$)  \\
\hline
ASASSN\,J180946.70              & 58648.6242 & 0.872 & 5860 $\pm$ 170 & 1.9 $\pm$ 0.3 & 3.8 $\pm$ 0.3 & 15 $\pm$ 2 &  31.7 $\pm$ 0.3  \\
~~~~~~~~~~~~~~~~~-182238.2      & 58662.5124 & 0.777 & 5480 $\pm$ 200 & 1.6 $\pm$ 0.3 & 4.3 $\pm$ 0.3 & 18 $\pm$ 3 &  51.2 $\pm$ 0.6  \\
                                & 58677.5251 & 0.836 & 5750 $\pm$ 240 & 1.5 $\pm$ 0.4 & 3.8 $\pm$ 0.2 & 18 $\pm$ 2 &  41.0 $\pm$ 0.3  \\
\hline                                                                                                                               
ASAS\,J052610+1151.3            & 58809.5630 & 0.366 & 5650 $\pm$ 220 & 1.2 $\pm$ 0.1 & 2.5 $\pm$ 0.3 & 13 $\pm$ 1 &  41.3 $\pm$ 0.1   \\
                                & 58814.4975 & 0.532 & 5590 $\pm$ 120 & 1.3 $\pm$ 0.1 & 2.6 $\pm$ 0.3 & 13 $\pm$ 1 &  50.8 $\pm$ 0.1   \\
                                & 58837.4555 & 0.957 & 6390 $\pm$ 310 & 1.9 $\pm$ 0.1 & 2.8 $\pm$ 0.3 & 13 $\pm$ 1 &  25.2 $\pm$ 0.2   \\
\hline                                                                                                                               
ASAS\,J061022+1438.6            & 58809.6053 & 0.608 & 5860 $\pm$ 140 & 1.0 $\pm$ 0.1 & 2.6 $\pm$ 0.3 & 13 $\pm$ 1 &  56.6 $\pm$ 0.1   \\
                                & 58814.5464 & 0.629 & 5900 $\pm$ 150 & 1.1 $\pm$ 0.1 & 2.7 $\pm$ 0.3 & 13 $\pm$ 1 &  56.9 $\pm$ 0.1   \\
                                & 58837.4976 & 0.373 & 5860 $\pm$ 170 & 0.9 $\pm$ 0.1 & 2.7 $\pm$ 0.3 & 13 $\pm$ 1 &  51.2 $\pm$ 0.1   \\
\hline                                                                                                                               
ASAS\,J063519+2117.8            & 59152.6212 & 0.418 & 6140 $\pm$ 180 & 2.0 $\pm$ 0.1 & 3.7 $\pm$ 0.3 & 15 $\pm$ 2 &  48.3 $\pm$ 0.2 \\
                                & 59171.5498 & 0.934 & 6650 $\pm$ 250 & 2.3 $\pm$ 0.1 & 3.6 $\pm$ 0.2 & 16 $\pm$ 2 &  26.6 $\pm$ 0.2 \\
                                & 59202.7366 & 0.555 & 6040 $\pm$ 210 & 1.5 $\pm$ 0.1 & 3.5 $\pm$ 0.3 & 18 $\pm$ 1 &  52.0 $\pm$ 0.2 \\
\hline                                                                                                                               
ASAS\,J065413+0756.5            & 59152.6940 & 0.408 & 6250 $\pm$ 290 & 2.1 $\pm$ 0.1 & 3.3 $\pm$ 0.3 & 13 $\pm$ 1 &  42.5 $\pm$ 0.1 \\
                                & 59172.6137 & 0.032 & 6540 $\pm$ 200 & 1.9 $\pm$ 0.1 & 3.3 $\pm$ 0.2 & 12 $\pm$ 2 &  20.8 $\pm$ 0.2 \\
                                & 59260.3476 & 0.887 & 6480 $\pm$ 260 & 2.1 $\pm$ 0.1 & 3.3 $\pm$ 0.2 & 14 $\pm$ 1 &  26.6 $\pm$ 0.2 \\
\hline                                                                                                                               
ASAS\,J070832-1454.5            & 58809.6666 & 0.392 & 5970 $\pm$ 190 & 1.0 $\pm$ 0.1 & 2.4 $\pm$ 0.3 & 15 $\pm$ 1 &  61.4 $\pm$ 0.1   \\
                                & 58814.6358 & 0.170 & 6130 $\pm$ 250 & 1.2 $\pm$ 0.1 & 2.6 $\pm$ 0.3 & 15 $\pm$ 1 &  56.8 $\pm$ 0.1   \\
                                & 58837.5954 & 0.764 & 6060 $\pm$ 120 & 1.2 $\pm$ 0.1 & 2.4 $\pm$ 0.3 & 15 $\pm$ 1 &  64.5 $\pm$ 0.2   \\
\hline                                                                                                                               
ASAS\,J070911-1217.2            & 58809.6873 & 0.955 & 6440 $\pm$ 240 & 1.9 $\pm$ 0.1 & 2.5 $\pm$ 0.3 & 18 $\pm$ 1 &  63.4 $\pm$ 0.2   \\
                                & 58814.6554 & 0.015 & 6410 $\pm$ 170 & 1.8 $\pm$ 0.1 & 2.5 $\pm$ 0.3 & 18 $\pm$ 1 &  60.7 $\pm$ 0.2   \\
                                & 58837.6146 & 0.532 & 6170 $\pm$ 180 & 1.8 $\pm$ 0.1 & 2.6 $\pm$ 0.3 & 18 $\pm$ 1 &  76.2 $\pm$ 0.1   \\
\hline                                                                                                                               
ASAS\,J072424-0751.3            & 58809.7098 & 0.027 & 6390 $\pm$ 210 & 2.0 $\pm$ 0.1 & 2.6 $\pm$ 0.3 & 22 $\pm$ 1 &  63.3 $\pm$ 0.3   \\
                                & 58814.6787 & 0.426 & 6030 $\pm$ 280 & 1.7 $\pm$ 0.1 & 2.7 $\pm$ 0.3 & 22 $\pm$ 1 &  78.8 $\pm$ 0.3   \\
                                & 58837.6373 & 0.510 & 6060 $\pm$ 130 & 1.7 $\pm$ 0.1 & 2.5 $\pm$ 0.3 & 22 $\pm$ 1 &  81.2 $\pm$ 0.4   \\
\hline                                                                                                                               
ASAS\,J074401-1707.8            & 59152.7297 & 0.853 & 6180 $\pm$ 250 & 1.5 $\pm$ 0.1 & 3.2 $\pm$ 0.2 & 10 $\pm$ 1 & 108.1 $\pm$ 0.1   \\
                                & 59172.7003 & 0.743 & 6530 $\pm$ 310 & 1.3 $\pm$ 0.1 & 3.2 $\pm$ 0.2 & 12 $\pm$ 1 &  94.9 $\pm$ 0.2   \\
                                & 59264.4987 & 0.818 & 6440 $\pm$ 150 & 1.4 $\pm$ 0.1 & 3.3 $\pm$ 0.2 & 14 $\pm$ 2 &  97.4 $\pm$ 0.2   \\
\hline                                                                                                                               
ASAS\,J074412-1704.9            & 58809.7553 & 0.709 & 5580 $\pm$ 240 & 0.7 $\pm$ 0.1 & 1.8 $\pm$ 0.9 & 19 $\pm$ 1 &  89.7 $\pm$ 0.2   \\
                                & 58814.7347 & 0.768 & 5810 $\pm$ 190 & 1.3 $\pm$ 0.1 & 2.0 $\pm$ 0.9 & 19 $\pm$ 1 &  86.6 $\pm$ 0.2   \\
                                & 58837.6921 & 0.651 & 5550 $\pm$ 140 & 0.6 $\pm$ 0.1 & 1.7 $\pm$ 0.8 & 19 $\pm$ 1 &  90.8 $\pm$ 0.2   \\
\hline                                                                                                                               
ASAS\,J162326-0941.0            & 58648.4844 & 0.675 & 6050 $\pm$ 170 & 1.8 $\pm$ 0.4 & 2.9 $\pm$ 0.1 & 11 $\pm$ 1 &  48.2 $\pm$ 0.2  \\
                                & 58662.4401 & 0.801 & 6170 $\pm$ 230 & 1.8 $\pm$ 0.5 & 2.9 $\pm$ 0.1 & 10 $\pm$ 1 &  40.2 $\pm$ 0.2  \\
                                & 58677.4015 & 0.656 & 6080 $\pm$ 170 & 2.0 $\pm$ 0.4 & 2.8 $\pm$ 0.1 & 11 $\pm$ 1 &  49.5 $\pm$ 0.2  \\
\hline                                                                                                                               
ASAS\,J180342-2211.0            & 59029.5474 & 0.453 & 4820 $\pm$ 180 & 1.5 $\pm$ 0.1 & 4.8 $\pm$ 0.7 & 13 $\pm$ 5 &  13.4 $\pm$ 0.1  \\
                                & 59046.5040 & 0.851 & 4930 $\pm$ 180 & 1.6 $\pm$ 0.1 & 4.5 $\pm$ 0.2 & 20 $\pm$ 3 &  27.7 $\pm$ 1.9  \\
\hline                                                                                                                                
ASAS\,J182714-1507.1            & 59029.5667 & 0.283 & 5600 $\pm$ 260 & 0.7 $\pm$ 0.1 & 2.3 $\pm$ 0.3 & 17 $\pm$ 1 &  20.8 $\pm$ 0.1  \\
                                & 59046.5280 & 0.342 & 5550 $\pm$ 200 & 1.2 $\pm$ 0.1 & 2.6 $\pm$ 0.2 & 17 $\pm$ 1 &  24.1 $\pm$ 0.1  \\
\hline                                                                                                                               
ASAS\,J183347-0448.6            & 58648.5014 & 0.918 & 6450 $\pm$ 170 & 1.7 $\pm$ 0.5 & 3.0 $\pm$ 0.4 & 14 $\pm$ 1 &$-$8.1 $\pm$ 0.3  \\
                                & 58662.5939 & 0.461 & 6120 $\pm$ 180 & 1.5 $\pm$ 0.5 & 2.8 $\pm$ 0.4 & 13 $\pm$ 1 &$-$1.4 $\pm$ 0.2  \\
                                & 58677.4421 & 0.247 & 6280 $\pm$ 180 & 1.1 $\pm$ 0.6 & 2.4 $\pm$ 0.1 & 14 $\pm$ 1 &$-$8.4 $\pm$ 0.2  \\
\hline                                                                                                                               
ASAS\,J183652-0907.1            & 58648.5855 & 0.409 & 5990 $\pm$ 130 & 2.2 $\pm$ 0.3 & 2.5 $\pm$ 0.3 & 10 $\pm$ 1 &$-$3.9 $\pm$ 0.2  \\
                                & 58662.6195 & 0.827 & 6250 $\pm$ 160 & 2.0 $\pm$ 0.5 & 2.8 $\pm$ 0.3 & 15 $\pm$ 1 &$-$6.8 $\pm$ 0.2  \\
                                & 58677.6301 & 0.623 & 6090 $\pm$ 180 & 2.7 $\pm$ 0.4 & 2.5 $\pm$ 0.3 & 13 $\pm$ 1 &$-$0.1 $\pm$ 0.2  \\
\hline                                                                                                                               
ASAS\,J183904-1049.3            & 58648.6433 & 0.132 & 6240 $\pm$ 240 & 1.8 $\pm$ 0.5 & 2.5 $\pm$ 0.3 & 20 $\pm$ 1 &$-$6.9 $\pm$ 0.4  \\
                                & 58662.6544 & 0.714 & 6140 $\pm$ 160 & 1.6 $\pm$ 0.5 & 2.6 $\pm$ 0.3 & 20 $\pm$ 1 &   5.8 $\pm$ 0.3  \\
                                & 58677.5867 & 0.597 & 6090 $\pm$ 150 & 1.7 $\pm$ 0.4 & 2.7 $\pm$ 0.3 & 20 $\pm$ 1 &   9.3 $\pm$ 0.3  \\
 \hline
\end{tabular}
\end{table*}

\begin{table*}
\addtocounter{table}{-1}
\caption{{\it ...continued.}}
\begin{tabular}{lcccccrr}
\hline \hline
~~~~~~ID                        &      HJD   & $\phi$ &T$_{\rm eff}$ & $\log g$ &    $\xi$      & v$_{br}$~~~~   & v$_{\rm rad}$~~~~~~ \\
                                & 2400000.+  &        & (K)          &          & (km s$^{-1}$) & (km s$^{-1}$) & (km s$^{-1}$)  \\
\hline                              
ASAS\,J192007+1247.7            &59029.47931 & 0.619 & 5600 $\pm$ 130 & 0.7 $\pm$ 0.1 & 2.4 $\pm$ 0.1 & 18 $\pm$ 2 &    7.1 $\pm$ 0.3  \\
                                &59046.45261 & 0.586 & 5500 $\pm$ 220 & 0.7 $\pm$ 0.1 & 2.2 $\pm$ 0.1 & 18 $\pm$ 2 &    9.8 $\pm$ 0.1  \\
                                &59053.65998 & 0.421 & 5250 $\pm$ 125 & 0.8 $\pm$ 0.1 & 2.6 $\pm$ 0.2 & 14 $\pm$ 1 &    6.2 $\pm$ 0.1  \\
\hline                                                                                                                                 
ASAS\,J192310+1351.4            &59029.49690 & 0.234 & 6250 $\pm$ 130 & 1.1 $\pm$ 0.1 & 2.1 $\pm$ 0.2 & 12 $\pm$ 2 & $-$2.1 $\pm$ 0.1  \\
                                &59053.57714 & 0.334 & 6250 $\pm$ 220 & 1.5 $\pm$ 0.1 & 2.1 $\pm$ 0.2 & 11 $\pm$ 1 &    1.0 $\pm$ 0.1  \\
\hline                                                                                                                                 
BD+59\,12                       &59152.34090 & 0.772 & 6350 $\pm$ 250 & 1.5 $\pm$ 0.1 & 3.5 $\pm$ 0.3 & 24 $\pm$ 1 &$-$28.8 $\pm$ 0.1  \\
                                &59167.44631 & 0.088 & 6380 $\pm$ 320 & 1.1 $\pm$ 0.1 & 3.3 $\pm$ 0.2 & 22 $\pm$ 2 &$-$44.8 $\pm$ 0.3  \\
                                &59200.33040 & 0.660 & 6270 $\pm$ 180 & 1.5 $\pm$ 0.1 & 3.6 $\pm$ 0.4 & 22 $\pm$ 2 &$-$31.0 $\pm$ 0.1  \\
\hline                                                                                                                                 
CF\,Cam                         &59152.50388 & 0.274 & 6060 $\pm$ 150 & 1.7 $\pm$ 0.1 & 3.2 $\pm$ 0.4 & 10 $\pm$ 1 &$-$39.3 $\pm$ 0.1  \\
                                &59215.53402 & 0.953 & 6400 $\pm$ 154 & 1.9 $\pm$ 0.1 & 3.3 $\pm$ 0.2 & 12 $\pm$ 2 &$-$45.2 $\pm$ 0.1  \\
                                &59259.32798 & 0.595 & 5585 $\pm$ 140 & 0.8 $\pm$ 0.1 & 3.0 $\pm$ 0.2 & 11 $\pm$ 1 &$-$26.5 $\pm$ 0.1  \\
\hline                                                                                                                                 
DR2468646563398354176           &59152.54715 & 0.053 & 6430 $\pm$ 200 & 0.0 $\pm$ 0.1 & 3.1 $\pm$ 0.2 & 11 $\pm$ 1 &   29.1 $\pm$ 0.1  \\
                                &59170.59596 & 0.088 & 6670 $\pm$ 230 & 0.5 $\pm$ 0.1 & 3.3 $\pm$ 0.2 & 11 $\pm$ 2 &   30.0 $\pm$ 0.1  \\
                                &59242.45968 & 0.114 & 6520 $\pm$ 210 & 1.2 $\pm$ 0.1 & 3.1 $\pm$ 0.4 & 10 $\pm$ 1 &   30.5 $\pm$ 0.1  \\
\hline                                                                                                                                 
DR2514736269771300224           &59152.39425 & 0.876 & 6450 $\pm$ 280 & 1.4 $\pm$ 0.1 & 3.7 $\pm$ 0.4 & 14 $\pm$ 1 &$-$66.3 $\pm$ 0.1  \\
                                &59170.30991 & 0.307 & 6050 $\pm$ 190 & 1.1 $\pm$ 0.1 & 3.2 $\pm$ 0.2 &  9 $\pm$ 1 &$-$63.4 $\pm$ 0.1  \\
                                &59185.62302 & 0.949 & 6470 $\pm$ 280 & 1.3 $\pm$ 0.1 & 3.2 $\pm$ 0.2 & 13 $\pm$ 1 &$-$69.6 $\pm$ 0.1  \\
\hline                                                                                                                                 
HD\,160473                      &59029.53160 & 0.732 & 6150 $\pm$ 150 & 1.0 $\pm$ 0.1 & 2.3 $\pm$ 0.2 & 14 $\pm$ 1 &    5.1 $\pm$ 0.1  \\
                                &59046.48443 & 0.217 & 6000 $\pm$  80 & 0.9 $\pm$ 0.1 & 2.5 $\pm$ 0.2 & 12 $\pm$ 1 &    2.9 $\pm$ 0.1  \\
\hline                                                                                                                                 
HO\,Vul                         &59029.61249 & 0.211 & 5950 $\pm$ 290 & 1.9 $\pm$ 0.1 & 2.7 $\pm$ 0.2 & 19 $\pm$ 2 &$-$12.0 $\pm$ 0.1  \\
                                &59046.72463 & 0.250 & 5950 $\pm$ 190 & 1.9 $\pm$ 0.1 & 2.9 $\pm$ 0.5 & 19 $\pm$ 2 & $-$9.2 $\pm$ 0.1  \\
                                &59053.68819 & 0.486 & 5500 $\pm$ 140 & 1.3 $\pm$ 0.1 & 2.6 $\pm$ 0.3 & 19 $\pm$ 2 &    5.8 $\pm$ 0.2  \\
                                &59058.66606 & 0.370 & 5640 $\pm$ 250 & 1.3 $\pm$ 0.1 & 2.5 $\pm$ 0.3 & 20 $\pm$ 2 & $-$1.3 $\pm$ 0.2  \\
\hline                                                                                                                                 
OGLE-GD-CEP-0066                &59152.65764 & 0.186 & 6460 $\pm$ 200 & 1.4 $\pm$ 0.1 & 3.2 $\pm$ 0.2 & 11 $\pm$ 1 &   40.2 $\pm$ 0.1  \\
                                &59172.65751 & 0.835 & 6470 $\pm$ 280 & 1.6 $\pm$ 0.1 & 3.3 $\pm$ 0.2 & 12 $\pm$ 1 &   53.1 $\pm$ 0.2  \\
                                &59259.37833 & 0.342 & 6410 $\pm$ 175 & 1.8 $\pm$ 0.1 & 3.2 $\pm$ 0.2 &  9 $\pm$ 1 &   46.1 $\pm$ 0.1  \\
\hline                                                                                                                                 
OGLE-GD-CEP-0104                &59197.70252 & 0.799 & 6320 $\pm$ 210 & 1.6 $\pm$ 0.1 & 3.4 $\pm$ 0.2 & 22 $\pm$ 1 &   87.2 $\pm$ 0.2  \\
                                &59242.53786 & 0.387 & 6010 $\pm$ 220 & 1.3 $\pm$ 0.1 & 3.1 $\pm$ 0.4 & 16 $\pm$ 1 &   89.5 $\pm$ 0.2  \\
                                &59264.45418 & 0.451 & 6130 $\pm$ 225 & 1.4 $\pm$ 0.1 & 3.0 $\pm$ 0.3 & 16 $\pm$ 1 &   93.0 $\pm$ 0.2  \\
\hline                                                                                                                                 
OO\,Pup                         &59242.58199 & 0.065 & 5820 $\pm$ 210 & 0.7 $\pm$ 0.1 & 3.8 $\pm$ 0.4 & 14 $\pm$ 1 &   81.2 $\pm$ 0.1  \\
                                &59258.52041 & 0.516 & 5205 $\pm$ 160 & 0.9 $\pm$ 0.1 & 4.2 $\pm$ 0.5 & 15 $\pm$ 2 &  109.3 $\pm$ 0.1  \\
                                &59259.42289 & 0.598 & 5460 $\pm$ 200 & 0.9 $\pm$ 0.1 & 4.1 $\pm$ 0.5 & 18 $\pm$ 2 &  111.2 $\pm$ 0.1  \\
                                &59263.50787 & 0.970 & 5880 $\pm$ 145 & 0.9 $\pm$ 0.1 & 3.6 $\pm$ 0.4 & 11 $\pm$ 2 &   85.1 $\pm$ 0.1  \\
                                &59264.54255 & 0.065 & 5900 $\pm$ 275 & 1.2 $\pm$ 0.1 & 3.8 $\pm$ 0.4 & 14 $\pm$ 1 &   81.1 $\pm$ 0.1  \\
\hline                                                                                                                                 
OP\,Pup                         & 58809.7328 & 0.710 & 6230 $\pm$ 160 & 1.6 $\pm$ 0.1 & 2.1 $\pm$ 0.3 & 11 $\pm$ 1 &   76.0 $\pm$ 0.2  \\
                                & 58814.7059 & 0.624 & 6200 $\pm$ 230 & 1.8 $\pm$ 0.1 & 2.2 $\pm$ 0.3 & 11 $\pm$ 1 &   76.0 $\pm$ 0.1  \\
                                & 58837.6642 & 0.458 & 6030 $\pm$ 200 & 1.4 $\pm$ 0.1 & 2.1 $\pm$ 0.3 & 11 $\pm$ 1 &   70.0 $\pm$ 0.1  \\
\hline                                                                                                                                 
OR\,Cam                         & 58809.5398 & 0.272 & 6000 $\pm$ 200 & 1.3 $\pm$ 0.1 & 2.2 $\pm$ 0.3 & 15 $\pm$ 1 &$-$44.1 $\pm$ 0.1  \\
                                & 58814.5224 & 0.624 & 5960 $\pm$ 140 & 1.5 $\pm$ 0.1 & 2.5 $\pm$ 0.3 & 15 $\pm$ 1 &$-$34.5 $\pm$ 0.1  \\
                                & 58837.4260 & 0.839 & 6180 $\pm$ 270 & 1.6 $\pm$ 0.1 & 2.6 $\pm$ 0.3 & 15 $\pm$ 1 &$-$48.7 $\pm$ 0.1  \\
\hline                                                                                                                                 
TX\,Sct                         & 58648.5236 & 0.949 & 5730 $\pm$ 170 & 1.4 $\pm$ 0.3 & 3.5 $\pm$ 0.3 & 14 $\pm$ 1 &   18.0 $\pm$ 0.2  \\
                                & 58662.5583 & 0.525 & 4800 $\pm$ 150 & 1.1 $\pm$ 0.1 & 4.1 $\pm$ 0.4 & 13 $\pm$ 1 &   45.9 $\pm$ 0.6  \\
                                & 58677.5638 & 0.142 & 5200 $\pm$ 170 & 1.0 $\pm$ 0.3 & 3.5 $\pm$ 0.3 & 12 $\pm$ 1 &   18.3 $\pm$ 0.5  \\
\hline                                                                                                                                 
V1495\,Aql                      & 59029.4584 & 0.113 & 5750 $\pm$ 200 & 0.4 $\pm$ 0.1 & 2.6 $\pm$ 0.2 & 14 $\pm$ 1 &   37.1 $\pm$ 0.1  \\
                                & 59046.5736 & 0.058 & 6000 $\pm$ 190 & 0.2 $\pm$ 0.1 & 2.5 $\pm$ 0.2 & 13 $\pm$ 1 &   36.2 $\pm$ 0.1  \\
                                & 59053.6351 & 0.861 & 6000 $\pm$ 230 & 0.7 $\pm$ 0.1 & 2.7 $\pm$ 0.2 & 13 $\pm$ 1 &   33.8 $\pm$ 0.1  \\
\hline                                                                                                                                 
V1496\,Aql                      & 58648.6816 & 0.350 & 5200 $\pm$ 140 & 0.2 $\pm$ 0.3 & 4.4 $\pm$ 0.5 & 13 $\pm$ 2 &   78.4 $\pm$ 0.3  \\
                                & 58662.6680 & 0.564 & 5050 $\pm$ 150 & 0.1 $\pm$ 0.4 & 4.2 $\pm$ 0.5 & 14 $\pm$ 2 &   87.0 $\pm$ 0.5  \\
                                & 58677.4830 & 0.792 & 5230 $\pm$ 240 & 0.3 $\pm$ 0.4 & 4.3 $\pm$ 0.6 & 16 $\pm$ 3 &   93.8 $\pm$ 0.7  \\
\hline                                                                                                                                 
V1788\,Cyg                      & 59029.6752 & 0.412 & 4950 $\pm$ 150 & 0.9 $\pm$ 0.1 & 4.5 $\pm$ 0.3 & 10 $\pm$ 2 &$-$13.1 $\pm$ 0.1  \\
                                & 59046.6059 & 0.613 & 4800 $\pm$ 120 & 1.0 $\pm$ 0.1 & 5.0 $\pm$ 0.5 & 10 $\pm$ 2 &    4.0 $\pm$ 0.2  \\
                                & 59053.5364 & 0.105 & 5650 $\pm$ 190 & 1.1 $\pm$ 0.1 & 2.9 $\pm$ 0.2 & 16 $\pm$ 1 &$-$40.7 $\pm$ 0.2  \\
                                & 59066.7076 & 0.040 & 6000 $\pm$ 230 & 1.3 $\pm$ 0.1 & 3.1 $\pm$ 0.2 & 18 $\pm$ 2 &$-$40.5 $\pm$ 0.2  \\
\hline
\end{tabular}
\end{table*}

\begin{table*}
\addtocounter{table}{-1}
\caption{{\it ...continued.}}
\begin{tabular}{lcccccrr}
\hline \hline
~~~~~~ID                        &      HJD   & $\phi$ &T$_{\rm eff}$ & $\log g$ &    $\xi$      & v$_{br}$~~~~   & v$_{\rm rad}$~~~~~~\\
                                & 2400000.+  &        & (K)          &          & (km s$^{-1}$) & (km s$^{-1}$) & (km s$^{-1}$)    \\  
\hline                                                                   
\hline
V2475\,Cyg                      & 59029.6422 & 0.926 & 6180 $\pm$ 130 & 1.4 $\pm$ 0.1 & 3.6 $\pm$ 0.3 & 13 $\pm$ 2 &$-$12.3 $\pm$ 0.1 \\
                                & 59053.4596 & 0.987 & 6450 $\pm$ 190 & 0.2 $\pm$ 0.1 & 2.5 $\pm$ 0.2 & 13 $\pm$ 1 &$-$16.2 $\pm$ 0.3 \\
                                & 59102.3996 & 0.222 & 5700 $\pm$ 250 & 0.9 $\pm$ 0.1 & 2.9 $\pm$ 0.3 & 12 $\pm$ 1 &$-$19.0 $\pm$ 0.1 \\
\hline
V355\,Sge                       & 58648.7036 & 0.579 & 4730 $\pm$ 140 & 0.9 $\pm$ 0.2 & 4.1 $\pm$ 0.3 & 14 $\pm$ 2 &   46.0 $\pm$ 1.0  \\
                                & 58662.6872 & 0.015 & 5340 $\pm$ 120 & 0.7 $\pm$ 0.3 & 3.4 $\pm$ 0.2 & 15 $\pm$ 1 &   16.4 $\pm$ 0.4  \\
                                & 58677.4521 & 0.475 & 4730 $\pm$ 100 & 0.7 $\pm$ 0.1 & 3.6 $\pm$ 0.5 & 12 $\pm$ 2 &   40.6 $\pm$ 0.6  \\
\hline
V371\,Gem                       & 58809.5859 & 0.154 & 6500 $\pm$ 200 & 1.9 $\pm$ 0.1 & 2.2 $\pm$ 0.3 & 10 $\pm$ 1 &   21.5 $\pm$ 0.2   \\
                                & 58814.5662 & 0.484 & 6150 $\pm$ 160 & 1.6 $\pm$ 0.1 & 2.3 $\pm$ 0.3 & 10 $\pm$ 1 &   33.6 $\pm$ 0.1   \\
                                & 58837.4785 & 0.203 & 6420 $\pm$ 300 & 1.7 $\pm$ 0.1 & 2.4 $\pm$ 0.3 & 10 $\pm$ 1 &   23.7 $\pm$ 0.1   \\
\hline
V383\,Cyg                       & 59029.7032 & 0.402 & 5550 $\pm$ 270 & 1.3 $\pm$ 0.1 & 2.9 $\pm$ 0.5 & 18 $\pm$ 2 & $-$9.3 $\pm$ 0.1  \\
                                & 59053.4980 & 0.090 & 5950 $\pm$ 280 & 1.6 $\pm$ 0.1 & 2.9 $\pm$ 0.2 & 18 $\pm$ 2 &$-$15.3 $\pm$ 0.1  \\
                                & 59067.6913 & 0.098 & 6270 $\pm$ 170 & 1.6 $\pm$ 0.1 & 2.9 $\pm$ 0.2 & 16 $\pm$ 1 &$-$32.1 $\pm$ 0.1  \\
\hline
V389\,Sct                       & 59029.5895 & 0.997 & 6300 $\pm$ 300 & 0.3 $\pm$ 0.1 & 2.2 $\pm$ 0.2 & 11 $\pm$ 1 &   25.6 $\pm$ 0.1  \\
                                & 59046.5486 & 0.467 & 6000 $\pm$ 230 & 1.0 $\pm$ 0.1 & 2.4 $\pm$ 0.2 & 10 $\pm$ 1 &   39.3 $\pm$ 0.1  \\
                                & 59053.6068 & 0.328 & 6000 $\pm$ 230 & 0.8 $\pm$ 0.1 & 2.4 $\pm$ 0.2 & 10 $\pm$ 1 &   33.2 $\pm$ 0.1  \\
\hline                                                                                                                
V536\,Ser                       & 58648.5635 & 0.034 & 5740 $\pm$ 100 & 1.0 $\pm$ 0.4 & 4.2 $\pm$ 0.3 & 12 $\pm$ 1 &    1.2 $\pm$ 0.2  \\
                                & 58662.4887 & 0.579 & 5340 $\pm$ 120 & 1.2 $\pm$ 0.3 & 3.7 $\pm$ 0.1 & 12 $\pm$ 2 &   24.1 $\pm$ 0.5  \\
                                & 58677.5026 & 0.246 & 4730 $\pm$ 100 & 0.6 $\pm$ 0.1 & 3.4 $\pm$ 0.9 & 12 $\pm$ 1 &   40.6 $\pm$ 0.5  \\
\hline
V598\,Per                       & 59152.4608 & 0.752 & 5540 $\pm$ 140 & 2.4 $\pm$ 0.1 & 4.4 $\pm$ 0.5 & 16 $\pm$ 2 & $-$2.8 $\pm$ 0.1 \\
                                & 59200.4233 & 0.214 & 6040 $\pm$ 200 & 1.1 $\pm$ 0.1 & 3.2 $\pm$ 0.3 & 13 $\pm$ 1 &$-$33.8 $\pm$ 0.1 \\
                                & 59242.4884 & 0.635 & 5510 $\pm$ 140 & 1.0 $\pm$ 0.1 & 3.5 $\pm$ 0.4 & 14 $\pm$ 1 &$-$11.6 $\pm$ 0.1 \\
\hline
V824\,Cas                       & 59152.3638 & 0.427 & 5970 $\pm$ 180 & 0.8 $\pm$ 0.1 & 3.4 $\pm$ 0.2 & 11 $\pm$ 1 &$-$68.6 $\pm$ 0.1 \\
                                & 59167.4690 & 0.250 & 6160 $\pm$ 190 & 0.9 $\pm$ 0.1 & 3.5 $\pm$ 0.2 & 11 $\pm$ 1 &$-$76.0 $\pm$ 0.1 \\
                                & 59200.3533 & 0.396 & 6020 $\pm$ 220 & 0.8 $\pm$ 0.1 & 3.5 $\pm$ 0.2 & 10 $\pm$ 1 &$-$69.9 $\pm$ 0.1 \\
\hline
V912\,Aql                       & 58648.6623 & 0.452 & 5620 $\pm$ 150 & 2.0 $\pm$ 0.2 & 2.9 $\pm$ 0.3 &  8 $\pm$ 1 &   14.4 $\pm$ 0.2  \\
                                & 58662.4650 & 0.589 & 5570 $\pm$ 100 & 1.6 $\pm$ 0.3 & 3.3 $\pm$ 0.3 & 10 $\pm$ 1 &   21.8 $\pm$ 0.2  \\
                                & 58677.4225 & 0.988 & 6500 $\pm$ 200 & 1.6 $\pm$ 0.5 & 2.5 $\pm$ 0.3 & 14 $\pm$ 1 & $-$5.5 $\pm$ 0.2  \\
\hline                                                                                                                                
V914\,Mon                       & 58809.6472 & 0.069 & 6130 $\pm$ 170 & 1.4 $\pm$ 0.1 & 2.5 $\pm$ 0.3 &~~9 $\pm$ 1 &   44.5 $\pm$ 0.1   \\
                                & 58814.6125 & 0.921 & 6210 $\pm$ 220 & 1.3 $\pm$ 0.1 & 2.5 $\pm$ 0.3 &~~9 $\pm$ 1 &   38.5 $\pm$ 0.2   \\
                                & 58837.5770 & 0.488 & 6360 $\pm$ 260 & 1.7 $\pm$ 0.1 & 2.5 $\pm$ 0.3 &~~9 $\pm$ 1 &   45.9 $\pm$ 0.2   \\
\hline
V946\,Cas                       & 59152.4241 & 0.531 & 5610 $\pm$ 270 & 0.6 $\pm$ 0.1 & 3.6 $\pm$ 0.5 & 12 $\pm$ 1 & $-$4.3 $\pm$ 0.1 \\
                                & 59170.3837 & 0.768 & 5820 $\pm$ 240 & 1.1 $\pm$ 0.1 & 4.0 $\pm$ 0.3 & 18 $\pm$ 2 &    0.7 $\pm$ 0.1 \\
                                & 59200.3861 & 0.846 & 6150 $\pm$ 230 & 1.7 $\pm$ 0.1 & 4.2 $\pm$ 0.4 & 18 $\pm$ 2 & $-$7.6 $\pm$ 0.1 \\
\hline
V966\,Mon                       & 59187.7141 & 0.404 & 5580 $\pm$ 150 & 1.2 $\pm$ 0.1 & 2.5 $\pm$ 0.2 & 13 $\pm$ 2 &   51.4 $\pm$ 0.1 \\
                                & 59216.6039 & 0.453 & 5530 $\pm$ 230 & 1.4 $\pm$ 0.1 & 2.7 $\pm$ 0.2 & 13 $\pm$ 2 &   53.4 $\pm$ 0.1 \\
                                & 59242.6160 & 0.808 & 5570 $\pm$ 240 & 0.8 $\pm$ 0.1 & 2.7 $\pm$ 0.4 & 15 $\pm$ 3 &   50.2 $\pm$ 0.2 \\
\hline                                                                                                               
V5567\,Sgr                      & 58648.6052 & 0.103 & 5670 $\pm$~~80 & 1.5 $\pm$ 0.4 & 3.4 $\pm$ 0.4 & 12 $\pm$ 1 & $-$1.4 $\pm$ 0.2  \\
                                & 58662.5328 & 0.529 & 5400 $\pm$ 160 & 1.7 $\pm$ 0.3 & 3.7 $\pm$ 0.4 & 15 $\pm$ 2 &   24.8 $\pm$ 0.2  \\
                                & 58677.5445 & 0.067 & 5700 $\pm$ 140 & 2.1 $\pm$ 0.3 & 3.2 $\pm$ 0.3 &  8 $\pm$ 1 &    0.1 $\pm$ 0.1  \\
\hline
X\,Sct                          & 58648.5420 & 0.690 & 5750 $\pm$ 200 & 2.1 $\pm$ 0.3 & 3.2 $\pm$ 0.3 & 18 $\pm$ 2 &   28.5 $\pm$ 0.3  \\
                                & 58662.5769 & 0.033 & 6700 $\pm$ 240 & 2.1 $\pm$ 0.4 & 2.8 $\pm$ 0.3 & 15 $\pm$ 2 & $-$8.2 $\pm$ 0.3  \\
                                & 58677.6049 & 0.613 & 5700 $\pm$ 220 & 2.1 $\pm$ 0.3 & 3.0 $\pm$ 0.3 & 17 $\pm$ 1 &   24.5 $\pm$ 0.2  \\ 
\hline                                                                                                                                
ZTF\,J000234.99+650517.9        & 59152.3112 & 0.676 & 5360 $\pm$ 210 & 0.3 $\pm$ 0.1 & 4.3 $\pm$ 0.8 & 16 $\pm$ 1 &$-$56.7 $\pm$ 0.1 \\
                                & 59167.4165 & 0.058 & 5400 $\pm$ 230 & 0.7 $\pm$ 0.1 & 4.1 $\pm$ 0.5 & 13 $\pm$ 1 &$-$70.8 $\pm$ 0.1 \\
                                & 59200.3003 & 0.066 & 5480 $\pm$ 200 & 0.6 $\pm$ 0.1 & 4.1 $\pm$ 0.8 & 13 $\pm$ 1 &$-$71.0 $\pm$ 0.1 \\
\hline
\end{tabular}
\end{table*}


\section{Observations and data reduction}

Multi-phase spectroscopic observations for our targets 
 were obtained at the 3.5m Telescopio Nazionale Galileo (TNG) equipped with the HARPS-N instrument, during four different periods between summer 2019 and fall 2020 (TNG AOT 39 to AOT 42). 
HARPS-N features an echelle spectrograph covering the wavelength range between 3830 to 6930 {\AA}, with a spectral resolution R=115,000. We collected at least 3 spectra for each source. 
The signal-to-noise ratio (SNR) varies from 50 to 100 at $\lambda$ = 5000 {\AA}. 
Coordinates of our targets and the observation log are provided in Table~\ref{Tab:log}. 

Reduction of all the spectra, which included  bias subtraction, spectrum extraction, flat fielding and wavelength calibration,  was done using the HARPS-N reduction pipeline. 
Radial velocities were  measured by cross-correlating each spectrum with a synthetic template. The cross-correlation was performed using the IRAF task {\it FXCOR} and excluding Balmer lines as well as wavelength ranges  with telluric lines. The IRAF package RVCORRECT was used to determine the heliocentric velocity by  correcting the spectra for the Earth’s motion.

    
\section{Abundance analysis}
\label{spectro}
\subsection{Stellar parameters}
\label{sect:param}
To measure the iron abundance and,
more in general, the chemical pattern of our targets, we first need to estimate the main atmospheric parameters such as effective temperature (T$_{\rm eff}$), surface gravity ($\log g$), microturbulent ($\xi$) and line broadening parameter (v$_{br}$), which represents the combined effects of macroturbulence and rotational velocity (though, in the case of DCEPs, it is essentially dominated by macroturbulence).

An effective temperature for each spectrum of our targets was estimated by using the line depth ratios (LDRs) method \citep{Kovtyukh2000}.  LDRs  have the advantage of being sensitive to temperature variations, but not to abundances and interstellar reddening. Typically, we measured about 32 LDRs in each spectrum. 


An iterative procedure was then applied to determine microturbulent velocities, iron abundances and gravities. The $\xi$ values were estimated by demanding  the slope of [\ion {Fe}{I}/H] as a function of equivalent width (EW) be null, that is, the iron abundance does not depend on EWs. For this purpose we used a sample of 145 \ion{Fe}{I} spectral lines, extracted from the line list published by \citet{Romaniello2008}.
The iron content was  estimated by converting the measured EWs into abundances through the WIDTH9 code  \citep{kur81}, after 
generating the appropriate model atmosphere with ATLAS9. At this stage, we neglected the 
$\log g$, as the \ion{Fe}{I} lines are insensitive to this parameter. EWs were  measured using an {\it IDL}\footnote{IDL (Interactive Data Language) is a registered trademark of Harris Geospatial Solutions} semi-automatic custom routine which allowed us to minimize errors in 
the continuum estimation on the wings of the spectral lines.
Then, we estimated the surface gravity by imposing the ionization balance between \ion{Fe}{I} and \ion{Fe}{II} lines. For the  
\ion{Fe}{II} 
we used a list of 24 lines extracted from the compilation by \citet{Romaniello2008}.

The atmospheric parameters estimated as described above  are summarized in Table~\ref{table_sum_spectro}.  They were used as input values for the abundance analysis presented in the next section. In particular, for  the iron abundance  we used the weighted average of the values derived from the individual spectra available for each target. 


 \begin{figure*}
   \centering
   \includegraphics[width=\hsize]{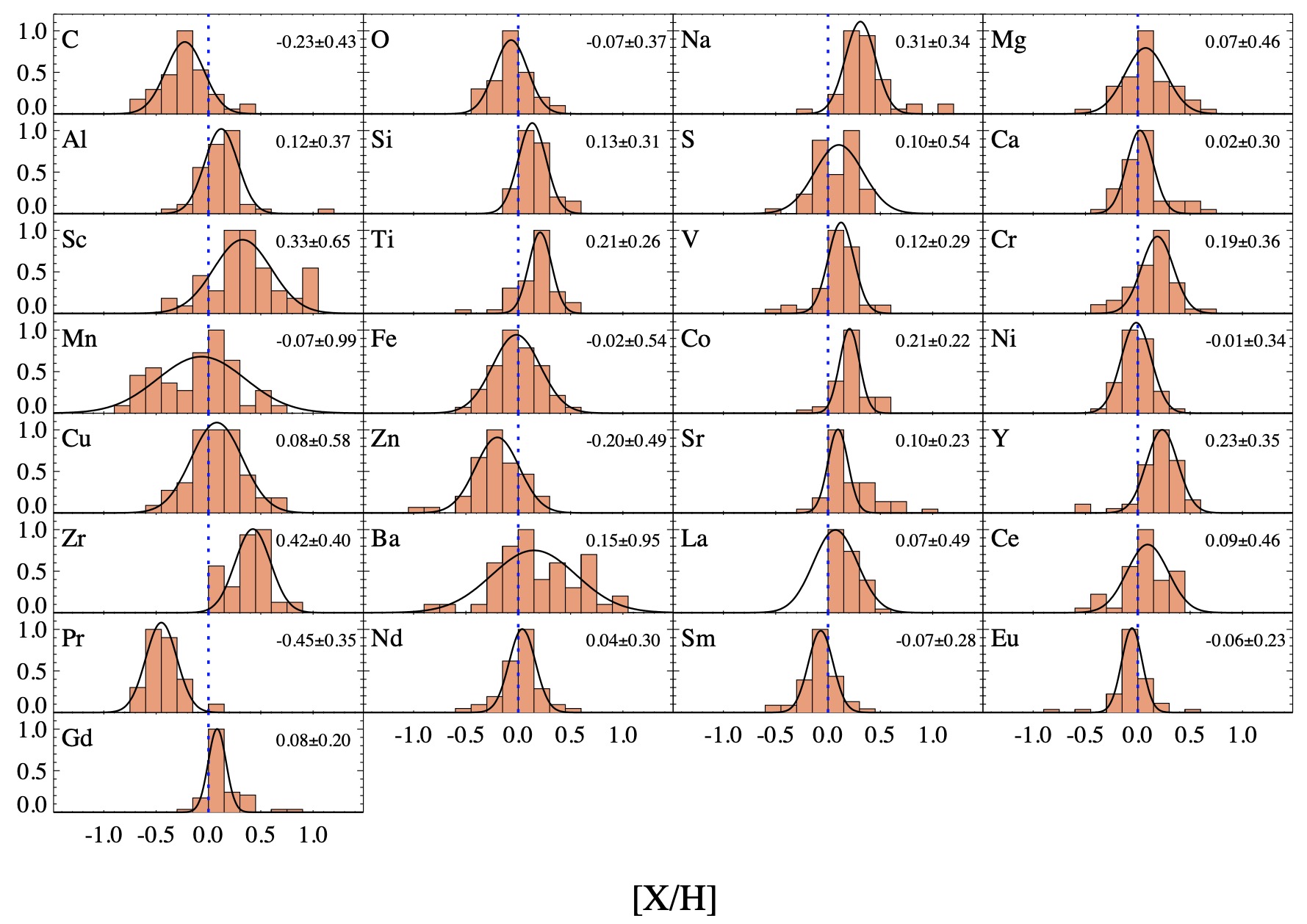}
   \caption{Histograms of the abundance distribution for each chemical element. In each panel, a Gaussian fit of the abundances has been over-plotted (solid line) and the blue dotted line indicates the solar value. For clarity, histograms have been normalized to the unity. The center of the Gaussian and its FWHM are also reported in the top-right corner of each panel. } 
   \label{fig:histograms}
    \end{figure*}

\subsection{Abundances}

The stellar parameters in  Table~\ref{table_sum_spectro} were used to compute stellar atmosphere models and synthetic spectra to analyze the spectroscopic observations of our targets.
We applied a spectral synthesis technique to avoid problems arising  from spectral line  blending caused by  line broadening. Synthetic spectra were generated in three steps: i)  plane-parallel local thermodynamic equilibrium (LTE) atmosphere models were computed using the ATLAS9 code \citep{kur93,kur93b}; ii)  stellar spectra were synthesized by using SYNTHE \citep{kur81}; iii) the synthetic spectra were convoluted for instrumental and line broadening. This was  evaluated by matching the synthetic line profiles to a selected set of the observed metal lines.

In our targets we detected spectral lines for 29 different chemical species. For all elements we performed the following analysis: we divided the observed spectra into  intervals, 25~{\AA} or 50~{\AA} wide, and derived the abundances in each interval by performing a $\chi^2$ minimization of the differences between the observed and synthetic spectra. The minimization algorithm was written in {\it IDL} language, using the {\it amoeba} routine. We adopted lists of spectral lines and atomic parameters from \citet{castelli04}, who have updated the original parameters of  \citet{kur95}. When necessary we also checked the NIST database \citep{kramida19}. 

The derived abundances are listed in Table~\ref{table_abund}. We considered several sources of errors in our abundances. First, we evaluated the expected errors caused by  variations in the fundamental stellar parameters of ($\delta T_{\rm eff} = \pm$\,150~K, $\delta \log g = \pm$\,0.2~dex, and $\delta \xi = \pm$\,0.3~km s$^{-1}$). According to our simulations, the  uncertainties in the atmospheric parameters contribute $\pm$\,0.1~dex to the total error budgets. Total errors were evaluated by summing in quadrature  this value to the standard deviations obtained from the average abundances.

Fig.~\ref{fig:histograms} shows the derived chemical elements distributions in the form of histograms, where bins have been fixed to 0.15~dex (that is representative of the experimental errors). A Gaussian fit has been overimposed on each histogram. In the upper right corner of each box, we also report the center of the Gaussian and its FWHM. All the abundances are referred to the solar value \citep{grevesse10}.
In the following we will comment on the distribution of the chemical abundance for each element or group of elements.

\begin{landscape}
\begin{table}
\label{table_abund}
\caption{Abundances expressed in solar terms \citep{grevesse10} for the chemical elements detected in our targets.}
\footnotesize\setlength{\tabcolsep}{3pt}
\begin{tabular}{lrrrrrrrrrrr}
\hline \hline                                                                                                                                                                                                                                                                                                                                                                                      
~~~~~~~~ID             &        [C/H]~~~~    &          [O/H]~~~~  &       [Na/H]~~~  &         [Mg/H]~~~  &        [Al/H]~~~   &        [Si/H]~~~   &           [S/H]~~~~ &        [Ca/H] ~~~   &          [Sc/H]~~~  &         [Ti/H]~~~   &        [V/H]~~~~     \\
\hline
\hline                                                                                                                                                                                                                                                             
ASASSN\,J180946.70     &  $-$0.14 $\pm$ 0.12 &  $-$0.03 $\pm$ 0.10 &  0.58 $\pm$ 0.15 &    0.04 $\pm$ 0.12 &    0.23 $\pm$ 0.20 &    0.19 $\pm$ 0.10 &     0.26 $\pm$ 0.05 &     0.09 $\pm$ 0.15 &     0.34 $\pm$ 0.09 &     0.17 $\pm$ 0.15 &     0.17 $\pm$ 0.11  \\
ASAS\,J052610+1151.3   &  $-$0.26 $\pm$ 0.18 &  $-$0.13 $\pm$ 0.09 &  0.29 $\pm$ 0.13 &    0.21 $\pm$ 0.13 & $-$0.08 $\pm$ 0.10 & $-$0.04 $\pm$ 0.16 &     0.10 $\pm$ 0.15 &     0.08 $\pm$ 0.14 &     0.17 $\pm$ 0.13 &     0.09 $\pm$ 0.19 &     0.14 $\pm$ 0.16  \\
ASAS\,J061022+1438.6   &  $-$0.27 $\pm$ 0.10 &  $-$0.09 $\pm$ 0.15 &  0.33 $\pm$ 0.15 &    0.36 $\pm$ 0.13 &    0.10 $\pm$ 0.10 &    0.08 $\pm$ 0.15 &     0.10 $\pm$ 0.08 &  $-$0.01 $\pm$ 0.16 &     0.47 $\pm$ 0.18 &     0.21 $\pm$ 0.16 &     0.22 $\pm$ 0.12  \\
ASAS\,J063519+2117.8   &  $-$0.43 $\pm$ 0.18 &  $-$0.21 $\pm$ 0.15 &  0.23 $\pm$ 0.07 & $-$0.21 $\pm$ 0.18 &    0.18 $\pm$ 0.13 &    0.23 $\pm$ 0.17 &  $-$0.04 $\pm$ 0.14 &  $-$0.17 $\pm$ 0.13 &     0.20 $\pm$ 0.12 &     0.18 $\pm$ 0.11 &     0.50 $\pm$ 0.11  \\
ASAS\,J065413+0756.5   &  $-$0.45 $\pm$ 0.19 &  $-$0.01 $\pm$ 0.10 &  0.36 $\pm$ 0.13 & $-$0.07 $\pm$ 0.12 & $-$0.07 $\pm$ 0.13 &    0.10 $\pm$ 0.17 &  $-$0.19 $\pm$ 0.12 &  $-$0.16 $\pm$ 0.09 &  $-$0.18 $\pm$ 0.46 &     0.06 $\pm$ 0.20 &     0.06 $\pm$ 0.20  \\
ASAS\,J070832-1454.5   &  $-$0.11 $\pm$ 0.19 &  $-$0.34 $\pm$ 0.22 &  0.25 $\pm$ 0.15 &    0.23 $\pm$ 0.10 &    0.15 $\pm$ 0.11 &    0.01 $\pm$ 0.16 &     0.14 $\pm$ 0.14 &     0.02 $\pm$ 0.15 &     0.37 $\pm$ 0.15 &     0.17 $\pm$ 0.10 &     0.17 $\pm$ 0.18  \\
ASAS\,J070911-1217.2   &  $-$0.17 $\pm$ 0.12 &  $-$0.19 $\pm$ 0.10 &  0.12 $\pm$ 0.07 &    0.09 $\pm$ 0.14 & $-$0.11 $\pm$ 0.14 &    0.00 $\pm$ 0.10 &     0.20 $\pm$ 0.16 &  $-$0.06 $\pm$ 0.16 &     0.44 $\pm$ 0.18 &     0.16 $\pm$ 0.16 &     0.14 $\pm$ 0.14  \\
ASAS\,J072424-0751.3   &  $-$0.69 $\pm$ 0.15 &     0.10 $\pm$ 0.10 &  0.27 $\pm$ 0.15 &    0.36 $\pm$ 0.15 & $-$0.21 $\pm$ 0.12 & $-$0.04 $\pm$ 0.12 &  $-$0.14 $\pm$ 0.13 &     0.08 $\pm$ 0.16 &     0.58 $\pm$ 0.20 &     0.17 $\pm$ 0.18 &     0.14 $\pm$ 0.17  \\
ASAS\,J074401-1707.8   &  $-$0.46 $\pm$ 0.15 &  $-$0.20 $\pm$ 0.15 &  0.17 $\pm$ 0.10 & $-$0.18 $\pm$ 0.18 &    0.12 $\pm$ 0.13 &    0.06 $\pm$ 0.18 &  $-$0.14 $\pm$ 0.12 &  $-$0.22 $\pm$ 0.05 &  $-$0.06 $\pm$ 0.17 &  $-$0.07 $\pm$ 0.14 &     0.25 $\pm$ 0.11  \\
ASAS\,J074412-1704.9   &  $-$0.26 $\pm$ 0.12 &  $-$0.34 $\pm$ 0.15 &  0.15 $\pm$ 0.12 &    0.15 $\pm$ 0.12 & $-$0.21 $\pm$ 0.18 &    0.01 $\pm$ 0.18 &     0.27 $\pm$ 0.09 &     0.01 $\pm$ 0.07 &     0.55 $\pm$ 0.18 &     0.19 $\pm$ 0.13 &     0.21 $\pm$ 0.09  \\
ASAS\,J162326-0941.0   &     0.42 $\pm$ 0.05 &  $-$0.09 $\pm$ 0.05 &  0.56 $\pm$ 0.10 & $-$0.04 $\pm$ 0.10 &    0.23 $\pm$ 0.10 &    0.19 $\pm$ 0.11 &     0.31 $\pm$ 0.15 &  $-$0.09 $\pm$ 0.10 &     0.36 $\pm$ 0.13 &     0.16 $\pm$ 0.15 &     0.04 $\pm$ 0.10  \\
ASAS\,J180342-2211.0   &   --- ~~~~~~         &     0.40 $\pm$ 0.17 &  1.06 $\pm$ 0.06 &    0.02 $\pm$ 0.25 &    1.08 $\pm$ 0.13 &    0.52 $\pm$ 0.17 &     0.25 $\pm$ 0.15 &     0.18 $\pm$ 0.06 &     0.63 $\pm$ 0.13 &     0.36 $\pm$ 0.19 &     0.24 $\pm$ 0.14  \\
ASAS\,J182714-1507.1   &  $-$0.02 $\pm$ 0.20 &     0.15 $\pm$ 0.15 &  0.88 $\pm$ 0.12 &    0.35 $\pm$ 0.18 &    0.25 $\pm$ 0.13 &    0.40 $\pm$ 0.12 &     0.25 $\pm$ 0.13 &     0.59 $\pm$ 0.05 &     1.02 $\pm$ 0.24 &     0.36 $\pm$ 0.12 &  $-$0.04 $\pm$ 0.19  \\
ASAS\,J183347-0448.6   &  $-$0.20 $\pm$ 0.12 &  $-$0.04 $\pm$ 0.10 &  0.49 $\pm$ 0.13 &    0.26 $\pm$ 0.09 &    0.23 $\pm$ 0.10 &    0.19 $\pm$ 0.13 &     0.25 $\pm$ 0.05 &     0.03 $\pm$ 0.10 &     0.35 $\pm$ 0.15 &     0.23 $\pm$ 0.14 &     0.14 $\pm$ 0.12  \\
ASAS\,J183652-0907.1   &  $-$0.28 $\pm$ 0.05 &  $-$0.32 $\pm$ 0.10 &  0.42 $\pm$ 0.12 &    0.26 $\pm$ 0.12 &    0.10 $\pm$ 0.10 &    0.27 $\pm$ 0.14 &     0.26 $\pm$ 0.05 &     0.04 $\pm$ 0.10 &     0.22 $\pm$ 0.09 &     0.21 $\pm$ 0.10 &     0.14 $\pm$ 0.13  \\
ASAS\,J183904-1049.3   &  $-$0.24 $\pm$ 0.14 &  $-$0.07 $\pm$ 0.10 &  0.38 $\pm$ 0.08 &    0.14 $\pm$ 0.12 &    0.07 $\pm$ 0.10 &    0.12 $\pm$ 0.15 &     0.25 $\pm$ 0.13 &     0.01 $\pm$ 0.10 &     0.31 $\pm$ 0.09 &     0.21 $\pm$ 0.15 &     0.14 $\pm$ 0.09  \\
ASAS\,J192007+1247.7   &     0.02 $\pm$ 0.15 &  $-$0.07 $\pm$ 0.14 &  0.31 $\pm$ 0.15 &    0.56 $\pm$ 0.16 &    0.12 $\pm$ 0.13 &    0.24 $\pm$ 0.13 &  $-$0.14 $\pm$ 0.15 &     0.58 $\pm$ 0.13 &     1.03 $\pm$ 0.25 &     0.31 $\pm$ 0.18 &  $-$0.03 $\pm$ 0.21  \\
ASAS\,J192310+1351.4   &  $-$0.39 $\pm$ 0.08 &     0.11 $\pm$ 0.11 &  0.25 $\pm$ 0.15 &    0.22 $\pm$ 0.18 &    0.06 $\pm$ 0.13 &    0.16 $\pm$ 0.09 &  $-$0.04 $\pm$ 0.15 &     0.12 $\pm$ 0.10 &     0.55 $\pm$ 0.37 &     0.26 $\pm$ 0.20 &     0.12 $\pm$ 0.10  \\
     BD+59\,12         &  $-$0.40 $\pm$ 0.12 &  $-$0.01 $\pm$ 0.10 &  0.25 $\pm$ 0.08 & $-$0.19 $\pm$ 0.20 &    0.25 $\pm$ 0.13 &    0.18 $\pm$ 0.18 &  $-$0.04 $\pm$ 0.15 &  $-$0.09 $\pm$ 0.06 &     0.05 $\pm$ 0.07 &     0.05 $\pm$ 0.10 &     0.26 $\pm$ 0.09  \\
      CF\,Cam          &  $-$0.25 $\pm$ 0.13 &  $-$0.20 $\pm$ 0.11 &  0.19 $\pm$ 0.08 & $-$0.28 $\pm$ 0.18 &    0.00 $\pm$ 0.13 &    0.20 $\pm$ 0.16 &  $-$0.19 $\pm$ 0.12 &  $-$0.11 $\pm$ 0.04 &     0.24 $\pm$ 0.20 &     0.29 $\pm$ 0.18 &     0.42 $\pm$ 0.08  \\
DR2468646563398354176  &  $-$0.62 $\pm$ 0.13 &  $-$0.23 $\pm$ 0.15 &  0.12 $\pm$ 0.05 & $-$0.21 $\pm$ 0.27 &    0.25 $\pm$ 0.13 &    0.07 $\pm$ 0.14 &  $-$0.14 $\pm$ 0.15 &  $-$0.25 $\pm$ 0.10 &  $-$0.43 $\pm$ 0.17 &  $-$0.17 $\pm$ 0.16 &     0.23 $\pm$ 0.12  \\
DR2514736269771300224  &  $-$0.27 $\pm$ 0.15 &     0.24 $\pm$ 0.32 &  0.33 $\pm$ 0.10 &    0.01 $\pm$ 0.13 &    0.18 $\pm$ 0.13 &    0.21 $\pm$ 0.12 &     0.06 $\pm$ 0.15 &  $-$0.01 $\pm$ 0.07 &     0.16 $\pm$ 0.14 &     0.15 $\pm$ 0.11 &     0.19 $\pm$ 0.12  \\
   HD\,160473          &  $-$0.17 $\pm$ 0.09 &     0.05 $\pm$ 0.12 &  0.41 $\pm$ 0.14 &    0.51 $\pm$ 0.23 &    0.18 $\pm$ 0.13 &    0.24 $\pm$ 0.18 &     0.15 $\pm$ 0.11 &     0.30 $\pm$ 0.08 &     0.95 $\pm$ 0.20 &     0.25 $\pm$ 0.19 &     0.01 $\pm$ 0.14  \\
     HO\,Vul           &     0.38 $\pm$ 0.10 &     0.11 $\pm$ 0.12 &  0.41 $\pm$ 0.13 &    0.19 $\pm$ 0.22 &    0.06 $\pm$ 0.13 &    0.35 $\pm$ 0.18 &     0.15 $\pm$ 0.15 &     0.47 $\pm$ 0.10 &     0.96 $\pm$ 0.21 &     0.30 $\pm$ 0.16 &     0.10 $\pm$ 0.10  \\
OGLE-GD-CEP-0066       &  $-$0.46 $\pm$ 0.11 &  $-$0.01 $\pm$ 0.15 &  0.25 $\pm$ 0.08 & $-$0.12 $\pm$ 0.14 & $-$0.13 $\pm$ 0.13 &    0.09 $\pm$ 0.15 &  $-$0.14 $\pm$ 0.15 &  $-$0.14 $\pm$ 0.08 &  $-$0.06 $\pm$ 0.13 &  $-$0.09 $\pm$ 0.11 &     0.26 $\pm$ 0.09  \\
OGLE-GD-CEP-0104       &  $-$0.47 $\pm$ 0.07 &  $-$0.40 $\pm$ 0.15 &  0.22 $\pm$ 0.15 & $-$0.09 $\pm$ 0.26 &    0.12 $\pm$ 0.13 &    0.07 $\pm$ 0.18 &  $-$0.24 $\pm$ 0.10 &  $-$0.08 $\pm$ 0.08 &  $-$0.01 $\pm$ 0.08 &  $-$0.02 $\pm$ 0.20 &     0.09 $\pm$ 0.21  \\
     OO\,Pup           &  $-$0.53 $\pm$ 0.16 &  $-$0.40 $\pm$ 0.12 &  0.26 $\pm$ 0.20 &    0.04 $\pm$ 0.14 &    0.12 $\pm$ 0.13 &    0.03 $\pm$ 0.19 &  $-$0.14 $\pm$ 0.15 &     0.03 $\pm$ 0.05 &     0.17 $\pm$ 0.16 &     0.05 $\pm$ 0.10 &  $-$0.08 $\pm$ 0.08  \\
     OP\,Pup           &  $-$0.31 $\pm$ 0.09 &  $-$0.17 $\pm$ 0.12 &  0.21 $\pm$ 0.13 &    0.10 $\pm$ 0.12 & $-$0.05 $\pm$ 0.13 & $-$0.05 $\pm$ 0.13 &  $-$0.14 $\pm$ 0.12 &  $-$0.12 $\pm$ 0.12 &     0.44 $\pm$ 0.14 &     0.17 $\pm$ 0.17 &  $-$0.01 $\pm$ 0.10  \\
     OR\,Cam           &  $-$0.17 $\pm$ 0.11 &     0.00 $\pm$ 0.10 &  0.26 $\pm$ 0.17 &    0.04 $\pm$ 0.14 & $-$0.03 $\pm$ 0.13 &    0.03 $\pm$ 0.13 &     0.19 $\pm$ 0.15 &  $-$0.07 $\pm$ 0.12 &     0.24 $\pm$ 0.20 &     0.13 $\pm$ 0.19 &     0.21 $\pm$ 0.14  \\
     TX\,Sct           &  $-$0.23 $\pm$ 0.10 &  $-$0.19 $\pm$ 0.10 &  0.66 $\pm$ 0.14 & $-$0.04 $\pm$ 0.10 &    0.18 $\pm$ 0.05 &    0.17 $\pm$ 0.07 &     0.31 $\pm$ 0.10 &     0.10 $\pm$ 0.10 &     0.30 $\pm$ 0.11 &     0.18 $\pm$ 0.08 &     0.14 $\pm$ 0.14  \\
  V1495\,Aql           &     0.10 $\pm$ 0.14 &     0.05 $\pm$ 0.15 &  1.06 $\pm$ 0.15 &    0.71 $\pm$ 0.25 &    0.56 $\pm$ 0.13 &    0.48 $\pm$ 0.10 &     0.35 $\pm$ 0.11 &     0.65 $\pm$ 0.04 &     0.99 $\pm$ 0.25 &     0.57 $\pm$ 0.18 &     0.44 $\pm$ 0.20  \\
  V1496\,Aql           &     0.12 $\pm$ 0.09 &     0.31 $\pm$ 0.10 &  0.76 $\pm$ 0.10 &    0.14 $\pm$ 0.12 &    0.22 $\pm$ 0.10 &    0.39 $\pm$ 0.15 &     0.41 $\pm$ 0.10 &     0.28 $\pm$ 0.10 &     0.68 $\pm$ 0.15 &     0.26 $\pm$ 0.15 &     0.04 $\pm$ 0.11  \\
  V1788\,Cyg           &     0.16 $\pm$ 0.07 &  $-$0.24 $\pm$ 0.15 &  0.44 $\pm$ 0.03 &    0.40 $\pm$ 0.16 &    0.18 $\pm$ 0.13 &    0.50 $\pm$ 0.13 &     0.05 $\pm$ 0.15 &     0.44 $\pm$ 0.05 &     1.05 $\pm$ 0.23 &     0.53 $\pm$ 0.20 &     0.24 $\pm$ 0.10  \\
  V2475\,Cyg           &  $-$0.10 $\pm$ 0.10 &  $-$0.07 $\pm$ 0.07 &  0.53 $\pm$ 0.14 &    0.57 $\pm$ 0.22 &    0.37 $\pm$ 0.13 &    0.39 $\pm$ 0.10 &     0.25 $\pm$ 0.15 &     0.30 $\pm$ 0.08 &     0.90 $\pm$ 0.38 &     0.46 $\pm$ 0.15 &     0.52 $\pm$ 0.15  \\
   V355\,Sge           &  $-$0.33 $\pm$ 0.11 &  $-$0.11 $\pm$ 0.10 &  0.46 $\pm$ 0.15 &    0.06 $\pm$ 0.10 &    0.21 $\pm$ 0.05 &    0.17 $\pm$ 0.11 &     0.16 $\pm$ 0.10 &     0.13 $\pm$ 0.10 &     0.58 $\pm$ 0.15 &     0.18 $\pm$ 0.15 &     0.19 $\pm$ 0.16  \\
   V371\,Gem           &  $-$0.29 $\pm$ 0.08 &  $-$0.03 $\pm$ 0.13 &  0.33 $\pm$ 0.11 &    0.10 $\pm$ 0.13 & $-$0.08 $\pm$ 0.11 &    0.04 $\pm$ 0.12 &     0.14 $\pm$ 0.15 &  $-$0.02 $\pm$ 0.15 &     0.07 $\pm$ 0.15 &     0.14 $\pm$ 0.13 &     0.10 $\pm$ 0.05  \\
   V383\,Cyg           &  $-$0.09 $\pm$ 0.14 &  $-$0.01 $\pm$ 0.17 &  0.13 $\pm$ 0.15 &    0.41 $\pm$ 0.18 &    0.03 $\pm$ 0.13 &    0.20 $\pm$ 0.17 &  $-$0.04 $\pm$ 0.15 &     0.09 $\pm$ 0.16 &     0.75 $\pm$ 0.35 &     0.34 $\pm$ 0.19 &     0.25 $\pm$ 0.13  \\
   V389\,Sct           &  $-$0.27 $\pm$ 0.08 &     0.21 $\pm$ 0.15 &  0.53 $\pm$ 0.13 &    0.41 $\pm$ 0.18 &    0.18 $\pm$ 0.13 &    0.26 $\pm$ 0.18 &     0.15 $\pm$ 0.15 &     0.24 $\pm$ 0.07 &     0.65 $\pm$ 0.26 &     0.39 $\pm$ 0.17 &     0.29 $\pm$ 0.15  \\
   V536\,Ser           &  $-$0.14 $\pm$ 0.05 &  $-$0.14 $\pm$ 0.10 &  0.60 $\pm$ 0.06 &    0.08 $\pm$ 0.05 &    0.25 $\pm$ 0.05 &    0.19 $\pm$ 0.10 &     0.35 $\pm$ 0.05 &     0.13 $\pm$ 0.10 &     0.31 $\pm$ 0.05 &     0.21 $\pm$ 0.10 &     0.11 $\pm$ 0.13  \\
  V5567\,Sgr           &  $-$0.15 $\pm$ 0.12 &     0.10 $\pm$ 0.10 &  0.25 $\pm$ 0.10 & $-$0.06 $\pm$ 0.07 &    0.10 $\pm$ 0.06 &    0.13 $\pm$ 0.10 &     0.19 $\pm$ 0.08 &     0.00 $\pm$ 0.07 &     0.11 $\pm$ 0.07 &     0.14 $\pm$ 0.12 &     0.04 $\pm$ 0.09  \\
   V598\,Per           &  $-$0.31 $\pm$ 0.11 &     0.05 $\pm$ 0.14 &  0.39 $\pm$ 0.20 & $-$0.04 $\pm$ 0.25 &    0.37 $\pm$ 0.13 &    0.18 $\pm$ 0.14 &  $-$0.04 $\pm$ 0.16 &     0.03 $\pm$ 0.05 &     0.31 $\pm$ 0.23 &     0.13 $\pm$ 0.20 &  $-$0.13 $\pm$ 0.21  \\
   V824\,Cas           &  $-$0.39 $\pm$ 0.08 &     0.18 $\pm$ 0.12 &  0.45 $\pm$ 0.17 &    0.09 $\pm$ 0.24 &    0.25 $\pm$ 0.13 &    0.14 $\pm$ 0.19 &  $-$0.04 $\pm$ 0.12 &     0.02 $\pm$ 0.04 &     0.26 $\pm$ 0.07 &     0.11 $\pm$ 0.20 &     0.07 $\pm$ 0.15  \\
   V912\,Aql           &  $-$0.10 $\pm$ 0.10 &  $-$0.05 $\pm$ 0.04 &  0.32 $\pm$ 0.10 &    0.01 $\pm$ 0.12 &    0.20 $\pm$ 0.06 &    0.14 $\pm$ 0.10 &     0.24 $\pm$ 0.13 &     0.04 $\pm$ 0.10 &     0.17 $\pm$ 0.07 &     0.23 $\pm$ 0.15 &     0.09 $\pm$ 0.13  \\
   V914\,Mon           &  $-$0.18 $\pm$ 0.10 &  $-$0.09 $\pm$ 0.15 &  0.24 $\pm$ 0.17 & $-$0.13 $\pm$ 0.15 & $-$0.05 $\pm$ 0.07 &    0.06 $\pm$ 0.18 &     0.09 $\pm$ 0.09 &  $-$0.17 $\pm$ 0.17 &  $-$0.13 $\pm$ 0.10 &  $-$0.02 $\pm$ 0.18 &     0.16 $\pm$ 0.13  \\
   V946\,Cas           &  $-$0.65 $\pm$ 0.08 &  $-$0.15 $\pm$ 0.15 &  0.22 $\pm$ 0.15 & $-$0.20 $\pm$ 0.18 & $-$0.07 $\pm$ 0.13 & $-$0.04 $\pm$ 0.15 &  $-$0.54 $\pm$ 0.15 &  $-$0.22 $\pm$ 0.05 &     0.10 $\pm$ 0.16 &  $-$0.15 $\pm$ 0.19 &  $-$0.29 $\pm$ 0.12  \\
   V966\,Mon           &  $-$0.49 $\pm$ 0.13 &  $-$0.39 $\pm$ 0.15 &  0.00 $\pm$ 0.08 & $-$0.53 $\pm$ 0.18 & $-$0.13 $\pm$ 0.13 & $-$0.08 $\pm$ 0.13 &  $-$0.19 $\pm$ 0.15 &  $-$0.38 $\pm$ 0.10 &  $-$0.37 $\pm$ 0.28 &  $-$0.46 $\pm$ 0.08 &  $-$0.42 $\pm$ 0.17  \\
      X\,Sct           &  $-$0.14 $\pm$ 0.12 &  $-$0.04 $\pm$ 0.10 &  0.30 $\pm$ 0.18 &    0.06 $\pm$ 0.13 &    0.09 $\pm$ 0.10 &    0.09 $\pm$ 0.15 &     0.16 $\pm$ 0.10 &  $-$0.09 $\pm$ 0.10 &     0.47 $\pm$ 0.10 &     0.22 $\pm$ 0.15 &     0.14 $\pm$ 0.15  \\
ZTF\,J000234.99+650517.9 &  $-$0.30 $\pm$ 0.19 &     0.11 $\pm$ 0.15 &  0.36 $\pm$ 0.23 & $-$0.12 $\pm$ 0.17 &    0.00 $\pm$ 0.13 &    0.11 $\pm$ 0.15 &  $-$0.14 $\pm$ 0.15 &     0.06 $\pm$ 0.05 &     0.28 $\pm$ 0.20 &  $-$0.12 $\pm$ 0.18 &  $-$0.33 $\pm$ 0.08  \\
\end{tabular}
\end{table}
\end{landscape}

\begin{landscape}
\begin{table}
\addtocounter{table}{-1}
\caption{{\it ...continued.}}
\footnotesize\setlength{\tabcolsep}{3pt}
\begin{tabular}{lrrrrrrrrrrr}
\hline \hline                                                                                                                                                                                                                                                                                                                                                                                      
~~~~~~~~ID              &         [Cr/H]~~~   &         [Mn/H]~~~   &         [Fe/H] ~~~  &         [Co/H]~~~   &         [Ni/H]~~~   &         [Cu/H]~~~   &        [Zn/H]~~~    &          [Sr/H]~~~  &          [Y/H]~~~~  &       [Zr/H]~~~   &          [Ba/H]~~~   \\
\hline
\hline                                        
ASASSN\,J180946.70      &     0.23 $\pm$ 0.16 &     0.20 $\pm$ 0.15 &     0.15 $\pm$ 0.11 &     0.10 $\pm$ 0.10 &     0.12 $\pm$ 0.10 &     0.24 $\pm$ 0.10 &     0.17 $\pm$ 0.13 &     0.12 $\pm$ 0.09 &     0.30 $\pm$ 0.19 &   0.40 $\pm$ 0.40 &     0.08 $\pm$ 0.15  \\
ASAS\,J052610+1151.3    &     0.12 $\pm$ 0.18 &     0.01 $\pm$ 0.16 &  $-$0.06 $\pm$ 0.10 &     0.12 $\pm$ 0.09 &  $-$0.08 $\pm$ 0.20 &  $-$0.07 $\pm$ 0.10 &     0.06 $\pm$ 0.14 &  $-$0.10 $\pm$ 0.15 &     0.30 $\pm$ 0.15 &   0.34 $\pm$ 0.14 &     1.04 $\pm$ 0.07  \\
ASAS\,J061022+1438.6    &     0.15 $\pm$ 0.10 &     0.12 $\pm$ 0.19 &     0.01 $\pm$ 0.11 &     0.17 $\pm$ 0.10 &     0.02 $\pm$ 0.18 &     0.13 $\pm$ 0.10 &  $-$0.10 $\pm$ 0.13 &     0.10 $\pm$ 0.12 &     0.32 $\pm$ 0.13 &   0.50 $\pm$ 0.14 &     0.49 $\pm$ 0.40  \\
ASAS\,J063519+2117.8    &     0.09 $\pm$ 0.13 &  $-$0.51 $\pm$ 0.08 &  $-$0.18 $\pm$ 0.15 &     0.53 $\pm$ 0.26 &  $-$0.04 $\pm$ 0.16 &     0.27 $\pm$ 0.15 &  $-$0.31 $\pm$ 0.15 &     0.44 $\pm$ 0.21 &     0.04 $\pm$ 0.12 &   0.56 $\pm$ 0.15 &     0.10 $\pm$ 0.16  \\
ASAS\,J065413+0756.5    &  $-$0.08 $\pm$ 0.10 &  $-$0.64 $\pm$ 0.14 &  $-$0.27 $\pm$ 0.15 &     0.15 $\pm$ 0.15 &  $-$0.12 $\pm$ 0.20 &     0.11 $\pm$ 0.13 &  $-$0.43 $\pm$ 0.15 &     0.30 $\pm$ 0.19 &     0.23 $\pm$ 0.12 &   0.30 $\pm$ 0.04 &  $-$0.23 $\pm$ 0.16  \\
ASAS\,J070832-1454.5    &     0.06 $\pm$ 0.10 &     0.24 $\pm$ 0.12 &  $-$0.11 $\pm$ 0.11 &     0.20 $\pm$ 0.10 &  $-$0.07 $\pm$ 0.10 &     0.16 $\pm$ 0.12 &  $-$0.19 $\pm$ 0.11 &     0.09 $\pm$ 0.11 &     0.18 $\pm$ 0.12 &   0.53 $\pm$ 0.16 &     0.66 $\pm$ 0.10  \\
ASAS\,J070911-1217.2    &     0.10 $\pm$ 0.13 &     0.32 $\pm$ 0.11 &  $-$0.11 $\pm$ 0.11 &  $-$0.08 $\pm$ 0.15 &  $-$0.08 $\pm$ 0.15 &     0.23 $\pm$ 0.10 &  $-$0.26 $\pm$ 0.12 &     0.07 $\pm$ 0.10 &     0.17 $\pm$ 0.14 &   0.40 $\pm$ 0.13 &     0.90 $\pm$ 0.15  \\
ASAS\,J072424-0751.3    &     0.15 $\pm$ 0.19 &  $-$0.06 $\pm$ 0.16 &  $-$0.04 $\pm$ 0.11 &     0.19 $\pm$ 0.14 &  $-$0.03 $\pm$ 0.13 &     0.15 $\pm$ 0.07 &     0.03 $\pm$ 0.10 &     0.17 $\pm$ 0.12 &     0.19 $\pm$ 0.08 &   0.74 $\pm$ 0.12 &     0.65 $\pm$ 0.13  \\
ASAS\,J074401-1707.8    &  $-$0.20 $\pm$ 0.11 &  $-$0.74 $\pm$ 0.06 &  $-$0.41 $\pm$ 0.15 &     0.19 $\pm$ 0.16 &  $-$0.30 $\pm$ 0.13 &     0.02 $\pm$ 0.15 &  $-$0.57 $\pm$ 0.11 &     0.25 $\pm$ 0.21 &  $-$0.10 $\pm$ 0.14 &   0.30 $\pm$ 0.15 &  $-$0.22 $\pm$ 0.22  \\
ASAS\,J074412-1704.9    &     0.04 $\pm$ 0.11 &  $-$0.02 $\pm$ 0.12 &  $-$0.18 $\pm$ 0.10 &     0.25 $\pm$ 0.14 &  $-$0.17 $\pm$ 0.12 &  $-$0.24 $\pm$ 0.10 &  $-$0.25 $\pm$ 0.07 &     0.07 $\pm$ 0.15 &     0.00 $\pm$ 0.07 &   0.54 $\pm$ 0.06 &     0.71 $\pm$ 0.19  \\
ASAS\,J162326-0941.0    &     0.15 $\pm$ 0.15 &     0.05 $\pm$ 0.10 &  $-$0.03 $\pm$ 0.11 &     0.10 $\pm$ 0.10 &     0.12 $\pm$ 0.15 &     0.32 $\pm$ 0.10 &     0.11 $\pm$ 0.03 &     0.10 $\pm$ 0.15 &     0.30 $\pm$ 0.15 &   0.42 $\pm$ 0.12 &     0.26 $\pm$ 0.15  \\
ASAS\,J180342-2211.0    &     0.37 $\pm$ 0.15 &  $-$0.31 $\pm$ 0.10 &     0.10 $\pm$ 0.15 &     0.27 $\pm$ 0.23 &     0.14 $\pm$ 0.12 &  $-$0.06 $\pm$ 0.15 &     0.22 $\pm$ 0.11 &     0.68 $\pm$ 0.07 &     0.42 $\pm$ 0.10 &   0.78 $\pm$ 0.18 &  $-$0.05 $\pm$ 0.09  \\
ASAS\,J182714-1507.1    &     0.33 $\pm$ 0.11 &     0.52 $\pm$ 0.09 &     0.36 $\pm$ 0.14 &     0.30 $\pm$ 0.23 &     0.23 $\pm$ 0.11 &     0.14 $\pm$ 0.10 &     0.08 $\pm$ 0.19 &     0.20 $\pm$ 0.10 &     0.23 $\pm$ 0.05 &   0.30 $\pm$ 0.04 &  $-$0.71 $\pm$ 0.10  \\
ASAS\,J183347-0448.6    &     0.27 $\pm$ 0.09 &  $-$0.05 $\pm$ 0.13 &     0.20 $\pm$ 0.11 &     0.22 $\pm$ 0.11 &     0.11 $\pm$ 0.14 &     0.38 $\pm$ 0.10 &  $-$0.04 $\pm$ 0.10 &     0.08 $\pm$ 0.10 &     0.22 $\pm$ 0.07 &   0.60 $\pm$ 0.18 &     0.41 $\pm$ 0.15  \\
ASAS\,J183652-0907.1    &     0.29 $\pm$ 0.15 &     0.15 $\pm$ 0.10 &     0.16 $\pm$ 0.11 &     0.12 $\pm$ 0.10 &     0.06 $\pm$ 0.15 &     0.28 $\pm$ 0.15 &  $-$0.01 $\pm$ 0.10 &     0.27 $\pm$ 0.10 &     0.25 $\pm$ 0.07 &   0.48 $\pm$ 0.10 &     0.61 $\pm$ 0.15  \\
ASAS\,J183904-1049.3    &     0.19 $\pm$ 0.17 &  $-$0.02 $\pm$ 0.15 &     0.04 $\pm$ 0.11 &     0.22 $\pm$ 0.15 &     0.00 $\pm$ 0.09 &     0.13 $\pm$ 0.10 &  $-$0.16 $\pm$ 0.08 &     0.07 $\pm$ 0.10 &     0.25 $\pm$ 0.14 &   0.48 $\pm$ 0.10 &     0.61 $\pm$ 0.15  \\
ASAS\,J192007+1247.7    &     0.29 $\pm$ 0.13 &     0.49 $\pm$ 0.07 &     0.31 $\pm$ 0.14 &     0.07 $\pm$ 0.32 &     0.00 $\pm$ 0.15 &  $-$0.04 $\pm$ 0.09 &     0.06 $\pm$ 0.23 &     0.30 $\pm$ 0.15 &     0.54 $\pm$ 0.24 &   0.00 $\pm$ 0.04 &     0.10 $\pm$ 0.24  \\
ASAS\,J192310+1351.4    &     0.17 $\pm$ 0.12 &  $-$0.24 $\pm$ 0.20 &     0.03 $\pm$ 0.15 &     0.11 $\pm$ 0.16 &  $-$0.08 $\pm$ 0.15 &  $-$0.50 $\pm$ 0.15 &  $-$0.27 $\pm$ 0.08 &     0.57 $\pm$ 0.17 &     0.14 $\pm$ 0.06 &   0.10 $\pm$ 0.09 &  $-$0.15 $\pm$ 0.16  \\
     BD+59\,12          &  $-$0.01 $\pm$ 0.15 &  $-$0.54 $\pm$ 0.05 &  $-$0.10 $\pm$ 0.14 &     0.46 $\pm$ 0.18 &  $-$0.13 $\pm$ 0.13 &     0.27 $\pm$ 0.15 &  $-$0.42 $\pm$ 0.14 &     0.22 $\pm$ 0.09 &     0.36 $\pm$ 0.06 &   0.43 $\pm$ 0.15 &  $-$0.30 $\pm$ 0.29  \\
      CF\,Cam           &     0.32 $\pm$ 0.13 &  $-$0.40 $\pm$ 0.16 &  $-$0.17 $\pm$ 0.15 &     0.57 $\pm$ 0.10 &  $-$0.05 $\pm$ 0.14 &     0.65 $\pm$ 0.15 &  $-$0.37 $\pm$ 0.10 &  $-$0.20 $\pm$ 0.07 &     0.17 $\pm$ 0.04 &   0.60 $\pm$ 0.29 &     0.42 $\pm$ 0.16  \\
DR2468646563398354176   &  $-$0.38 $\pm$ 0.17 &  $-$0.80 $\pm$ 0.07 &  $-$0.45 $\pm$ 0.11 &     0.32 $\pm$ 0.27 &  $-$0.19 $\pm$ 0.19 &     0.15 $\pm$ 0.12 &  $-$0.57 $\pm$ 0.16 &  $-$0.07 $\pm$ 0.36 &  $-$0.50 $\pm$ 0.10 &   0.40 $\pm$ 0.04 &  $-$0.16 $\pm$ 0.30  \\
DR2514736269771300224   &     0.09 $\pm$ 0.13 &  $-$0.32 $\pm$ 0.05 &  $-$0.09 $\pm$ 0.15 &     0.38 $\pm$ 0.23 &  $-$0.04 $\pm$ 0.10 &     0.71 $\pm$ 0.15 &  $-$0.33 $\pm$ 0.14 &     0.39 $\pm$ 0.15 &     0.06 $\pm$ 0.31 &   0.30 $\pm$ 0.16 &     0.07 $\pm$ 0.24  \\
   HD\,160473           &     0.24 $\pm$ 0.19 &     0.21 $\pm$ 0.04 &     0.23 $\pm$ 0.15 &     0.21 $\pm$ 0.28 &     0.12 $\pm$ 0.08 &  $-$0.04 $\pm$ 0.27 &  $-$0.09 $\pm$ 0.12 &     0.40 $\pm$ 0.15 &     0.29 $\pm$ 0.10 &   0.30 $\pm$ 0.04 &     0.49 $\pm$ 0.18  \\
     HO\,Vul            &     0.32 $\pm$ 0.13 &     0.46 $\pm$ 0.04 &     0.20 $\pm$ 0.15 &     0.18 $\pm$ 0.21 &     0.11 $\pm$ 0.10 &  $-$0.09 $\pm$ 0.12 &  $-$0.28 $\pm$ 0.23 &     0.55 $\pm$ 0.07 &     0.35 $\pm$ 0.05 &   0.10 $\pm$ 0.15 &     0.33 $\pm$ 0.20  \\
OGLE-GD-CEP-0066        &     0.03 $\pm$ 0.18 &  $-$0.53 $\pm$ 0.06 &  $-$0.25 $\pm$ 0.15 &     0.23 $\pm$ 0.20 &  $-$0.10 $\pm$ 0.14 &     0.15 $\pm$ 0.15 &  $-$0.43 $\pm$ 0.10 &     0.20 $\pm$ 0.23 &     0.23 $\pm$ 0.19 &   0.15 $\pm$ 0.05 &     0.03 $\pm$ 0.18  \\
OGLE-GD-CEP-0104        &  $-$0.07 $\pm$ 0.08 &  $-$0.75 $\pm$ 0.10 &  $-$0.27 $\pm$ 0.13 &     0.03 $\pm$ 0.17 &  $-$0.25 $\pm$ 0.13 &  $-$0.10 $\pm$ 0.15 &  $-$1.03 $\pm$ 0.15 &     0.30 $\pm$ 0.07 &  $-$0.20 $\pm$ 0.14 &   0.52 $\pm$ 0.04 &  $-$0.16 $\pm$ 0.22  \\
     OO\,Pup            &  $-$0.12 $\pm$ 0.14 &  $-$0.51 $\pm$ 0.12 &  $-$0.18 $\pm$ 0.15 &     0.15 $\pm$ 0.12 &  $-$0.20 $\pm$ 0.15 &  $-$0.10 $\pm$ 0.15 &  $-$0.44 $\pm$ 0.11 &     0.10 $\pm$ 0.20 &  $-$0.10 $\pm$ 0.15 &   0.05 $\pm$ 0.04 &  $-$0.11 $\pm$ 0.28  \\
     OP\,Pup            &     0.11 $\pm$ 0.19 &     0.06 $\pm$ 0.20 &  $-$0.11 $\pm$ 0.13 &     0.22 $\pm$ 0.14 &  $-$0.08 $\pm$ 0.15 &  $-$0.17 $\pm$ 0.20 &  $-$0.27 $\pm$ 0.12 &     0.09 $\pm$ 0.15 &     0.21 $\pm$ 0.10 &   0.50 $\pm$ 0.11 &     0.41 $\pm$ 0.20  \\
     OR\,Cam            &     0.19 $\pm$ 0.19 &  $-$0.01 $\pm$ 0.16 &  $-$0.06 $\pm$ 0.09 &     0.17 $\pm$ 0.12 &  $-$0.03 $\pm$ 0.18 &     0.48 $\pm$ 0.11 &  $-$0.16 $\pm$ 0.10 &     0.17 $\pm$ 0.15 &     0.27 $\pm$ 0.08 &   0.59 $\pm$ 0.11 &     0.47 $\pm$ 0.17  \\
     TX\,Sct            &     0.32 $\pm$ 0.06 &     0.21 $\pm$ 0.10 &     0.11 $\pm$ 0.11 &     0.22 $\pm$ 0.09 &     0.02 $\pm$ 0.09 &     0.32 $\pm$ 0.10 &  $-$0.03 $\pm$ 0.10 &     0.09 $\pm$ 0.15 &     0.30 $\pm$ 0.10 &   0.30 $\pm$ 0.12 &     0.31 $\pm$ 0.15  \\
  V1495\,Aql            &     0.47 $\pm$ 0.10 &     0.64 $\pm$ 0.10 &     0.55 $\pm$ 0.10 &     0.50 $\pm$ 0.25 &     0.38 $\pm$ 0.19 &     0.35 $\pm$ 0.12 &     0.24 $\pm$ 0.14 &     0.29 $\pm$ 0.15 &     0.45 $\pm$ 0.09 &   0.52 $\pm$ 0.04 &     0.06 $\pm$ 0.26  \\
  V1496\,Aql            &     0.37 $\pm$ 0.11 &     0.17 $\pm$ 0.08 &  $-$0.07 $\pm$ 0.11 &     0.32 $\pm$ 0.13 &     0.13 $\pm$ 0.11 &     0.08 $\pm$ 0.10 &  $-$0.06 $\pm$ 0.07 &     0.11 $\pm$ 0.15 &     0.40 $\pm$ 0.13 &   0.48 $\pm$ 0.10 &     0.00 $\pm$ 0.15  \\
  V1788\,Cyg            &     0.61 $\pm$ 0.13 &     0.02 $\pm$ 0.08 &     0.37 $\pm$ 0.15 &     0.48 $\pm$ 0.20 &     0.24 $\pm$ 0.17 &     0.36 $\pm$ 0.15 &  $-$0.23 $\pm$ 0.42 &     0.55 $\pm$ 0.07 &     0.50 $\pm$ 0.15 &   --- ~~~~~~       &     0.71 $\pm$ 0.14  \\
  V2475\,Cyg            &     0.34 $\pm$ 0.12 &  $-$0.03 $\pm$ 0.14 &     0.25 $\pm$ 0.15 &     0.17 $\pm$ 0.20 &     0.20 $\pm$ 0.15 &  $-$0.23 $\pm$ 0.15 &  $-$0.09 $\pm$ 0.22 &     0.10 $\pm$ 0.15 &     0.29 $\pm$ 0.06 &   0.10 $\pm$ 0.09 &  $-$0.02 $\pm$ 0.10  \\
   V355\,Sge            &     0.23 $\pm$ 0.15 &     0.00 $\pm$ 0.14 &     0.03 $\pm$ 0.11 &     0.22 $\pm$ 0.10 &     0.02 $\pm$ 0.18 &     0.17 $\pm$ 0.15 &  $-$0.04 $\pm$ 0.14 &     0.12 $\pm$ 0.10 &     0.49 $\pm$ 0.19 &   0.30 $\pm$ 0.15 &     0.41 $\pm$ 0.10  \\
   V371\,Gem            &     0.06 $\pm$ 0.19 &     0.06 $\pm$ 0.17 &  $-$0.06 $\pm$ 0.13 &     0.23 $\pm$ 0.12 &  $-$0.09 $\pm$ 0.09 &     0.11 $\pm$ 0.17 &  $-$0.19 $\pm$ 0.19 &     0.07 $\pm$ 0.15 &     0.19 $\pm$ 0.17 &   0.54 $\pm$ 0.16 &     0.67 $\pm$ 0.06  \\
   V383\,Cyg            &     0.27 $\pm$ 0.13 &  $-$0.08 $\pm$ 0.20 &     0.17 $\pm$ 0.15 &     0.42 $\pm$ 0.36 &     0.06 $\pm$ 0.12 &  $-$0.37 $\pm$ 0.12 &  $-$0.27 $\pm$ 0.13 &     0.74 $\pm$ 0.10 &     0.11 $\pm$ 0.06 &   0.15 $\pm$ 0.04 &  $-$0.08 $\pm$ 0.10  \\
   V389\,Sct            &     0.27 $\pm$ 0.15 &     0.11 $\pm$ 0.20 &     0.22 $\pm$ 0.15 &     0.27 $\pm$ 0.23 &     0.17 $\pm$ 0.19 &  $-$0.01 $\pm$ 0.31 &  $-$0.15 $\pm$ 0.08 &     0.61 $\pm$ 0.17 &     0.29 $\pm$ 0.18 &   0.77 $\pm$ 0.04 &     0.10 $\pm$ 0.16  \\
   V536\,Ser            &     0.21 $\pm$ 0.12 &  $-$0.06 $\pm$ 0.10 &     0.11 $\pm$ 0.12 &     0.22 $\pm$ 0.13 &     0.15 $\pm$ 0.15 &     0.13 $\pm$ 0.10 &     0.09 $\pm$ 0.10 &     0.10 $\pm$ 0.10 &     0.20 $\pm$ 0.05 &   0.37 $\pm$ 0.13 &  $-$0.02 $\pm$ 0.10  \\
  V5567\,Sgr            &     0.10 $\pm$ 0.10 &     0.00 $\pm$ 0.14 &     0.05 $\pm$ 0.11 &     0.17 $\pm$ 0.10 &     0.07 $\pm$ 0.09 &  $-$0.02 $\pm$ 0.12 &     0.04 $\pm$ 0.10 &     0.11 $\pm$ 0.10 &     0.27 $\pm$ 0.15 &   0.32 $\pm$ 0.11 &     0.00 $\pm$ 0.10  \\
   V598\,Per            &     0.12 $\pm$ 0.12 &  $-$0.24 $\pm$ 0.12 &  $-$0.07 $\pm$ 0.15 &     0.19 $\pm$ 0.20 &     0.02 $\pm$ 0.11 &     0.27 $\pm$ 0.15 &  $-$0.38 $\pm$ 0.15 &     0.39 $\pm$ 0.18 &     0.07 $\pm$ 0.44 &   0.02 $\pm$ 0.04 &  $-$0.04 $\pm$ 0.09  \\
   V824\,Cas            &     0.16 $\pm$ 0.18 &  $-$0.28 $\pm$ 0.06 &  $-$0.08 $\pm$ 0.15 &     0.27 $\pm$ 0.23 &  $-$0.02 $\pm$ 0.11 &     0.46 $\pm$ 0.14 &  $-$0.37 $\pm$ 0.15 &  $-$0.01 $\pm$ 0.11 &     0.04 $\pm$ 0.27 &   0.15 $\pm$ 0.04 &     0.07 $\pm$ 0.20  \\
   V912\,Aql            &     0.22 $\pm$ 0.18 &     0.15 $\pm$ 0.11 &     0.14 $\pm$ 0.11 &     0.10 $\pm$ 0.10 &  $-$0.01 $\pm$ 0.18 &     0.08 $\pm$ 0.08 &  $-$0.16 $\pm$ 0.15 &     0.12 $\pm$ 0.11 &     0.30 $\pm$ 0.14 &   0.39 $\pm$ 0.15 &     0.28 $\pm$ 0.15  \\
   V914\,Mon            &     0.01 $\pm$ 0.12 &  $-$0.45 $\pm$ 0.14 &  $-$0.16 $\pm$ 0.10 &     0.11 $\pm$ 0.14 &  $-$0.05 $\pm$ 0.20 &     0.11 $\pm$ 0.07 &  $-$0.29 $\pm$ 0.12 &     0.07 $\pm$ 0.19 &     0.12 $\pm$ 0.20 &   0.39 $\pm$ 0.14 &     0.17 $\pm$ 0.11  \\
   V946\,Cas            &  $-$0.30 $\pm$ 0.10 &  $-$0.75 $\pm$ 0.10 &  $-$0.43 $\pm$ 0.12 &  $-$0.06 $\pm$ 0.20 &  $-$0.24 $\pm$ 0.17 &  $-$0.17 $\pm$ 0.11 &  $-$0.46 $\pm$ 0.13 &  $-$0.05 $\pm$ 0.19 &     0.04 $\pm$ 0.15 &   0.02 $\pm$ 0.04 &  $-$0.17 $\pm$ 0.09  \\
   V966\,Mon            &  $-$0.33 $\pm$ 0.17 &  $-$0.65 $\pm$ 0.14 &  $-$0.60 $\pm$ 0.15 &  $-$0.22 $\pm$ 0.26 &  $-$0.37 $\pm$ 0.10 &  $-$0.35 $\pm$ 0.13 &  $-$0.82 $\pm$ 0.15 &     0.99 $\pm$ 0.20 &  $-$0.60 $\pm$ 0.10 &   0.02 $\pm$ 0.04 &  $-$0.86 $\pm$ 0.12  \\
      X\,Sct            &     0.26 $\pm$ 0.11 &     0.07 $\pm$ 0.15 &     0.05 $\pm$ 0.11 &     0.27 $\pm$ 0.10 &  $-$0.01 $\pm$ 0.15 &  $-$0.06 $\pm$ 0.10 &  $-$0.16 $\pm$ 0.10 &     0.10 $\pm$ 0.04 &     0.35 $\pm$ 0.15 &   0.54 $\pm$ 0.15 &  $-$0.06 $\pm$ 0.20  \\
ZTF\,J000234+650517.9   &  $-$0.14 $\pm$ 0.14 &  $-$0.51 $\pm$ 0.05 &  $-$0.15 $\pm$ 0.12 &     0.00 $\pm$ 0.20 &  $-$0.21 $\pm$ 0.14 &     0.08 $\pm$ 0.15 &  $-$0.41 $\pm$ 0.11 &     0.42 $\pm$ 0.24 &     0.23 $\pm$ 0.15 &   0.52 $\pm$ 0.04 &  $-$0.36 $\pm$ 0.28  \\
\end{tabular}
\end{table}
\end{landscape}
 
\begin{landscape}
\begin{table}
\addtocounter{table}{-1}
\caption{{\it ...continued.}}
\footnotesize\setlength{\tabcolsep}{3pt}
\begin{tabular}{lrrrrrrr}
\hline \hline                                                                                                                                                                                                                                                                                                                                                                                      
~~~~~~~~ID               &       [La/H]~~~   &         [Ce/H]~~~   &         [Pr/H]~~~   &          [Nd/H]~~~  &         [Sm/H]~~~   &         [Eu/H]~~~   &         [Gd/H]~~~ \\
\hline
\hline                  
ASASSN\,J180946.70       &   0.18 $\pm$ 0.06 &     0.41 $\pm$ 0.15 &     --- ~~~~~~       &     0.12 $\pm$ 0.16 &     0.00 $\pm$ 0.20 &  $-$0.01 $\pm$ 0.04 &     0.37 $\pm$ 0.25 \\
ASAS\,J052610+1151.3     &   0.28 $\pm$ 0.24 &     0.32 $\pm$ 0.12 &  $-$0.33 $\pm$ 0.31 &     0.14 $\pm$ 0.10 &     0.07 $\pm$ 0.16 &  $-$0.04 $\pm$ 0.14 &     0.03 $\pm$ 0.06 \\
ASAS\,J061022+1438.6     &   0.07 $\pm$ 0.08 &     0.16 $\pm$ 0.14 &     --- ~~~~~~       &     0.04 $\pm$ 0.15 &  $-$0.18 $\pm$ 0.08 &  $-$0.08 $\pm$ 0.02 &     0.00 $\pm$ 0.10 \\
ASAS\,J063519+2117.8     &   0.23 $\pm$ 0.23 &     0.41 $\pm$ 0.13 &  $-$0.33 $\pm$ 0.12 &     0.34 $\pm$ 0.13 &  $-$0.02 $\pm$ 0.23 &  $-$0.05 $\pm$ 0.30 &     0.30 $\pm$ 0.14 \\
ASAS\,J065413+0756.5     &   0.08 $\pm$ 0.13 &  $-$0.07 $\pm$ 0.15 &     --- ~~~~~~&  $-$0.04 $\pm$ 0.17 &  $-$0.25 $\pm$ 0.18 &  $-$0.27 $\pm$ 0.13 &     0.20 $\pm$ 0.10 \\
ASAS\,J070832-1454.5     &   0.18 $\pm$ 0.04 &     0.16 $\pm$ 0.14 &  $-$0.60 $\pm$ 0.05 &     0.09 $\pm$ 0.12 &  $-$0.06 $\pm$ 0.05 &  $-$0.03 $\pm$ 0.04 &     0.03 $\pm$ 0.06 \\
ASAS\,J070911-1217.2     &   0.41 $\pm$ 0.12 &     0.26 $\pm$ 0.15 &     --- ~~~~~~       &     0.18 $\pm$ 0.11 &     0.02 $\pm$ 0.13 &     0.18 $\pm$ 0.07 &     0.10 $\pm$ 0.10 \\
ASAS\,J072424-0751.3     &   0.17 $\pm$ 0.15 &     0.06 $\pm$ 0.20 &     --- ~~~~~~       &     0.14 $\pm$ 0.10 &  $-$0.08 $\pm$ 0.12 &     0.05 $\pm$ 0.13 &     0.10 $\pm$ 0.10 \\
ASAS\,J074401-1707.8     &   0.09 $\pm$ 0.20 &  $-$0.15 $\pm$ 0.04 &  $-$0.30 $\pm$ 0.17 &  $-$0.17 $\pm$ 0.13 &  $-$0.12 $\pm$ 0.13 &  $-$0.15 $\pm$ 0.15 &     0.10 $\pm$ 0.10 \\
ASAS\,J074412-1704.9     &   0.19 $\pm$ 0.19 &     0.25 $\pm$ 0.13 &     --- ~~~~~~       &     0.04 $\pm$ 0.15 &  $-$0.50 $\pm$ 0.26 &  $-$0.49 $\pm$ 0.21 &  $-$0.27 $\pm$ 0.32 \\
ASAS\,J162326-0941.0     &   0.23 $\pm$ 0.08 &     0.10 $\pm$ 0.15 &  $-$0.50 $\pm$ 0.10 &  $-$0.07 $\pm$ 0.18 &  $-$0.05 $\pm$ 0.05 &  $-$0.13 $\pm$ 0.13 &     0.90 $\pm$ 0.10 \\
ASAS\,J180342-2211.0     &   0.57 $\pm$ 0.10 &     0.13 $\pm$ 0.08 &  $-$0.65 $\pm$ 1.06 &     0.28 $\pm$ 0.04 &     0.38 $\pm$ 0.04 &     0.56 $\pm$ 0.05 &     0.60 $\pm$ 0.71 \\
ASAS\,J182714-1507.1     &   0.07 $\pm$ 0.22 &  $-$0.32 $\pm$ 0.15 &     --- ~~~~~~       &  $-$0.12 $\pm$ 0.19 &  $-$0.40 $\pm$ 0.14 &  $-$0.12 $\pm$ 0.11 &  $-$0.05 $\pm$ 0.07 \\
ASAS\,J183347-0448.6     &   0.26 $\pm$ 0.07 &     0.04 $\pm$ 0.15 &  $-$0.47 $\pm$ 0.23 &  $-$0.01 $\pm$ 0.16 &  $-$0.05 $\pm$ 0.05 &     0.03 $\pm$ 0.06 &     0.40 $\pm$ 0.26 \\
ASAS\,J183652-0907.1     &   0.15 $\pm$ 0.05 &     0.10 $\pm$ 0.15 &     --- ~~~~~~       &     0.02 $\pm$ 0.16 &  $-$0.17 $\pm$ 0.06 &  $-$0.11 $\pm$ 0.07 &     0.10 $\pm$ 0.10 \\
ASAS\,J183904-1049.3     &   0.14 $\pm$ 0.13 &     0.05 $\pm$ 0.15 &     --- ~~~~~~       &  $-$0.22 $\pm$ 0.17 &  $-$0.15 $\pm$ 0.05 &  $-$0.03 $\pm$ 0.02 &     0.23 $\pm$ 0.23 \\
ASAS\,J192007+1247.7     &   0.08 $\pm$ 0.18 &  $-$0.07 $\pm$ 0.15 &     --- ~~~~~~       &     0.15 $\pm$ 0.13 &  $-$0.03 $\pm$ 0.25 &  $-$0.04 $\pm$ 0.17 &  $-$0.03 $\pm$ 0.15 \\
ASAS\,J192310+1351.4     &   0.12 $\pm$ 0.12 &     0.14 $\pm$ 0.09 &  $-$0.45 $\pm$ 0.07 &     0.05 $\pm$ 0.15 &  $-$0.10 $\pm$ 0.07 &  $-$0.15 $\pm$ 0.14 &     0.25 $\pm$ 0.21 \\
     BD+59\,12           &   0.11 $\pm$ 0.07 &  $-$0.27 $\pm$ 0.04 &  $-$0.33 $\pm$ 0.21 &  $-$0.13 $\pm$ 0.06 &  $-$0.23 $\pm$ 0.12 &  $-$0.05 $\pm$ 0.05 &     0.33 $\pm$ 0.21 \\
      CF\,Cam            &   0.32 $\pm$ 0.21 &     0.23 $\pm$ 0.15 &  $-$0.17 $\pm$ 0.21 &     0.38 $\pm$ 0.13 &     0.07 $\pm$ 0.21 &     0.03 $\pm$ 0.13 &     0.10 $\pm$ 0.10 \\
DR2468646563398354176    &   0.18 $\pm$ 0.33 &  $-$0.60 $\pm$ 0.15 &  $-$0.53 $\pm$ 0.42 &  $-$0.14 $\pm$ 0.08 &  $-$0.28 $\pm$ 0.23 &  $-$0.28 $\pm$ 0.25 &  $-$0.05 $\pm$ 0.05 \\
DR2514736269771300224    &   0.19 $\pm$ 0.10 &     0.10 $\pm$ 0.15 &     --- ~~~~~~       &  $-$0.07 $\pm$ 0.08 &  $-$0.10 $\pm$ 0.09 &     0.00 $\pm$ 0.13 &     0.10 $\pm$ 0.10 \\
   HD\,160473            &   0.14 $\pm$ 0.04 &  $-$0.07 $\pm$ 0.15 &  $-$0.50 $\pm$ 0.06 &     0.00 $\pm$ 0.19 &  $-$0.10 $\pm$ 0.07 &  $-$0.04 $\pm$ 0.05 &     0.10 $\pm$ 0.10 \\
     HO\,Vul             &   0.39 $\pm$ 0.20 &  $-$0.15 $\pm$ 0.04 &     --- ~~~~~~       &     0.06 $\pm$ 0.07 &     0.00 $\pm$ 0.17 &     0.20 $\pm$ 0.11 &     0.27 $\pm$ 0.29 \\
OGLE-GD-CEP-0066         &   0.31 $\pm$ 0.12 &     0.10 $\pm$ 0.04 &  $-$0.23 $\pm$ 0.35 &  $-$0.15 $\pm$ 0.12 &  $-$0.03 $\pm$ 0.12 &  $-$0.03 $\pm$ 0.18 &     0.23 $\pm$ 0.23 \\
OGLE-GD-CEP-0104         &   0.10 $\pm$ 0.15 &  $-$0.07 $\pm$ 0.15 &     --- ~~~~~~       &  $-$0.35 $\pm$ 0.18 &  $-$0.22 $\pm$ 0.18 &  $-$0.09 $\pm$ 0.17 &     0.03 $\pm$ 0.06 \\
     OO\,Pup             &   0.18 $\pm$ 0.13 &  $-$0.40 $\pm$ 0.04 &  $-$0.66 $\pm$ 0.31 &  $-$0.17 $\pm$ 0.20 &  $-$0.02 $\pm$ 0.04 &  $-$0.03 $\pm$ 0.13 &     0.04 $\pm$ 0.13 \\
     OP\,Pup             &   0.18 $\pm$ 0.12 &     0.36 $\pm$ 0.12 &  $-$0.33 $\pm$ 0.12 &     0.14 $\pm$ 0.11 &  $-$0.10 $\pm$ 0.10 &  $-$0.07 $\pm$ 0.13 &     0.10 $\pm$ 0.10 \\
     OR\,Cam             &   0.20 $\pm$ 0.04 &     0.31 $\pm$ 0.14 &     --- ~~~~~~       &     0.09 $\pm$ 0.12 &  $-$0.12 $\pm$ 0.08 &     0.00 $\pm$ 0.06 &     0.10 $\pm$ 0.10 \\
     TX\,Sct             &   0.02 $\pm$ 0.26 &     0.10 $\pm$ 0.15 &     --- ~~~~~~       &     0.14 $\pm$ 0.20 &     0.00 $\pm$ 0.16 &     0.00 $\pm$ 0.18 &     0.10 $\pm$ 0.10 \\
  V1495\,Aql             &   0.02 $\pm$ 0.17 &     0.01 $\pm$ 0.15 &     --- ~~~~~~       &     0.04 $\pm$ 0.14 &  $-$0.37 $\pm$ 0.12 &  $-$0.06 $\pm$ 0.04 &     0.00 $\pm$ 0.10 \\
  V1496\,Aql             &   0.07 $\pm$ 0.11 &     0.10 $\pm$ 0.15 &     --- ~~~~~~       &     0.15 $\pm$ 0.18 &  $-$0.14 $\pm$ 0.09 &     0.12 $\pm$ 0.08 &     0.07 $\pm$ 0.06 \\
  V1788\,Cyg             &   0.34 $\pm$ 0.10 &     0.40 $\pm$ 0.04 &  $-$0.63 $\pm$ 0.35 &     0.53 $\pm$ 0.14 &     0.17 $\pm$ 0.12 &     0.18 $\pm$ 0.08 &     0.10 $\pm$ 0.10 \\
  V2475\,Cyg             &   0.33 $\pm$ 0.17 &     0.10 $\pm$ 0.04 &  $-$0.33 $\pm$ 0.21 &  $-$0.05 $\pm$ 0.19 &     0.15 $\pm$ 0.31 &     0.12 $\pm$ 0.19 &     0.10 $\pm$ 0.10 \\
   V355\,Sge             &   0.10 $\pm$ 0.17 &     0.11 $\pm$ 0.15 &  $-$0.53 $\pm$ 0.31 &     0.18 $\pm$ 0.20 &     0.17 $\pm$ 0.15 &     0.00 $\pm$ 0.08 &     0.40 $\pm$ 0.36 \\
   V371\,Gem             &   0.25 $\pm$ 0.09 &     0.16 $\pm$ 0.13 &  $-$0.33 $\pm$ 0.12 &     0.24 $\pm$ 0.15 &  $-$0.03 $\pm$ 0.06 &  $-$0.03 $\pm$ 0.15 &     0.10 $\pm$ 0.10 \\
   V383\,Cyg             &   0.14 $\pm$ 0.07 &     0.10 $\pm$ 0.04 &     --- ~~~~~~       &  $-$0.10 $\pm$ 0.04 &  $-$0.07 $\pm$ 0.23 &  $-$0.10 $\pm$ 0.05 &     0.20 $\pm$ 0.23 \\
   V389\,Sct             &   0.12 $\pm$ 0.15 &     0.14 $\pm$ 0.09 &  $-$0.57 $\pm$ 0.06 &     0.04 $\pm$ 0.17 &  $-$0.20 $\pm$ 0.17 &  $-$0.06 $\pm$ 0.12 &     0.00 $\pm$ 0.17 \\
   V536\,Ser             &   0.14 $\pm$ 0.16 &  $-$0.04 $\pm$ 0.10 &     --- ~~~~~~       &  $-$0.07 $\pm$ 0.12 &     0.00 $\pm$ 0.17 &  $-$0.02 $\pm$ 0.12 &     0.30 $\pm$ 0.35 \\
  V5567\,Sgr             &   0.21 $\pm$ 0.12 &     0.31 $\pm$ 0.10 &  $-$0.55 $\pm$ 0.07 &     0.15 $\pm$ 0.10 &  $-$0.05 $\pm$ 0.07 &  $-$0.16 $\pm$ 0.05 &     0.05 $\pm$ 0.07 \\
   V598\,Per             &   0.14 $\pm$ 0.13 &  $-$0.07 $\pm$ 0.15 &  $-$0.37 $\pm$ 0.32 &     0.03 $\pm$ 0.16 &     0.00 $\pm$ 0.35 &     0.04 $\pm$ 0.31 &     0.10 $\pm$ 0.10 \\
   V824\,Cas             &   0.23 $\pm$ 0.08 &  $-$0.15 $\pm$ 0.04 &     --- ~~~~~~       &     0.07 $\pm$ 0.14 &  $-$0.07 $\pm$ 0.03 &     0.03 $\pm$ 0.05 &     0.10 $\pm$ 0.10 \\
   V912\,Aql             &   0.08 $\pm$ 0.11 &     0.31 $\pm$ 0.15 &  $-$0.53 $\pm$ 0.23 &     0.24 $\pm$ 0.20 &  $-$0.10 $\pm$ 0.10 &  $-$0.17 $\pm$ 0.10 &     0.13 $\pm$ 0.15 \\
   V914\,Mon             &   0.18 $\pm$ 0.17 &     0.32 $\pm$ 0.15 &  $-$0.60 $\pm$ 0.70 &  $-$0.06 $\pm$ 0.15 &  $-$0.08 $\pm$ 0.13 &  $-$0.06 $\pm$ 0.04 &     0.10 $\pm$ 0.10 \\
   V946\,Cas             &   0.34 $\pm$ 0.20 &  $-$0.15 $\pm$ 0.04 &  $-$0.40 $\pm$ 0.10 &     0.06 $\pm$ 0.11 &     0.08 $\pm$ 0.20 &  $-$0.12 $\pm$ 0.24 &     0.07 $\pm$ 0.06 \\
   V966\,Mon             &   0.42 $\pm$ 0.16 &  $-$0.40 $\pm$ 0.15 &     --- ~~~~~~       &  $-$0.60 $\pm$ 0.13 &  $-$0.49 $\pm$ 0.04 &  $-$0.90 $\pm$ 0.44 &  $-$0.13 $\pm$ 0.12 \\
      X\,Sct             &   0.12 $\pm$ 0.34 &     0.01 $\pm$ 0.20 &     --- ~~~~~~       &  $-$0.16 $\pm$ 0.21 &  $-$0.30 $\pm$ 0.26 &  $-$0.26 $\pm$ 0.27 &  $-$0.03 $\pm$ 0.15 \\
ZTF\,J000234+650517.9    &   0.12 $\pm$ 0.10 &  $-$0.40 $\pm$ 0.04 &     0.00 $\pm$ 0.10 &  $-$0.40 $\pm$ 0.15 &  $-$0.28 $\pm$ 0.06 &  $-$0.23 $\pm$ 0.06 &     0.00 $\pm$ 0.10 \\
\end{tabular}
\end{table}
\end{landscape}

\begin{itemize}

\item {\it C(N)O}: carbon was measured through the following spectral lines: $\lambda \lambda$\,4770.027, 4932.049, 5052.144, and 5380.325 {\AA}, the average abundance is about 0.23~dex under the solar value. For oxygen, whose abundance has been inferred via the forbidden line [\ion{O}{I}]\,6300.304 {\AA} and the \ion{O}{I} triplet at 6155-8 {\AA}, we obtained a distribution practically centered on the solar value. Unfortunately, in our spectral range no nitrogen lines have been detected.
\item {\it Sodium}: we measured four \ion{Na}{I} lines in each star: $\lambda \lambda$\,5682.647, 5688.217, 6154.230, and 6160.753 {\AA}. As expected, sodium is on average quite overabundant, its average value is $\approx$\,0.3~dex over the corresponding solar value. 
\item {\it Aluminum}: we detected two aluminum lines in our spectra, namely: \ion{Al}{I} $\lambda \lambda$\,6696.018 and 6698.667 {\AA}.  The obtained distribution is centered around 0.1\,dex.
\item {\it $\alpha$-elements: Mg, Si, S, Ca and Ti}: magnesium (inferred via the two spectral lines of \ion{Mg}{I} at $\lambda \lambda$\,4702.991 and 5528.405 {\AA}), silicon, sulfur (via the \ion{S}{II} line $\lambda$\,6757.153 {\AA}), calcium (three red lines generated by the \ion{Ca}{I}, $\lambda \lambda$\,6462.567, 6493.781, and 6499.650 {\AA}) and titanium (we detected ten lines of \ion{Ti}{I} as well as ten of \ion{Ti}{II}) show abundances consistent with  or slightly over than solar values.  
\item {\it Scandium}: three \ion{Sc}{II} lines were detected in our spectra, namely, $\lambda \lambda$\,5031.021, 5239.813, ans 6245.621 {\AA}. They lead to a distribution centered at 0.3~dex.
\item{\it Vanadium}:  five lines were measured: $\lambda \lambda$\,4379.246, 5698.482, 5703.569, 6090.194, and 6243.088 {\AA}. Vanadium is slight over-abundant in our sample of stars. 
\item{\it Iron peak elements: Cr, Mn, Fe, Co, Ni}: all the iron peak elements show distributions centered on standard (Mn, Fe, and Ni) or slightly over (Cr and Co) solar abundances. 
\item {\it Copper}: we measured two spectral lines (\ion{Cu}{I}\,5218.201 and 5782.170 {\AA}) obtaining an average abundance in agreement with the Sun.
\item {\it Zinc}: using the \ion{Zn}{I} spectral lines at $\lambda \lambda$\,4680.134, 4722.157, and 6362.338 {\AA}, we derived a distribution of abundances centered on about $-$0.20~dex.
\item {\it Heavier elements: Sr, Y, Zr} solar abundance was derived for strontium (using the \ion{Sr}{II} lines $\lambda \lambda$\,4077.709 and 4215.519 {\AA}), while for yttrium (\ion{Y}{II} lines $\lambda \lambda$\,4883.682, 4900.120, and 5087.418 {\AA}) and for zirconium (\ion{Zr}{II} lines $\lambda \lambda$\,4379.742 and 6114.853 {\AA}) slight and moderate over-abundances have been obtained, respectively.
\item{\it Barium:} abundances of barium (inferred by using five \ion{Ba}{II} spectral lines at $\lambda \lambda$\,4554.033, 4934.100, 5853.625, 6141.713, 6496.897 {\AA}) are spread out over a wide interval of values, ranging from approximately from $\approx\,-$1.0~dex to $\approx$\,+1.0~dex.
\item{\it Rare earth elements: La, Ce, Pr, Nd, Sm, Eu, Gd}: in our sample, lanthanum (\ion{La}{II} $\lambda \lambda$\,4921.775 and 5290.818 {\AA}), cerium (\ion{Ce}{II} $\lambda$\,4562.359\,{\AA}), neodymium (\ion{Nd}{II} $\lambda \lambda$\,4959.119, 4989.950, 5092.794 and 5293.163 {\AA}), samarium (\ion{Sm}{II} $\lambda \lambda$\,4537.941, 4577.688, 4642.228 and 4704.400 {\AA}), europium (\ion{Eu}{II} $\lambda$\,6645.114\,{\AA}) and gadolinum (\ion{Gd}{II} $\lambda$\,5092.249\,{\AA}) show almost solar value. On the contrary, praseodymium (\ion{Pr}{II} $\lambda$\,5219.045 {\AA}) has a distribution of abundances centered on $\approx\,-$0.45~dex.
\end{itemize}

\subsubsection{Comparison with literature}
\begin{figure}
   \centering
   \includegraphics[width=\hsize]{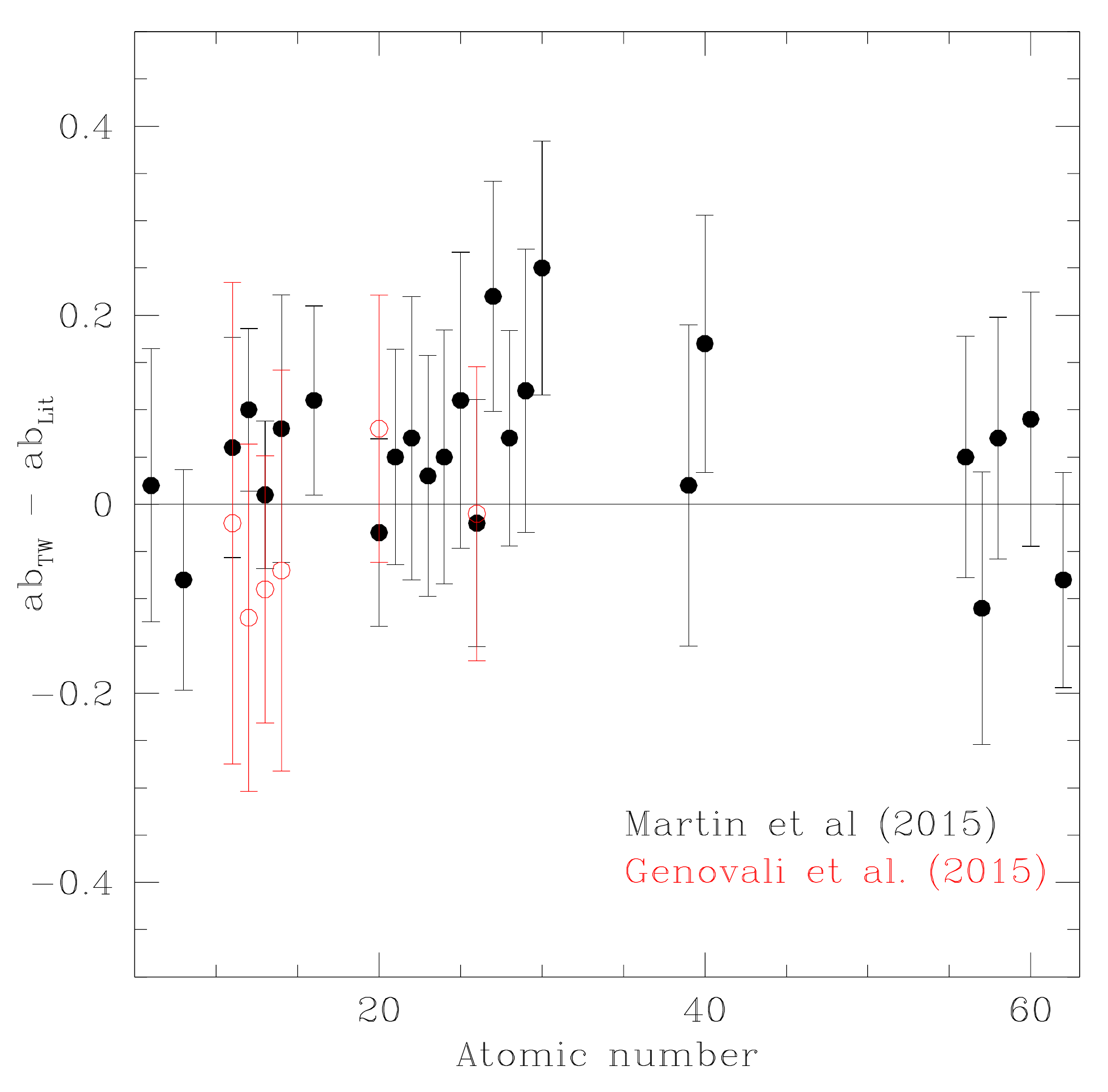}
   \caption{Comparison between the abundances derived in this study and those available in the literature for two of our targets. Filled black circles refer to V5567\,Sgr \citep{Martin2015} and open red circles to X\,Sct \citep{Genovali2015}.} 
   \label{fig:comp}
    \end{figure}

Abundances from high-resolution spectroscopy are available in the literature for two among our targets: V5567\,Sgr and X\,Sct, from the studies of \citet{Martin2015} and \citet{Genovali2015}, respectively. A comparison performed for  the elements in common is shown in Fig.~\ref{fig:comp}, where we have adopted as $\sigma_{TOT}$ the sum in quadrature between errors derived in this paper and those reported by the other authors. Almost all of our abundances are in agreement, within the errors, with the literature estimates.

\begin{figure*}
   \centering
   \includegraphics[width=\hsize]{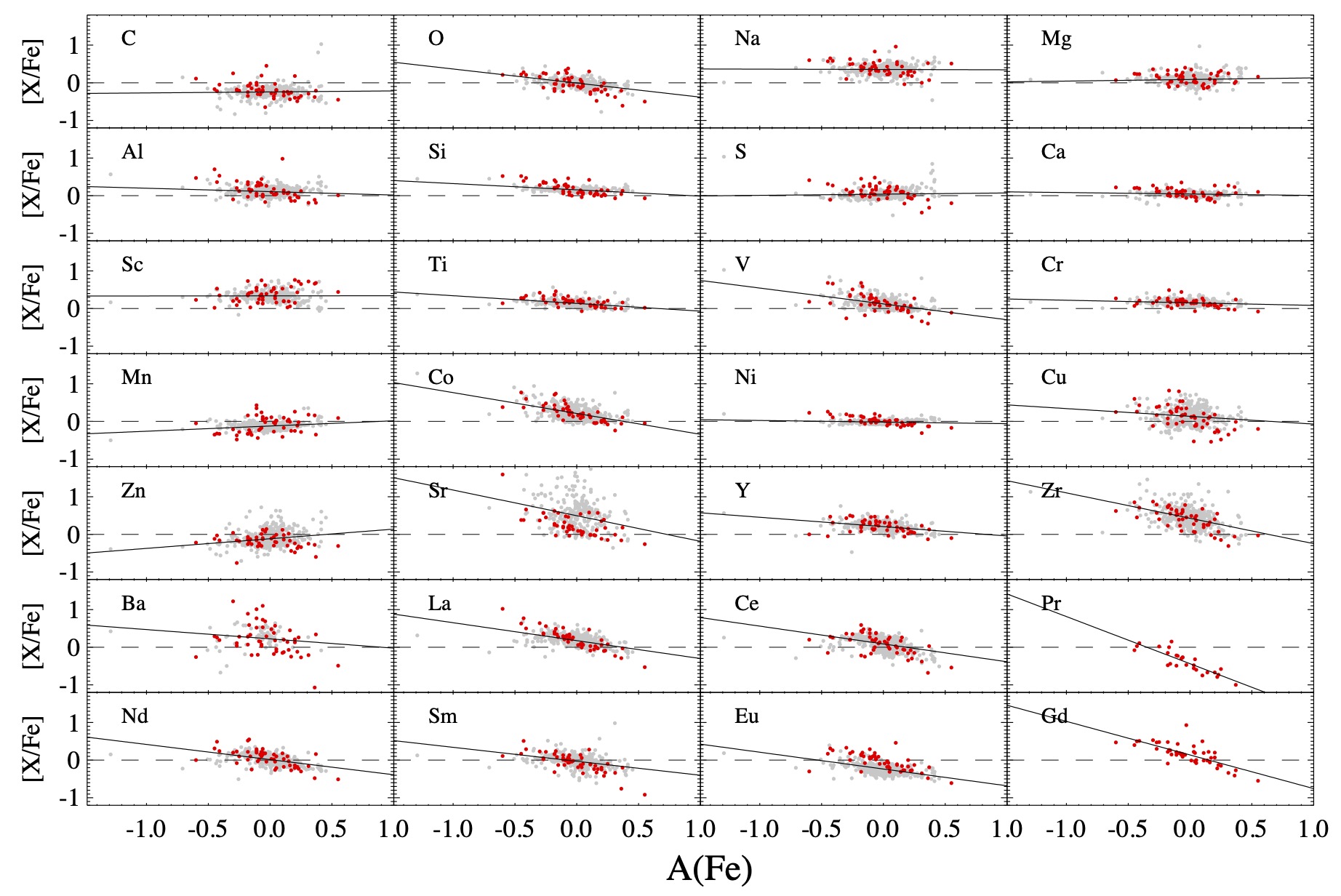}
   \caption{Comparison of the abundances derived in this work expressed as [X/Fe] (red filled dots) with those derived for 435 Galactic DCEPs by \citet{luck18} (light gray symbols). The solid line is the best fit computed considering the literature and our sample of stars (coefficients of the line fit are reported in Table~\ref{fitgrad}), while the dashed line represents the solar standard abundance. All abundances are plotted as a function of the iron abundance.} 
   \label{fig:patt_luck}
    \end{figure*}

A large data set of abundances derived for 435 DCEPs (based on an analysis of 1127 spectra), was more  recently published by \citet{luck18}. This author derived abundance gradients using Gaia DR2 parallax data and drew the reader's attention to a possible anti-correlation between [O/Fe] and iron abundances, i.e. an increase of iron corresponding to a decrease of [O/Fe]. In Fig.~\ref{fig:patt_luck}, we extended and compared our results with those of \citet{luck18} to all chemical species in common.

We confirm the anti-correlation found for oxygen. In general our data are in agreement with those of \citet{luck18}, even if some elements such as scandium, titanium, zinc, strontium, and zirconium are confined at the lower edge of Luck's  distribution. The coefficients of the linear fit plotted in Fig~\ref{fig:patt_luck} are reported in Table~\ref{fitgrad} as well as the scatter of the data.

\begin{figure*}
   \centering
   \includegraphics[width=\hsize]{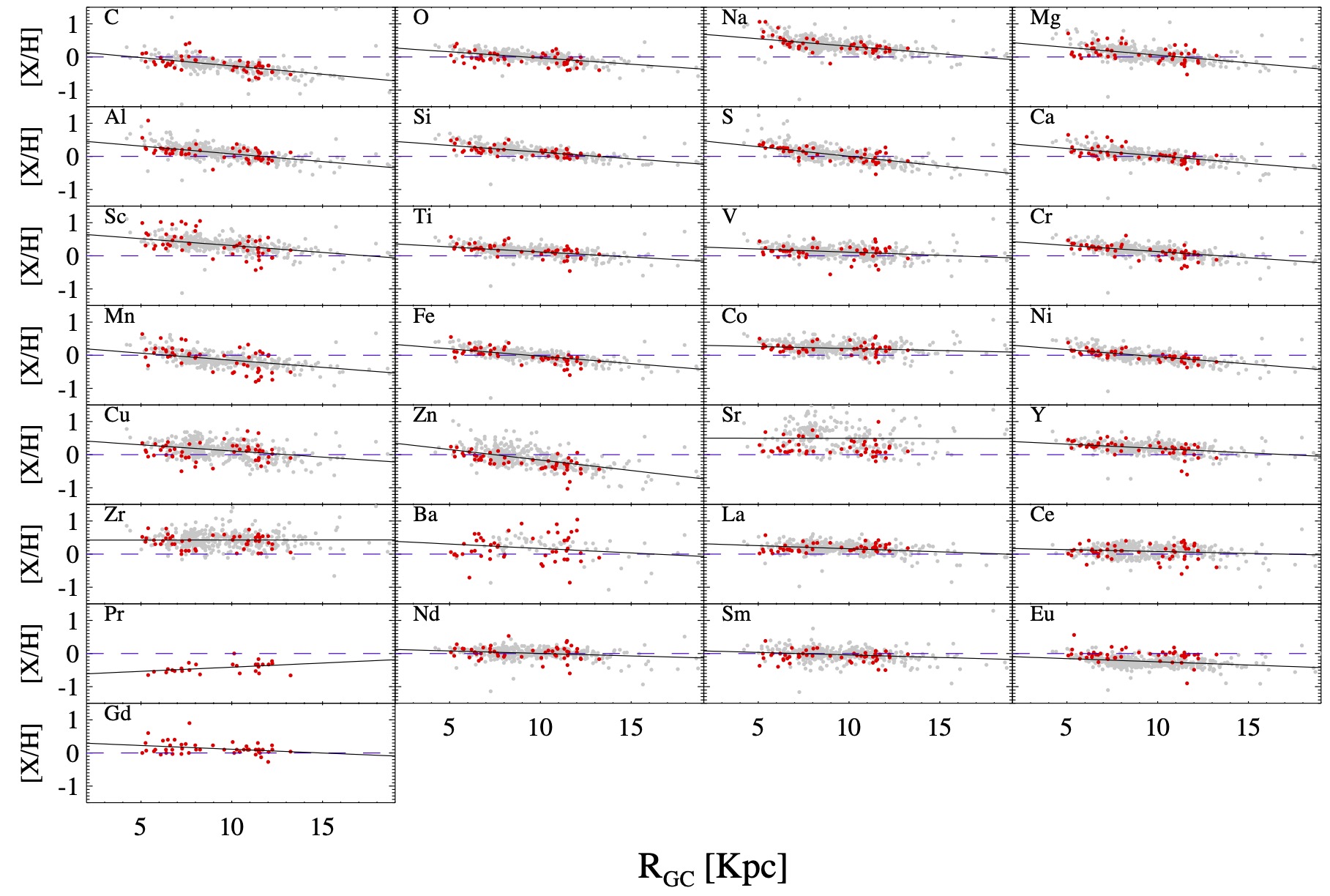}
   \caption{Galactic metallicity gradient for our sample of DECPs (red filled circles) compared with the gradient traced by 435 Galactic DCEPs studied in  \citet{luck18} (light gray symbols). The solid line is the best fit computed considering the literature and our sample of stars (coefficients of the line fit are reported in Table~\ref{fitgraddist}), while the dashed line represents the solar standard abundance.} 
   \label{fig:grad_luck}
    \end{figure*}

We also compared the distribution of our abundances in the context of the Galactic metallicity gradient. In Fig.~\ref{fig:grad_luck}, we reported the results by \citet{luck18} with over-imposed the abundances of our DCEP sample. Our abundances are in agreement with the general trend. The coefficients of the linear fit plotted in Fig~\ref{fig:grad_luck} are reported in Table~\ref{fitgraddist} as well as the scatter of the data.

\begin{table*}
\caption{Results of the photometric analysis of the targets. The meaning of the different columns is: (1) Stars name; (2) mode of pulsation; (3)-(4) Period of pulsation and relative error; (5)-(6) Epoch of maximum light and relative error; (7)-(8) Pulsation amplitude in the $V$ band and relative error; (9)-(22) intensity-averaged magnitude and relative errors in the $B,V,I,J,H,K_s$ bands, respectively; (23) literature sources for the optical photometry.}
\label{Tab:multiband}
\footnotesize\setlength{\tabcolsep}{3pt}
\centering
\begin{tabular}{cccccccccccc}
\hline\hline             
\noalign{\smallskip} 
 Star & Mode & P & P$_{\rm ERR}$  &  E$^{\rm MAX}$  & E$^{\rm MAX}_{\rm ERR}$  & $ampV$ & $ampV_{\rm ERR}$  &  $B$ & $B_{\rm ERR}$ & $V$ & $V_{\rm ERR}$\\
         &  & days & days & days & days & mag & mag & mag & mag & mag & mag \\
	(1) &  	(2) &  	(3) &  	(4) &  	(5) &  	(6) &  	(7) &  	(8) &  	(9) &  	(10) &  	(11) &  	(12) \\
\noalign{\smallskip} \hline \noalign{\smallskip}
ASASSN\_J180946.70-182238.2   &     DCEP\_F    &   7.28954       &  1.3e-04    &  57053.15      &  2.0e-02   &   0.692   &   0.010    &    ---    &     ---     &   11.456     &    0.010    \\
      ASAS\_J052610+1151.3    &     DCEP\_F    &   4.232016      &  1.2e-05    &  51964.8488    &  1.0e-04   &   0.604   &   0.010    &    12.046    &     0.019     &   10.946     &    0.010    \\
      ASAS\_J061022+1438.6    &     DCEP\_1O   &    4.83807      &   2.0e-05   &   52604.26     &   7.0e-02  &    0.258  &    0.010   &     ---   &      ---    &     9.880    &     0.010   \\	
      ASAS\_J063519+2117.8    &     DCEP\_1O   &    1.512365     &   2.0e-06   &   52617.06     &   2.0e-02  &    0.320  &    0.010   &     ---   &      ---    &    11.453    &     0.010   \\	
      ASAS\_J065413+0756.5    &     DCEP\_1O   &    0.7774053    &   4.0e-07   &   52547.541    &   6.0e-03  &    0.302  &    0.010   &     ---   &      ---    &    11.362    &     0.010   \\
      ASAS\_J070832-1454.5    &     DCEP\_1O   &    6.38771      &   4.0e-05   &   51844.56     &   1.3e-01  &    0.144  &    0.010   &     10.809   &      0.016    &     9.696    &     0.010   \\	
      ASAS\_J070911-1217.2    &     DCEP\_1O   &    2.412282     &   9.0e-06   &   51860.011    &   2.0e-03  &    0.280  &    0.010   &     11.544   &      0.019    &    10.603    &     0.010   \\	
      ASAS\_J072424-0751.3    &     DCEP\_1O   &    2.071284     &   4.0e-06   &   51862.56755  &   1.0e-05  &    0.224  &    0.010   &     11.092   &      0.020    &    10.352    &     0.010   \\	
      ASAS\_J074401-1707.8    &     DCEP\_1O   &    2.629941     &   3.0e-06   &   51859.00045  &   2.0e-05  &    0.445  &    0.010   &     ---   &      ---    &    11.510    &     0.010   \\
      ASAS\_J074412-1704.9    &     DCEP\_1O   &    4.7013       &   3.0e-05   &   51853.20     &   3.0e-02  &    0.267  &    0.010   &     ---   &      ---    &    10.196    &     0.010   \\
      ASAS\_J162326-0941.0    &     DCEP\_1O   &    1.378236     &   3.0e-06   &   51931.41005  &   1.0e-05  &    0.261  &    0.010   &     10.607   &      0.010    &     9.817    &     0.010   \\	
      ASAS\_J180342-2211.0    &     DCEP\_F    &   42.6272       &  1.2e-03    &  51806.225     &  1.4e-02   &   0.785   &   0.010    &    ---    &     ---     &   10.907     &    0.010    \\
      ASAS\_J182714-1507.1    &     DCEP\_F    &   5.54565       &  2.0e-05    &  51946.2010    &  3.0e-04   &   0.642   &   0.010    &    ---    &     ---     &   10.692     &    0.010    \\
      ASAS\_J183347-0448.6    &     DCEP\_1O   &    3.102301     &   8.0e-06   &   51953.99     &   2.0e-02  &    0.300  &    0.010   &     12.354   &      0.021    &    10.846    &     0.010   \\	
      ASAS\_J183652-0907.1    &     DCEP\_1O   &    2.589889     &   7.0e-06   &   51955.25     &   1.4e-02  &    0.236  &    0.010   &     12.973   &      0.013    &    11.600    &     0.010   \\	
      ASAS\_J183904-1049.3    &     DCEP\_1O   &    3.05774      &   1.0e-05   &   51951.79     &   5.0e-02  &    0.228  &    0.010   &     11.913   &      0.037    &    10.590    &     0.010   \\	
      ASAS\_J192007+1247.7    &     DCEP\_F    &   8.62797       &  4.0e-05    &  52673.95390   &  4.0e-05   &   0.423   &   0.010    &    ---    &     ---     &   10.359     &    0.010    \\
      ASAS\_J192310+1351.4    &     DCEP\_1O   &    2.972999     &   5.0e-06   &   52702.25806  &   7.0e-05  &    0.395  &    0.010   &     ---   &      ---    &    10.956    &     0.010   \\
                 BD+59\_12    &     DCEP\_1O   &    2.8417       &   7.0e-05   &   56998.98     &   4.0e-02  &    0.200  &    0.010   &     ---   &      ---    &    10.217    &     0.010   \\
                   CF\_Cam    &     DCEP\_F    &   9.4361        &  3.0e-04    &  57007.925     &  2.0e-03   &   0.573   &   0.010    &    ---    &     ---     &   12.118     &    0.010    \\
   DR2\_468646563398354176    &     DCEP\_1O   &    2.99108      &   5.0e-05   &   56998.81     &   3.0e-02  &    0.374  &    0.010   &     ---   &      ---    &    11.245    &     0.010   \\
   DR2\_514736269771300224    &     DCEP\_1O   &    3.29896      &   3.0e-05   &   56998.5814   &   2.0e-04  &    0.355  &    0.010   &     ---   &      ---    &    10.789    &     0.010   \\
                HD\_160473    &     DCEP\_1O   &    3.779702     &   1.4e-05   &   51928.4838   &   2.0e-04  &    0.180  &    0.010   &     10.680   &      0.005    &     9.490    &     0.010   \\	
                   HO\_Vul    &     DCEP\_F    &   5.630938      &  1.2e-05    &  52716.14412   &  8.0e-05   &   0.718   &   0.010    &    ---    &     ---     &   11.653     &    0.010    \\
         OGLE-GD-CEP-0066    &     DCEP\_1O   &    2.312167     &   4.0e-06   &   51972.95     &   3.0e-02  &    0.395  &    0.010   &     ---   &      ---    &    11.504    &     0.010   \\	
         OGLE-GD-CEP-0104    &     DCEP\_1O   &    2.177743     &   2.0e-06   &   51863.502    &   7.0e-03  &    0.336  &    0.010   &     ---   &      ---    &    11.508    &     0.010   \\	
                   OO\_Pup    &     DCEP\_F    &   10.98279      &  3.0e-05    &  51839.4660    &  3.0e-04   &   0.587   &   0.010    &    ---    &     ---     &   11.568     &    0.010    \\	
                   OP\_Pup    &     DCEP\_1O   &    2.598824     &   3.0e-06   &   51858.62722  &   1.0e-05  &    0.439  &    0.010   &     ---   &      ---    &    11.022    &     0.010   \\	
                   OR\_Cam    &     DCEP\_1O   &    3.6852       &   2.0e-04   &   57024.90     &   6.0e-02  &    0.286  &    0.010   &     ---   &      ---    &    10.402    &     0.010   \\
                   TX\_Sct    &     DCEP\_F    &   24.3522       &  2.0e-04    &  51879.853     &  2.0e-03   &   0.873   &   0.010    &    ---    &     ---     &   12.422     &    0.010    \\
                V1495\_Aql    &     DCEP\_F    &   8.79653       &  4.0e-05    &  51938.46232   &  1.0e-05   &   0.629   &   0.010    &    13.771    &     0.034     &   11.932     &    0.010    \\	
                V1496\_Aql    &     DCEP\_F    &   65.83         &  7.0e-02    &  56474.6       &  1.0e-01   &   0.543   &   0.010    &    12.449    &     0.028     &   10.226     &    0.010    \\
                V1788\_Cyg    &     DCEP\_F    &   14.09383      &  8.0e-05    &  52582.99      &  1.2e-02   &   0.753   &   0.010    &    ---    &     ---     &   12.709     &    0.010    \\
                V2475\_Cyg    &     DCEP\_F    &   11.5559       &  3.0e-04    &  57042.879     &  3.0e-03   &   0.954   &   0.010    &    ---    &     ---     &   12.549     &    0.010    \\
                 V355\_Sge    &     DCEP\_F    &   32.071        &  6.0e-06    &  52721.69      &  2.0e-02   &   0.660   &   0.012    &    14.966    &     0.010     &   12.280     &    0.010    \\	
                 V371\_Gem    &     DCEP\_1O   &    2.137457     &   9.0e-06   &   52614.91     &   3.0e-02  &    0.381  &    0.010   &     ---   &      ---    &    10.753    &     0.010   \\
                 V383\_Cyg    &     DCEP\_F    &   4.612297      &  6.0e-06    &  46974.62802   &  1.0e-05   &   0.547   &   0.010    &    12.512    &     0.010     &   10.883     &    0.010    \\
                 V389\_Sct    &     DCEP\_1O   &    3.793899     &   6.0e-06   &   51950.18481  &   6.0e-05  &    0.379  &    0.010   &     ---   &      ---    &    11.324    &     0.010   \\	
                 V536\_Ser    &     DCEP\_F    &   9.00985       &  5.0e-05    &  51917.9       &  1.1e-01   &   0.514   &   0.010    &    ---    &     ---     &   11.780     &    0.010    \\
                V5567\_Sgr    &     DCEP\_F    &   9.76289       &  5.0e-05    &  51920.97      &  7.0e-02   &   0.558   &   0.010    &    ---    &     ---     &   10.560     &    0.010    \\
                 V598\_Per    &     DCEP\_F    &   5.66819       &  4.0e-05    &  56988.6154    &  6.0e-04   &   0.809   &   0.010    &    ---    &     ---     &   12.396     &    0.010    \\
                 V824\_Cas    &     DCEP\_1O   &    5.35066      &   2.0e-05   &   50294.73783  &   1.0e-05  &    0.315  &    0.010   &     12.535   &      0.015    &    11.224    &     0.010   \\
                 V912\_Aql    &     DCEP\_F    &   4.4004        &  7.0e-06    &  48552.15605   &  1.0e-05   &   0.707   &   0.010    &    13.431    &     0.013     &   11.359     &    0.010    \\
                 V914\_Mon    &    DCEP\_1O    &   2.680631      &  1.5e-05    &  52542.15      &  4.0e-02   &   0.393   &   0.010    &    ---    &     ---     &   10.762     &    0.010    \\
                 V946\_Cas    &    DCEP\_1O    &   4.238789      &  1.0e-05    &  52715.69      &  3.0e-02   &   0.480   &   0.010    &    ---    &     ---     &   11.544     &    0.010    \\
                 V966\_Mon    &   DCEP\_1O/2O  &   1.068037      &  2.0e-06    &  52523.80      &  2.0e-02   &   0.281   &   0.010    &    ---    &     ---     &   11.728     &    0.010    \\
                    X\_Sct    &     DCEP\_F    &   4.198051      &  6.0e-06    &  51949.75476   &  1.0e-05   &   0.843   &   0.010    &    11.169    &     0.020     &   10.009     &    0.010    \\
  ZTF\_J000234.99+650517.9    &     DCEP\_F    &   10.9323       &  6.0e-04    &  57002.19      &  7.0e-02   &   0.340   &   0.010    &    ---    &     ---     &   11.519     &    0.010    \\
\noalign{\smallskip} \hline \noalign{\smallskip}
\end{tabular}
\end{table*}


\begin{table*}
\addtocounter{table}{-1}
\caption{{\it ...continued.}}
\footnotesize\setlength{\tabcolsep}{3pt}
\centering
\begin{tabular}{cccccccccccc} 
\hline\hline             
\noalign{\smallskip} 
 Star & $I$ & $I_{\rm ERR}$  &  $J$  & $J_{\rm ERR}$  & $H$ & $H_{\rm ERR}$ & $K_s$ & ${K_s}_{\rm ERR}$ & $E(B-V)$ & $E(B-V)_{\rm ERR}$ & Source\\
         & mag & mag & mag & mag & mag & mag & mag & mag & mag & mag & \\
	(13) &  (14) &  (15) &  (16) &  (17) &  (18) &  (19) &  (20) &  (21) & 	(22) &  	(23) & (24) \\
\noalign{\smallskip} \hline \noalign{\smallskip}

ASASSN\_J180946.70-182238.2    &      ---     &    ---     &    7.421     &    0.031     &    6.727    &     0.051     &    6.392     &    0.023     &    1.113     &    0.075     &    a              \\
      ASAS\_J052610+1151.3     &       9.706     &    0.019     &    8.724     &    0.020     &    8.334    &     0.033     &    8.118     &    0.020     &    0.506     &    0.036     &    a,b,c,d	   \\
      ASAS\_J061022+1438.6     &       ---    &     ---    &     7.643    &     0.030    &     7.222   &      0.047    &     7.129    &     0.029    &     0.400    &     0.075    &     a,c 	   \\	
      ASAS\_J063519+2117.8     &       ---    &     ---    &     9.997    &     0.029    &     9.676   &      0.028    &     9.578    &     0.026    &     0.234    &     0.075    &     a,c  	   \\	
      ASAS\_J065413+0756.5     &       ---    &     ---    &    10.060    &     0.029    &     9.732   &      0.031    &     9.604    &     0.029    &     0.322    &     0.075    &     a,c	   \\	
      ASAS\_J070832-1454.5     &        8.349    &     0.015    &     7.224    &     0.029    &     6.797   &      0.048    &     6.636    &     0.026    &     0.434    &     0.034    &     a,c,e	   \\	
      ASAS\_J070911-1217.2     &        9.474    &     0.015    &     8.476    &     0.028    &     8.111   &      0.039    &     7.922    &     0.033    &     0.405    &     0.039    &     a,c,e	   \\	
      ASAS\_J072424-0751.3     &        9.511    &     0.026    &     8.844    &     0.036    &     8.531   &      0.043    &     8.440    &     0.030    &     0.216    &     0.036    &     a,c,e	   \\	
      ASAS\_J074401-1707.8     &       ---    &     ---    &    10.078    &     0.031    &     9.758   &      0.030    &     9.651    &     0.028    &     0.141    &     0.075    &     a,c	   \\	
      ASAS\_J074412-1704.9     &       ---    &     ---    &     8.778    &     0.031    &     8.324   &      0.041    &     8.209    &     0.045    &     0.166    &     0.075    &     c		   \\
      ASAS\_J162326-0941.0     &        8.964    &     0.014    &     8.351    &     0.039    &     8.042   &      0.051    &     7.931    &     0.031    &     0.293    &     0.034    &     a,c,e	   \\	
      ASAS\_J180342-2211.0     &      ---     &    ---     &    5.779     &    0.021     &    4.966    &     0.042     &    4.484     &    0.033     &    1.492     &    0.075     &    a,c		   \\
      ASAS\_J182714-1507.1     &      ---     &    ---     &    7.850     &    0.020     &    7.308    &     0.020     &    7.124     &    0.024     &    0.706     &    0.075     &    a,c		   \\
      ASAS\_J183347-0448.6     &        8.979    &     0.019    &     7.449    &     0.033    &     7.005   &      0.035    &     6.765    &     0.029    &     0.955    &     0.036    &     a,c,e	   \\	
      ASAS\_J183652-0907.1     &        9.560    &     0.010    &     8.249    &     0.028    &     7.685   &      0.029    &     7.445    &     0.031    &     0.983    &     0.034    &     a,c,e	   \\	
      ASAS\_J183904-1049.3     &        8.925    &     0.040    &     7.793    &     0.029    &     7.282   &      0.027    &     7.087    &     0.026    &     0.779    &     0.040    &     a,c,e	   \\	
      ASAS\_J192007+1247.7     &      ---     &    ---     &    6.404     &    0.021     &    5.905    &     0.038     &    5.574     &    0.017     &    1.139     &    0.075     &    a,c		   \\
      ASAS\_J192310+1351.4     &       ---    &     ---    &     7.416    &     0.030    &     6.903   &      0.039    &     6.689    &     0.026    &     1.001    &     0.075    &     a,c	   \\	
                 BD+59\_12     &       ---    &     ---    &     8.149    &     0.037    &     7.759   &      0.029    &     7.620    &     0.030    &     0.419    &     0.075    &     a		   \\
                   CF\_Cam     &      ---     &    ---     &    8.445     &    0.022     &    7.819    &     0.024     &    7.543     &    0.024     &    1.003     &    0.075     &    a		   \\
   DR2\_468646563398354176     &       ---    &     ---    &     9.027    &     0.032    &     8.487   &      0.029    &     8.323    &     0.028    &     0.572    &     0.075    &     a		   \\
   DR2\_514736269771300224     &       ---    &     ---    &     8.486    &     0.035    &     8.038   &      0.028    &     7.864    &     0.029    &     0.513    &     0.075    &     a		   \\
                HD\_160473     &        7.901    &     0.010    &     6.756    &     0.030    &     6.254   &      0.045    &     6.048    &     0.027    &     0.651    &     0.033    &     a,c,e	   \\	
                   HO\_Vul     &      ---     &    ---     &    7.825     &    0.019     &    7.217    &     0.029     &    6.996     &    0.026     &    1.188     &    0.075     &    a,c		   \\
         OGLE-GD-CEP-0066     &       10.522    &     0.010    &     9.645    &     0.029    &     9.198   &      0.029    &     9.035    &     0.029    &     0.375    &     0.049    &     a,c,f	   \\	
         OGLE-GD-CEP-0104     &       10.390    &     0.010    &     9.791    &     0.033    &     9.364   &      0.030    &     9.188    &     0.029    &     0.397    &     0.046    &     a,c,f	   \\	
                   OO\_Pup     &      ---     &    ---     &    8.813     &    0.030     &    8.260    &     0.055     &    8.039     &    0.023     &    0.608     &    0.075     &    a,c,i	   \\	
                   OP\_Pup     &       ---    &     ---    &     9.046    &     0.031    &     8.697   &      0.030    &     8.606    &     0.033    &     0.294    &     0.075    &     a,c 	   \\	
                   OR\_Cam     &       ---    &     ---    &     7.014    &     0.028    &     6.455   &      0.026    &     6.211    &     0.026    &     1.035    &     0.075    &     a		   \\
                   TX\_Sct     &      ---     &    ---     &    7.004     &    0.029     &    6.049    &     0.027     &    5.638     &    0.021     &    1.635     &    0.075     &    c		   \\
                V1495\_Aql     &       9.765     &    0.017     &    8.230     &    0.026     &    7.571    &     0.053     &    7.302     &    0.024     &    1.127     &    0.044     &    a,c,g	   \\	
                V1496\_Aql     &       7.756     &    0.019     &    5.951     &    0.023     &    5.302    &     0.029     &    4.918     &    0.023     &    1.161     &    0.037     &    a,c,g,h	   \\
                V1788\_Cyg     &      ---     &    ---     &    7.639     &    0.039     &    6.722    &     0.026     &    6.402     &    0.017     &    1.553     &    0.075     &    a,i		   \\
                V2475\_Cyg     &      ---     &    ---     &    8.287     &    0.023     &    7.502    &     0.024     &    7.173     &    0.020     &    1.249     &    0.075     &    a		   \\
                 V355\_Sge     &       9.291     &    0.011     &    7.049     &    0.020     &    6.165    &     0.016     &    5.819     &    0.016     &    1.671     &    0.034     &    a,c,g	   \\	
                 V371\_Gem     &       ---    &     ---    &     8.774    &     0.028    &     8.484   &      0.029    &     8.358    &     0.027    &     0.328    &     0.075    &     c		   \\
                 V383\_Cyg     &       8.964     &    0.072     &    7.479     &    0.020     &    6.900    &     0.031     &    6.672     &    0.016     &    1.024     &    0.038     &    a,i,j,k	   \\
                 V389\_Sct     &       ---    &     ---    &     7.575    &     0.056    &     7.016   &      0.051    &     6.737    &     0.026    &     1.040    &     0.075    &     a,c,i	   \\	
                 V536\_Ser     &      ---     &    ---     &    7.782     &    0.023     &    7.091    &     0.040     &    6.775     &    0.020     &    1.249     &    0.075     &    a,c		   \\
                V5567\_Sgr     &      ---     &    ---     &    7.356     &    0.027     &    6.734    &     0.053     &    6.439     &    0.023     &    0.836     &    0.075     &    a,c    	   \\
                 V598\_Per     &      ---     &    ---     &    8.354     &    0.024     &    7.636    &     0.031     &    7.383     &    0.017     &    1.246     &    0.075     &    a		   \\
                 V824\_Cas     &       ---    &     ---    &     8.109    &     0.030    &     7.658   &      0.047    &     7.397    &     0.028    &     0.675    &     0.044    &     a,e	   \\	
                 V912\_Aql     &       8.879     &    0.012     &    7.118     &    0.026     &    6.448    &     0.042     &    6.149     &    0.018     &    1.473     &    0.034     &    a,c,g,j	   \\
                 V914\_Mon     &      ---     &    ---     &    8.981     &    0.027     &    8.573    &     0.033     &    8.430     &    0.031     &    0.330     &    0.075     &    c		   \\
                 V946\_Cas     &      ---     &    ---     &    8.563     &    0.044     &    8.075    &     0.034     &    7.899     &    0.035     &    0.709     &    0.075     &    a,i		   \\
                 V966\_Mon     &      ---     &    ---     &    9.443     &    0.031     &    9.069    &     0.031     &    8.920     &    0.030     &    0.575     &    0.075     &    a,c		   \\
                    X\_Sct     &       8.622     &    0.020     &    7.457     &    0.030     &    6.978    &     0.057     &    6.769     &    0.027     &    0.591     &    0.036     &    c,g		   \\
  ZTF\_J000234.99+650517.9     &      ---     &    ---     &    7.812     &    0.026     &    7.119    &     0.026     &    6.891     &    0.017     &    1.056     &    0.075     &    a              \\
\noalign{\smallskip} \hline \noalign{\smallskip}

\multicolumn{12}{l}{a=ASASSN \citep{Shappee2014,Kochanek2017};b=\citet{Berdnikov2011}; c=ASAS \citep{Pojmanski2002};d=\citet{Schmidt2011}} \\
\multicolumn{12}{l}{e=\citet{Berdnikov2009}; f=\citet{Udalski2018}; g=\citet{Berdnikov2015}; h=\citet{Berdnikov2004}; i=\citet{IOMC2012};}\\ 
\multicolumn{12}{l}{j=\citet{Schmidt2005}; k=\citet{Berdnikov2008}}
\end{tabular}
\end{table*}

\section{Complementary photometry and reddening}

We collected multiband photometry available in the  literature for our program stars, in order to accurately determine the pulsation period, time of maximum light (epoch), reddening 
$E(B-V)$ and intensity averaged magnitudes of each target. These quantities are needed to construct the PL/PW relations.
References for the literature photometry are provided in the last columns of Table~\ref{Tab:multiband}. Our use of the different photometric datasets is  discussed in the following sections.

\subsection{Period and epoch of maximum light estimates from optical photometry}
\label{Sect:period}
An accurate determination of the period and time of maximum light  
is crucial to correctly estimate the average magnitude of the program stars, particularly 
in the $J,H,K_s$ bands, as only single epoch NIR photometry is often available 
in the literature 
(see Sect.~\ref{Sect:NIR}). Periods are available in the literature for almost all our targets. 
However, more accurate period (and related error) estimates  can be obtained by merging the photometry of different surveys, in this way spanning  much longer time intervals than it is  covered by each single dataset. Johnson $V$-band photometry spanning more than 6000 days is available for the majority of our stars by 
combining the ASAS \citep[All Sky Automated Survey][]{Pojmanski2002} and ASASSN \citep{Shappee2014,Kochanek2017} surveys, which we  complemented with other literature data (see Table~\ref{Tab:multiband}), when possible. 
Small zero point differences on the order of hundredths of magnitude are present among different catalogues. They  were fixed as follows: when the Johnson $B,V,I$ photometry by \citet{Berdnikov2004,Berdnikov2009,Berdnikov2011,Berdnikov2015} was available, we used these studies as reference (to keep the information on color); in all other cases our reference was the ASAS catalogue. 


A new determination of the period for all our targets was made by applying to the $V$-band time-series data the same multi-step procedure  as the one devised for  the processing and specific characterisation of the DCEPs and RR Lyrae stars observed by the {\it Gaia} mission (\citealt{Clementini2016, Clementini2019}).
That is, a first estimate of the pulsation frequency was obtained with the Lomb-Scargle (LS) algorithm \citep{Lomb1976,Scargle1982}, by selecting the frequency with the highest power peak in the periodogram.
Then, the observations were folded using the selected frequency $\nu$  and modeled by means of a truncated Fourier series:
\begin{equation}\label{eq-LinFouModel}
V(\phi_k) = A_0 + \sum_{i=1}^{n_{max}}A_i \cdot \sin(i 2 \pi \nu \phi_k  + \Phi_i)
\end{equation}
where $\phi_k$ and $V_k$ are the phases and the magnitudes of the $k^{th}$ observation and the sum runs over the maximum number of harmonics of the model. To obtain the $A_0$, $A_i$ and $\Phi_i$ values we first run a linear  least square fitting to the data (since we assumed the frequency from the LS procedure). 
Starting from the linear estimate of  $A_0$, $A_i$ and $\Phi_i$, we then refined the value of the $\nu^{lin}$ frequency by fitting the following truncated Fourier series:
  \begin{equation}\label{eq-NonLinFouModel}
  V(t_k) = A_0 + \sum_{i=1}^{n_{max}}A_i \cdot \sin(i 2 \pi \nu t_k  + \Phi_i)
  \end{equation}

\noindent
where the frequency $\nu$ was now left free to vary. In this way, the least-square fit becames non--linear and was treated by means of a standard Levenberg--Marquardt algorithm \citep{Marquardt1963}, thus obtaining more accurate pulsation frequencies, amplitudes and phases.  
To estimate the uncertainty on the derived parameters, we performed 1000 bootstrap simulations, where the time series data were re-sampled and the non-linear fitting procedure was repeated for every simulation. The robust standard deviation ($\sigma_{robust}=1.4826 \cdot MAD$) of the resulting frequency, amplitudes and phase was adopted as an estimate of the uncertainty in these quantities.
Finally, the time of maximum light of each target  was calculated as the Heliocentric Julian day (HJD) of the brightest value of the V-band light curve which was closer to the HJD of the first observations minus three times the final pulsation period (inverse of the  best-fit frequency).   


The $V$-band light curves of our 48 program stars obtained by folding the literature photometry according to the new periods and epochs of maximum light derived in our study are shown in Fig.~\ref{fig:LC}. 

The procedure to estimate the pulsation period and the epoch of our targets  naturally provides  also their $V$-band intensity-averaged magnitudes and peak-to-peak amplitudes along with the associated errors. All these quantities are presented in Table~\ref{Tab:multiband}.

In addition to the $V$-band photometry, time-series in other bands of interest for the distance scale, i.e. Johnson-Cousins $B,I$ bands also available for several targets (see Table~\ref{Tab:multiband}). We  adopted the periods derived from the $V$ band photometry to model the light curves and derive the intensity-averaged $B,I$ magnitudes, which are provided in Table~\ref{Tab:multiband}.

\subsection{NIR photometry}
\label{Sect:NIR}

Time series $JHK_s$ photometry is available in the literature only for X Sct \citep{Monson2011} among our targets. For the remaining stars we  used the single-epoch photometry from  2MASS \citep{Skrutskie2006}\footnote{We note that 47 over 48 targets have an "AAA" quality flag in the 2MASS single-epoch photometry, indicating  fully usable measurements, with the exception of ASAS J180342-2211.0 that has an "E" flag in the $H$ band. Therefore, for this star the $H$ value is less accurate than for the others.}, along with a template fitting procedure as implemented by \citet{Sos2005}. A crucial point for the template fitting technique is the accuracy of the available periods and epochs. It is also very important that only a few pulsation cycles separate the epochs of the variable stars and the 2MASS observations, in order to minimise errors in phasing the NIR photometry.
We verified that phasing errors [$(P_{err}*n_{cycles})/P$] are negligible for all our stars (the maximum value being $\sim0.01$ for ASASSN J180946.70-182238.2). 

The other quantity required by the template procedure 
is the $V$-band amplitude, which we have derived for all our targets as described in Sect.~\ref{Sect:period}. We calculated intensity averaged $JHK_s$ magnitudes for the target stars, assigning an uncertainty equal to the sum in quadrature of the 2MASS uncertainty on the epoch photometry plus 0.02 mag for all the stars, to take into account the average uncertainty of the template fitting procedure \citep[][estimate an average error of 0.03 mag for their method. This value already includes the uncertainty of the single-epoch photometry. To avoid summing that latter contribution twice we adopted an average uncertainty of 0.02 mag for the template fitting procedure]{Sos2005}.  
Final values for the $J$, $H$ and $K_s$ mean magnitudes are summarised in Table~\ref{Tab:multiband}.

\subsection{Reddening}
\label{Sect:reddening}

Obtaining correct values of reddening for DCEPs is mandatory to construct and use the PL relations (the PW ones are reddening-free by definition). 
Many different methods have been developed in the past decades to face this problem \citep[see e.g.][and references therein]{Fernie1967,Dean1978,Pel1978,Tammann2003,Kovtyukh2008,Andrievsky2012,Kashuba2016,Turner2016}. 

Reddening estimates are present in the literature only for V5567 Sgr \citep[E(B-V)=0.960$\pm$0.096 mag\footnote{Value scaled by 0.97, according to \citet{Groenewegen2018}}, ][]{Martin2015} and X Sct \citep[E(B-V)=0.581$\pm$0.030 mag\footnote{Value scaled by 0.94, according to \citet{Groenewegen2018}}, ][]{Fernie1995}, among our 48  targets. For the remaining objects, the reddening was inferred from  the Period-Color (PC) relations, which are known to offer a reliable and straightforward method, widely used in the literature \citep[see e.g.][]{Tammann2003}. 

The most commonly used PC relations relying on the $(B-V)$ and $(V-I)$ colors were published by \citet{Tammann2003}. We decided to determine our own relations by using the moderately larger DCEP sample by \citet{Ngeow2012b} whose $E(B-V)$ values are mainly based on the compilation of \citet{Fernie1995},  multiplied by 0.95 as in \citet{Tammann2003}, so that our results will be basically compatible with those Authors. The sample by \citet{Ngeow2012a} consists of 349 usable objects, including first overtone pulsators, whose periods were fundamentalised using the relation $P_F = P_{1O}/(0.716-0.027 \log P_{1O})$, where $P_F$ and $P_{1O}$ are the fundamental  and first overtone DCEPs (DCEP\_F and DCEP\_1O) periods, respectively \citep[][]{Feast1997}. Finally, we performed a least-square fit to the data adopting  a $\sigma$-clipping technique with a $2.5\sigma$ threshold. The resulting PCs are:

\begin{eqnarray}
(B-V)_0=(0.395\pm0.010)(\log P-1)+(0.716\pm0.003) \label{eq1} \\
(V-I)_0=(0.318\pm0.012)(\log P-1)+(0.757\pm0.003) \label{eq2}
\end{eqnarray}

\noindent
where the colors were dereddened using the \citet{Cardelli1989} law with a ratio of total to selective absorption of $R_V$=3.1. Equations~\ref{eq1} and \ref{eq2} have rms=0.051 mag and 0.054 mag, respectively, i.e. are less dispersed than to the corresponding relations by \citet{Tammann2003}. Note also the use of the pivoting period at 10 days to reduce the correlation between  slopes and intercepts. These equations compare with the analogous relations by \citet{Tammann2003} as follows: $\Delta (B-V)_0=0.029 \log P -0.04$; $\Delta (V-I)_0=0.062 \log P -0.058$, where $\Delta$ stands for "this work $-$\citet{Tammann2003}". It can be seen that the results given in the two cases are compatible within the uncertainties, hence in the following we will use our relations. 

As shown in Table~\ref{Tab:multiband}, $B$ and $I$ photometry is missing for 32 and 31 targets, respectively. We can estimate the $(V-I)$ color of these sources (as well as of the whole data set) using the {\it Gaia} EDR3 magnitudes after properly converted them into $(V-I)$. To this aim, we used the \citet{Ngeow2012a} sample, along with the {\it Gaia} DR2  intensity-averaged $G$, $G_{BP}$ and $G_{RP}$ magnitudes \citep{Clementini2019,Ripepi2019} to derive the following conversion equation\footnote{The correct average magnitude of a DCEP is normally calculated by integrating the light curve in intensity and then transformed back into
magnitude. This is the procedure applied to derive the mean magnitudes of Cepheids and RR Lyrae stars published in the "vari" tables of Gaia \citep[][]{Clementini2016,Clementini2019}. The Gaia magnitudes of the "gaia\_source" general catalog are instead calculated through weighted arithmetic means and can differ from the "correct" ones by several hundredths of magnitude \citep[see e.g.][]{Caputo1999}. However, this effect becomes of the order of 1-2\% when using colors instead of magnitudes, because 
the "delta" magnitude, albeit decreasingly large from the bluest to reddest filters, has the same sign in all bands, therefore when subtracting two magnitudes to calculate a color, the value of "delta" becomes very small. 
and generally negligible with respect to the uncertainty introduced by the use of the Period-Color relation.}:

\begin{equation}
\begin{split}
(V-I)_G=G-G_{RP}+(0.5383\pm0.003)+(0.5492\pm0.006) \alpha \\
+(0.1126 \pm 0.014) \alpha^2    
\label{Eq:eqgaia}
\end{split}
\end{equation}

\noindent 
where $(V-I)_G$ is the color estimated from the {\it Gaia} photometry and  $\alpha=[(G_{BP}-G_{RP})-1.5]$. The rms of Eq.~\ref{Eq:eqgaia} is  0.022 mag. 

In the end, for 15 DCEPs with available $B,V,I$, we were able to obtain three estimates of the  $E(B-V)$, using the $(B-V)$, the $(V-I)$ and the $(V-I)_G$ colors, separately. For additional two sources we calculated two reddening values, depending on the availability of the $(B-V)$ and $(V-I)$ colors, respectively. For these 17 sources we adopted a weighted mean of the three (two)  measures as our best estimate of $E(B-V)$. 
For the remaining 31 stars, the $E(B-V)$ reddening could be estimated only from the $(V-I)_G$ color, hence carries errors about twice than for the previous 12 sources.
The individual reddening values are listed in Table~\ref{Tab:multiband}. 

   \begin{figure}
   \centering
   \includegraphics[width=8.5cm]{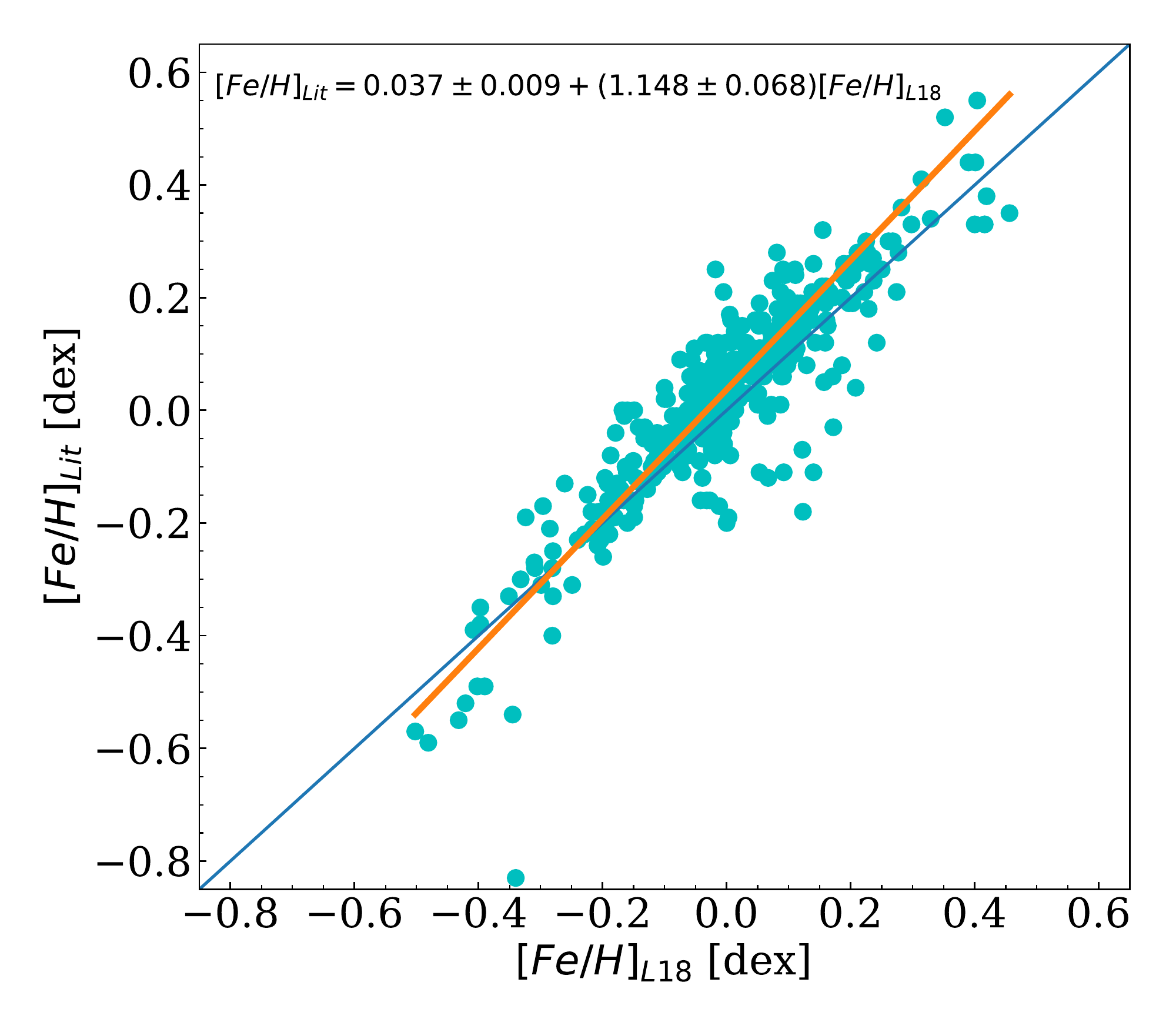}
      \caption{Comparison between the iron abundances from different literature sources and \citet{luck18}. The light blue and orange lines are the bisector and the fit to the data, respectively.              }
         \label{fig:compFeh}
   \end{figure}

   \begin{figure}
   \centering
   \includegraphics[width=8.5cm,angle=0]{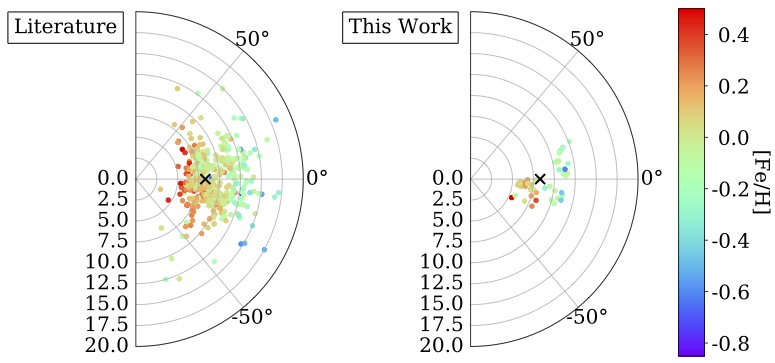}
      \caption{Position in the Galaxy of the Literature DCEPs (left panel) and those investigated in this work (right panel). The Sun is shown with a cross. The DCEPs symbols are color-coded according to their metallicity.   
              }
         \label{fig:position}
   \end{figure}

\section{Literature sample}
\label{Sect:Sample}

The main literature sources we adopt for our study is the updated catalog by \citet{Groenewegen2018}, from which we took the $V,J,H,K_s$ magnitudes as well as E(B-V) and [Fe/H] values on a uniform scale for 438 DCEPs\footnote{The original catalog includes 452 objects, 14 sources were not considered because their classification as DCEPs is highly uncertain}. This sample is complemented by $V,I,J,H,K_s$ magnitudes, E(B-V) and [Fe/H] values compiled by \citet{Clementini2017}. For a few stars we used photometry by \citet{vanLeeuwen2007} and \citet{Genovali2014}. For the $V,I$ magnitudes the main source was \citet{Ngeow2012a}\footnote{When the $V$ magnitudes was present in both \citet{Groenewegen2018} and \citet{Ngeow2012a}, we adopted this last work values to keep consistency with the $I$ band.}. The $J,H,K_s$ magnitudes listed by \citet{Groenewegen2018} and \citet{vanLeeuwen2007} were transformed in the 2MASS system \citep{Skrutskie2006} using the transformations outlined by \citet[][see Sect.2 for details]{Groenewegen2018}. The $J,H,K_s$ magnitudes obtained from the other sources were already in the 2MASS system. The uncertainties on the individual $J,H,K_s$ magnitudes by \citet{Groenewegen2018} were assigned according the criteria adopted in that paper. For the stars collected from other sources, we used the original individual uncertainties, or assigned the uncertainties in a similar way as in \citet{Groenewegen2018}. For the $V,I$ magnitudes we imposed a conservative uniform error of 0.02 mag for all the stars.
The periods and the sub-classification in F or 1O or multi-mode DCEPs were taken by  \citet{Ripepi2019,Ripepi2020a} and \citet{{Skowron2019}}. 
The total sample includes 453 DCEPs, and precisely, 391 DCEP\_F, 43 DCEP\_1O, 17 DCEP\_F/1O, and 2 DCEP\_1O/2O. For the multi-mode objects we used the longest period. 

An important ingredient of our analysis is represented by the iron abundance of  the investigated DCEPs. The values listed by \citet{Groenewegen2018} are based on a collection of literature values reported to a uniform scale by the Author. For comparison, we have also considered the work by \citet{luck18} that has homogeneously estimated the iron abundance for 435 pulsators. We have cross matched his catalog with our sample, finding 414 matches. The remaining 21 non-matching objects have not been included as they are mainly stars suspected of being type II or anomalous Cepheids. Among the 414 matches, we have 357 DCEP\_F, 38 DCEP\_1O, 17 DCEP\_F/1O, and 2 DCEP\_1O/2O, respectively. A comparison between the iron abundance of our literature compilation and \citet{luck18} sample is shown in Fig.~\ref{fig:compFeh}. We note a fair agreement between the two sets of metallicity estimates (apart the very discrepant object GP Per), even if there is a measurable trend: [Fe/H]$_{Lit}=0.037 \pm 0.009+(1.148 \pm 0.068)$[Fe/H]$_{L18}$, where with  [Fe/H]$_{Lit}$ and [Fe/H]$_{L18}$ we mean the samples built with our collection of different sources and the \citet{luck18} one, respectively. Given the small but detectable differences in the two quoted metallicity scales, we decided to carry out our analysis using both samples.

The resulting merged catalogues were cross-matched with {\it Gaia} EDR3 to extract parallaxes, parallax errors, ruwe parameter of the sources which provides an indication of goodness of the astrometric solution \footnote{Section 14.1.2 of "Gaia Data Release 2 Documentation release 1.2"; https://gea.esac.esa.int/archive/documentation/GDR2/}, as well as other parameters needed to calculate the individual zero point corrections to the parallaxes \citep[see][for full details]{Lindegren2021a}. 
The approximate distribution in the Galaxy of the considered objects is shown in Fig.~\ref{fig:position}, where we plot separately the literature and our samples. The different symbols are color coded according to the metallicity value.


\section{Gaia EDR3 parallaxes}

   \begin{figure}
   \centering
         \includegraphics[width=8.7cm]{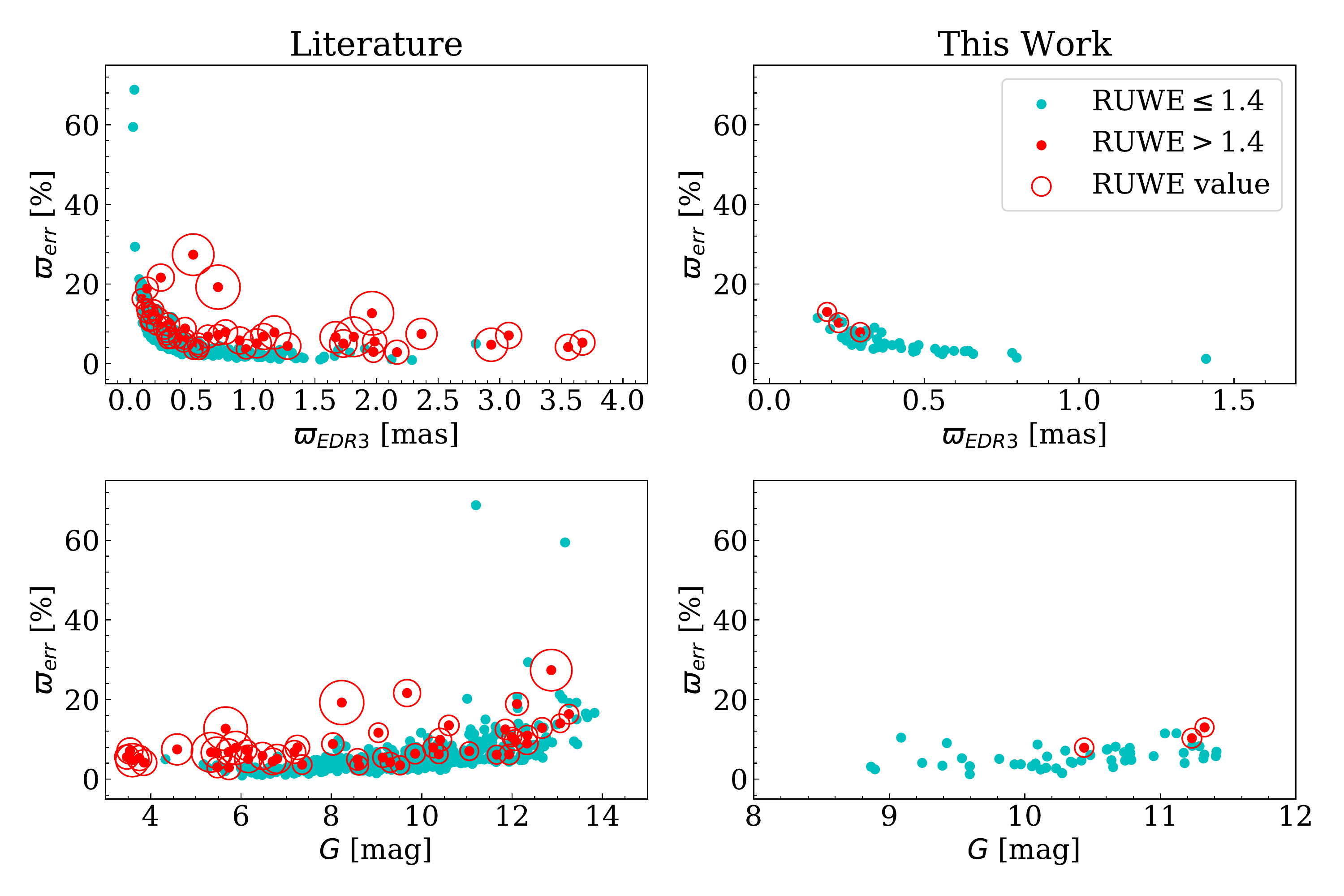}
      \caption{Top panels: parallax versus relative parallax error (\%/100) for the literature (top-left) and our (top-right) samples, respectively. Red and green filled circles represent DCEPs with ruwe parameters below/above the nominal safety threshold, respectively (see labels). The size of the red circles is proportional to the same ruwe parameter.  
      Bottom panels: as in the top panels, but in this case red and green filled circles represent DCEPs with $G$ mag above/below the 6th magnitude (see labels). The size of the circles is always proportional to the ruwe.  
              }
         \label{fig:parallax}
   \end{figure}

\subsection{Parallax values for literature and present work samples.}

The relative errors of the EDR3 parallaxes as a function of the parallax and $G$ magnitude are displayed in the top and bottom panels of Fig.~\ref{fig:parallax}, respectively. In the figure the DCEPs are plotted with different colors depending on the ruwe value, using $1.4$ as threshold \citep[see, e.g.][]{Lindegren2021b}. In the figure, the size of the circles surrounding the objects with ruwe$>$1.4 is proportional to the ruwe value. The figure shows that: i) the large majority of the investigated DCEPs has a relative precision on parallax better than 20\%; ii) the trend of relative precision with parallax and magnitude is the expected one, as it increases for smaller parallaxes and larger magnitudes, respectively; iii) the objects with large ruwe values do not follow the general trends, whereas many objects with 1.4$<$ruwe$<$2 do. This probably means that the parallaxes of these objects could be used; iv) stars with $G<6$ mag have generally high ruwe values, as expected by the fact that the Gaia parallaxes for sources brighter than 
$G\sim 6$ mag are still somewhat uncertain \citep{Lindegren2021b}. 

\begin{figure}
   \centering
   \includegraphics[width=8.5cm]{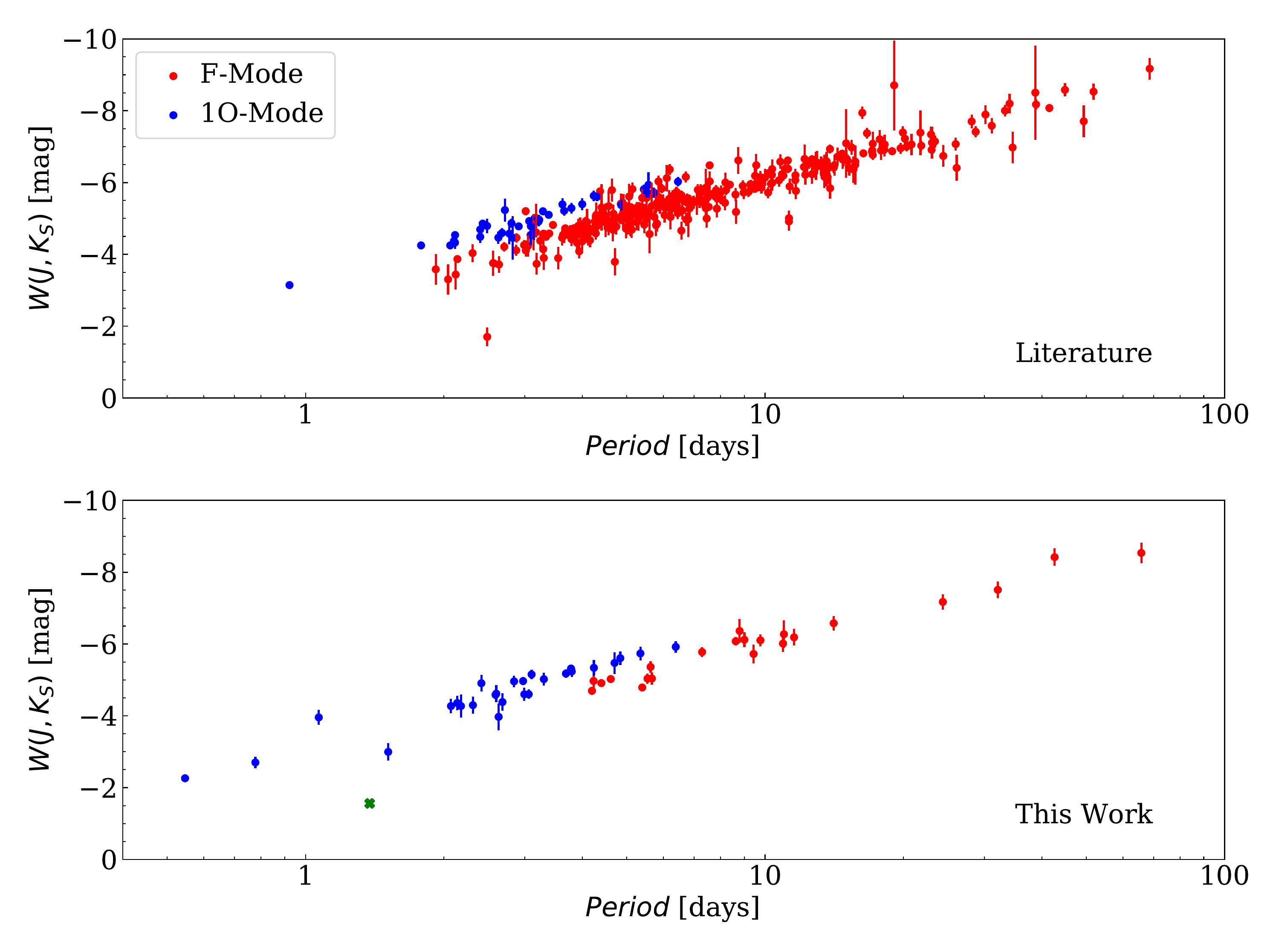}
      \caption{PW relation in $J,K_s$ bands for the literature and present work datasets (top and bottom panels, respectively). Red and blue symbols represent F and 1O mode DCEPs, respectively. The green symbol in the bottom panel is for ASAS J162326-0941.0, one of our targets that turned out to be a BLHER star.  
              }
         \label{fig:pw}
   \end{figure}

Considering the relatively large size of our samples, in the following we decided to be conservative and use the standard ruwe threshold=1.4 and cleaned  the samples from  objects with ruwe value exceeding this limit. 
The resulting literature sample is thus composed of 373 F and 50 1O pulsators, respectively. 
Figure~\ref{fig:pw} shows the $PWJK_s$ for the literature sample (top panel) and for our sample (bottom panel). The figure shows that the large majority of the literature DCEPs have very precise parallaxes and both DCEP\_F and DCEP\_1O show rather tight relations. The only relevant exception is the star V447 Mon which appears significantly fainter than the rest of the DCEPs. Looking at its light curve on the ASAS-SN survey site \citep{Shappee2014,Kochanek2017}, there are little doubts that this star is a F-mode DCEP. Since both the Gaia astrometry and the 2MASS photometry of V447 Mon appear good, we do not have currently an explanation for its under-luminosity in the $PWJK_s$ plane.

The tightness of the $PWJK_s$ relation is even more evident for our sample, with a few exceptions. The star ASAS J162326-0941.0 is clearly under-luminous with respect to the bulk of the other stars. Its parallax has high precision and the ruwe parameter is well below the 1.4 threshold. After inspecting its light-curve shown in Fig.~\ref{fig:LC}, we reckoned that ASAS J162326-0941.0 is a BLHER star misclassified as DCEP in the literature. We do not consider this star in the following analysis. 

The other exceptions are V966 Mon (mixed mode 1O/2O pulsator with dominant period $\sim$1.06 days) and ASAS J063519+2117.8 (1O DCEP with period$\sim$1.51 days) which are about 0.5-0.6 mag too bright and too faint than expected, respectively. The ruwe parameter for both stars is well below 1.4 and also the photometry does not show obvious problems. The two stars appear well isolated on the sky, without any bright companion. We do not have a clear explanation for the deviation of these two stars apart from an undetected problem with astrometry.      

It is important to note that Fig.~\ref{fig:pw} also shows the position of two  additional DCEPs that were observed during the same observing run as the 48 stars discussed above, but whose spectra and photometry were published in separate papers, as they are both objects peculiar in different ways. Indeed, V363 Cas (mixed mode 1O/2O pulsator with dominant period$\sim$0.546 days) is a  rare case of DCEP with Lithium in its spectra \citep[][]{Catanzaro2020} while HD 344787 (mixed mode F/1O pulsator with dominant period$\sim$5.4 days) is a Polaris analogue \citep[][]{Ripepi2021}. Both stars are included in the following analysis.  

   \begin{figure}
   \centering
   \includegraphics[width=8.8cm]{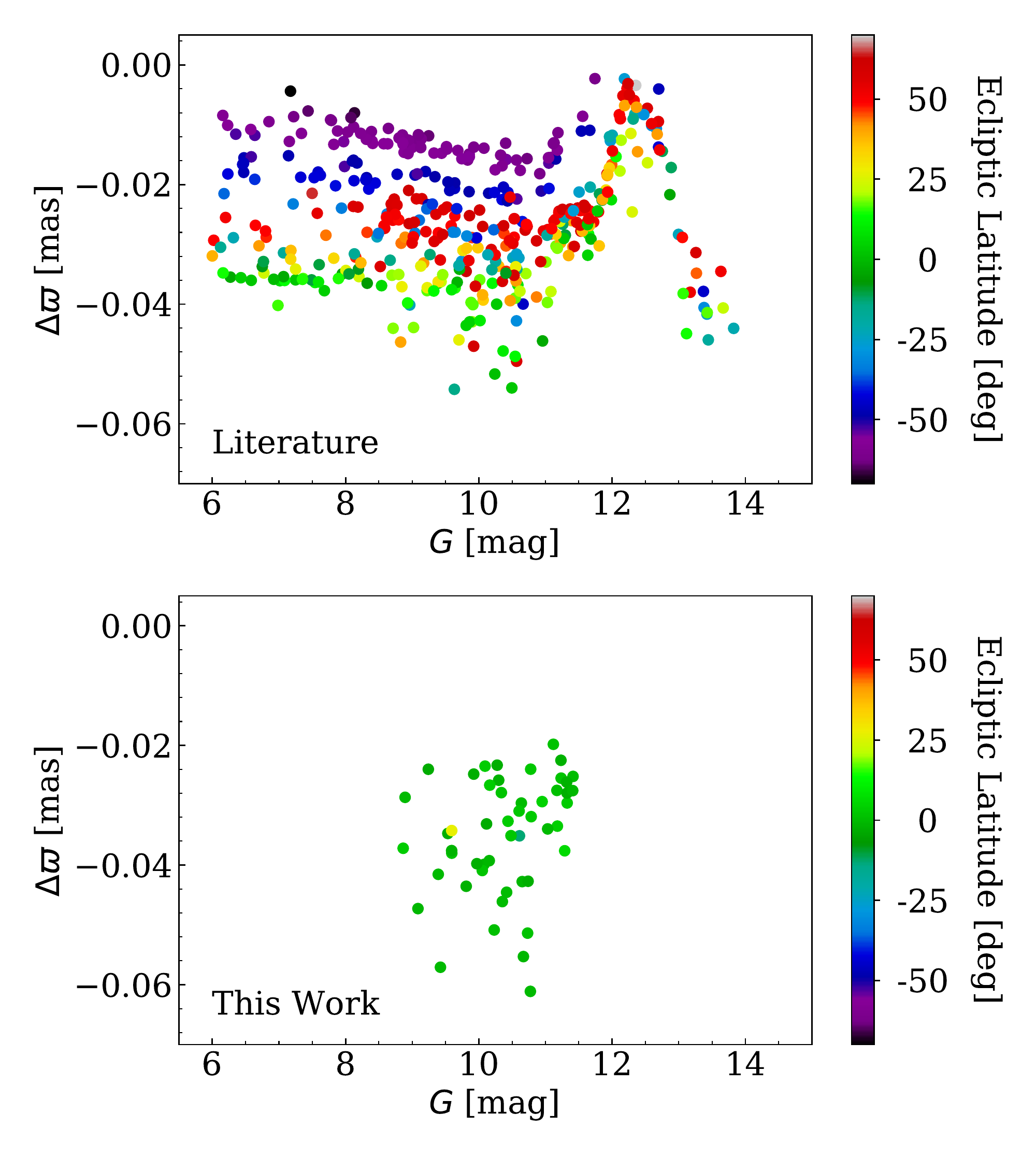}
      \caption{Corrections applied to the EDR3 parallaxes according to \citet{Lindegren2021a}
              }
         \label{fig:zpvals}
   \end{figure}

\subsection{EDR3 parallaxes zero point correction}

We adopted the correction for zero point parallaxes developed by \citet{Lindegren2021a} and made the actual  calculation through the python code accompanying that paper. As remarked by the Authors, the parallax zero-point offset depends in a complex way on (at least) the magnitude, colour, and ecliptic latitude of the target. This dependence in turn is different for the five- and six-parameter solutions in EDR3. 
As the corrections to the parallaxes add uncertainties on these values, we added in quadrature to each parallax error a fixed correction uncertainty of 0.013 mas. 
As explained in \citet{Lindegren2021a} the correction is only valid in the following intervals: $G$ magnitude: 6 $<$ phot\_g\_mean\_mag $<$ 21 mag; Colour: 1.1 $<$ nu\_eff\_used\_in\_astrometry $<$ 1.9 for sources with 5 parameter solution; 1.24 < pseudocolour < 1.72 for 6 parameter solution sources  \citep[see][for details on these parameters]{Lindegren2021a}. Beyond these values one can choose to both extrapolate the solution or neglect the correction. We decided to avoid using the extrapolated values and instead applied an average zero point offset equal to the average of the correction for all the other DCEPs, i.e. 0.025$\pm$0.011 mas. As expected, the variables corrected with the fixed value all have $G<6$ mag. The corrections adopted here are displayed in Fig.~\ref{fig:zpvals} as a function of {\it Gaia} $G$ magnitude and ecliptic latitude \citep[see][]{Lindegren2021a}. It can be seen that the correction depends largely  on the ecliptic latitude of the DCEP, but there is a minimum at $G\sim 12.5$ mag, in agreement with \citet{Lindegren2021a}. 

After the publication of \citet{Lindegren2021a} individual parallax ZPO, several investigations appeared in the literature finding that these corrections are on average overestimated. \citet[][]{Riess2021}, based on the same sample of DCEPs adopted to estimate the metallicity dependence of the $PWHVI$, found an over-correction of $-14\pm$6 $\mu$as. A similar result was obtained by \citet{Zinn2021} from asteroseismology ($-15\pm$3 $\mu$as for stars with G $>$ 10.8 mag), while \citet[][]{Gilligan2021} estimated an over-correction of $-10\pm$7 $\mu$uas from a study of RR Lyrae variables. More recently, \citet{Groenewegen2021} published individual parallax ZPO alternative to \citet{Lindegren2021a}'s but the data is still not available and the procedure rather complex. We will apply these new individual corrections in our future works.
In this paper we shall consider the over-correction by \citet[][]{Riess2021}. We expect that the results will be similar in the other cases.

\subsection{Comparison with DR2 and HST parallaxes} 

\begin{figure}
   \centering
   \includegraphics[width=8.5cm]{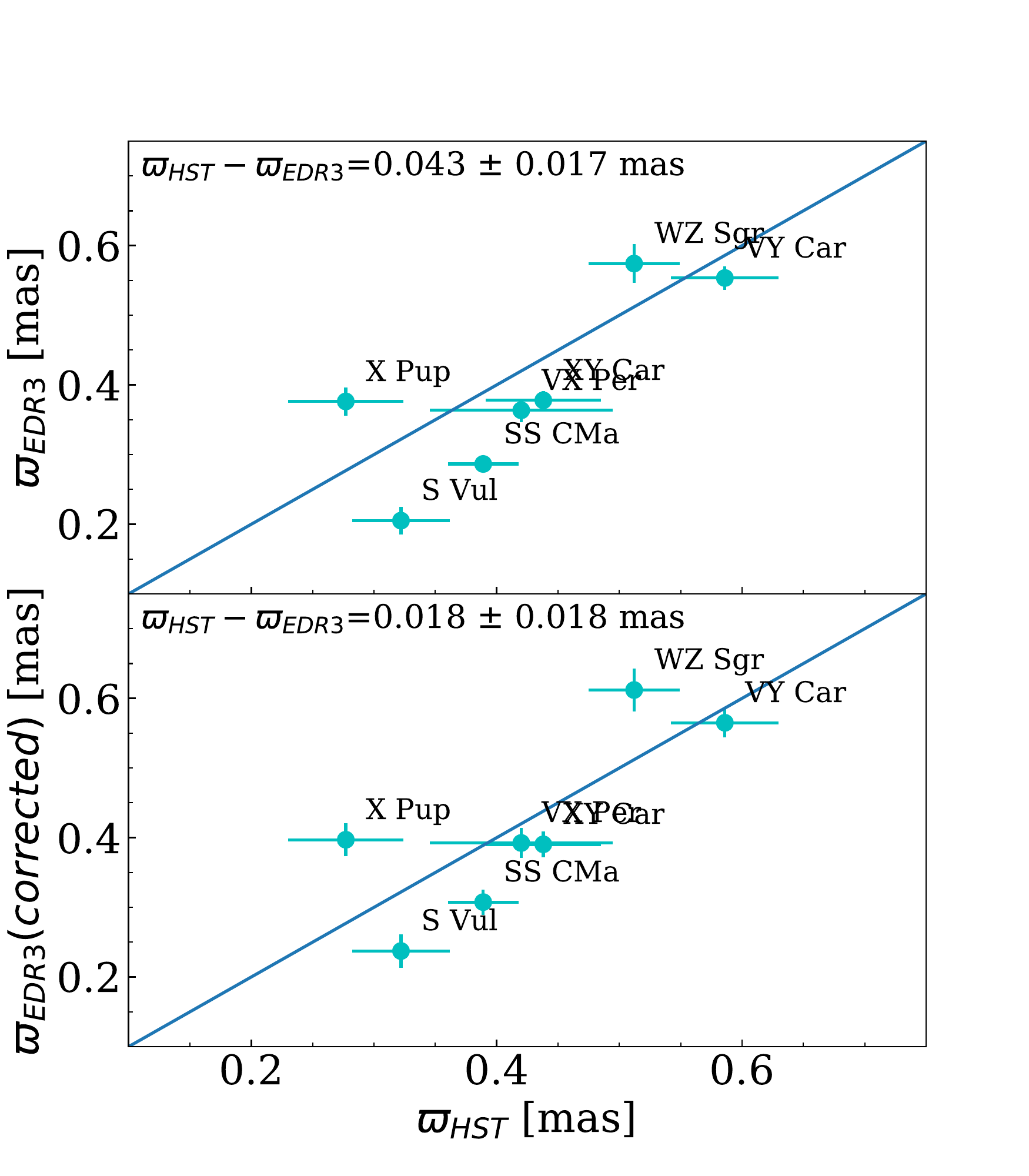}
      \caption{Comparison between the parallaxes in EDR3 and HST. Top and bottom panel show the difference without and with the individual ZPOs by \citet{Lindegren2021a}, respectively.
              }
         \label{fig:compHST}
   \end{figure}

There are seven stars in our sample with independent parallaxes measured with  the HST \citep{Riess2018}. The comparison between these values and those from EDR3 is reported in Fig.~\ref{fig:compHST} where the top and bottom panels show the comparison with the original and ZPO corrected parallaxes, respectively. The weighted averages of the parallax differences in the two cases quoted above are reported in the figure. It can be seen that the uncorrected EDR3 parallaxes have underestimated values by $\Delta \varpi = 0.043 \pm 0.017$ mas with respect to HST determinations. The zero point offset is reduced to $\Delta \varpi = 0.018 \pm 0.018$ mas when the corrected EDR3 values are considered. This result shows that the parallax correction is effective albeit not conclusive. However, we remark that the statistic is low and a few stars (namely X Pup and WZ Sgr) appear to have a different behaviour with  respect to the others. Since all the stars have ruwe value smaller than 1.4 we have no reason to decide whether or not the EDR3 parallaxes of these stars are problematic. 

\begin{figure}
   \centering
   \includegraphics[width=8.5cm]{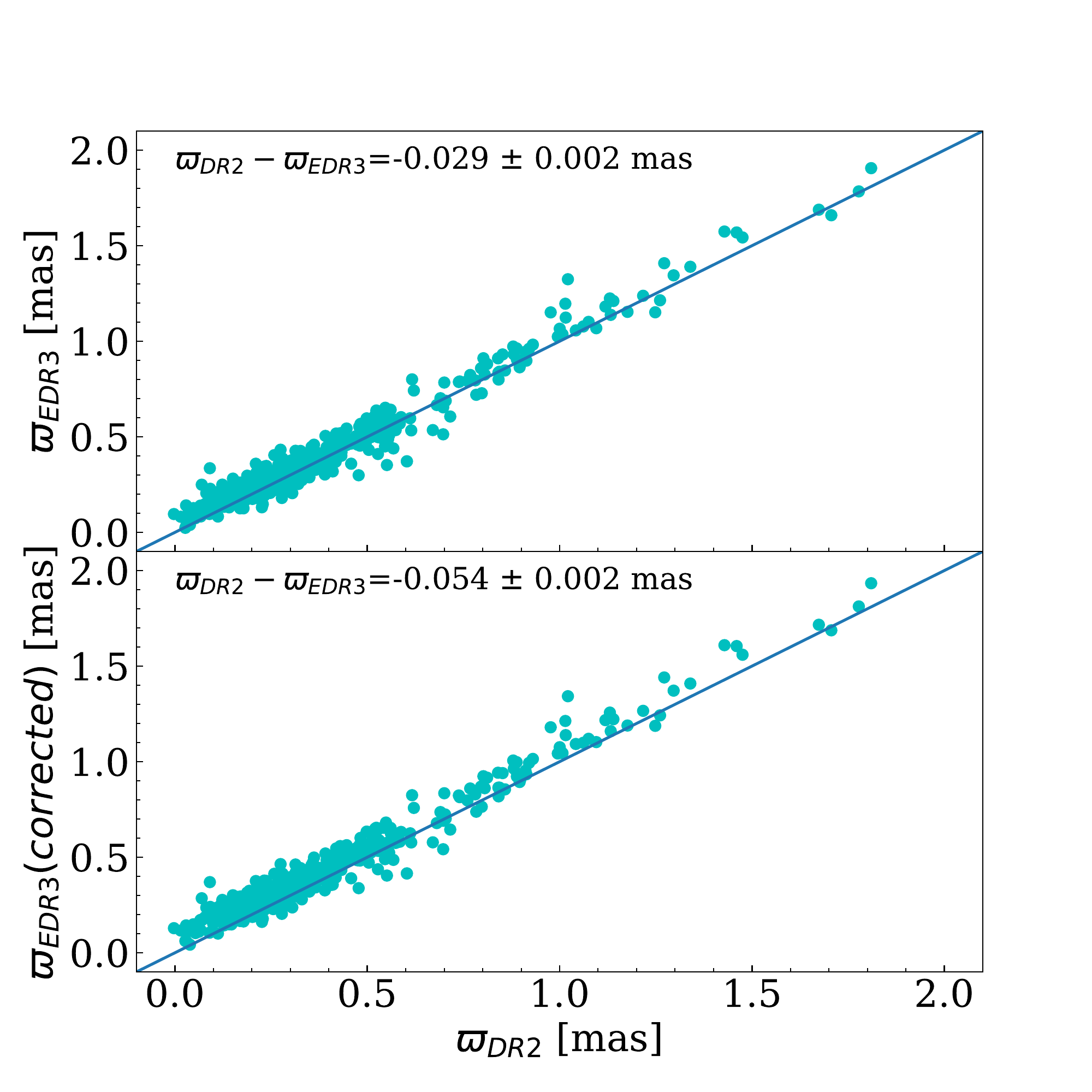}
      \caption{Comparison between the parallaxes in EDR3 and DR2.
              }
         \label{fig:compDR2}
   \end{figure}

The same comparison can be carried out for the DR2 parallaxes. The result is displayed in  Fig.~\ref{fig:compDR2}, where the top and bottom panels show again the comparison with uncorrected and corrected EDR3 parallaxes. As expected, the difference is larger in the case of corrected parallaxes which differ on average by $\Delta \varpi = 0.054 \pm 0.002$ mas with respect to EDR3.  

\section{Fitting of the data}

In this section we use the data described above to estimate the metallicity dependence of the DCEPs PL/PW relations. Based on the sample described in Sect.~\ref{Sect:Sample}, we investigated the PL in the $K_s$ ($PLK_s$) band and the PW for the following three Wesenheit apparent magnitudes: $w(J,K_s)=K_s-0.69(J-K_s)$, $w(V,K_s)=K_s-0.13(V-K_s)$, and $w(H,V,I)=H-0.386(V-I)$, where the coefficients of these relations were taken from \citet{Cardelli1989} for the first two Wesenheit magnitudes and from  \citet{Riess2021} for the last one. To determine the $PLK_s$ we have obtained dereddened $K_{s,0}$ magnitudes by adopting  $A_{K_s}=0.34 E(B-V)$ \citep[as derived by][using $R_V$=3.1]{Cardelli1989}, where the $E(B-V)$ values have been collected as described in Sect.~\ref{Sect:Sample}. 

To analyse the sample we followed the procedure adopted in \citet{Ripepi2020a}. As in that work, to avoid any bias, we considered the whole sample, irrespective of the precision of individual parallaxes or of the presence of negative values for this parameter. To manage the negative parallaxes and to keep the symmetry in the parallax uncertainty, we used the astrometry-based luminosity \citep[ABL,][]{Feast1997,Arenou1999}, which is defined as follows:
${\rm ABL}=10^{0.2 M}=\varpi10^{0.2 m -2} $, where $\varpi$ is the parallax, while $M$ and $m$ are the absolute and apparent generic magnitudes, respectively. 
Following \citet{Ripepi2020a}, we can write $M$ in the most general form, including the metallicity dependence on both the slope and intercept of the PL/PW relations: 

\begin{equation}
M=\alpha+(\beta+\delta {\rm [Fe/H])}(\log P- \log P_0) + \gamma {\rm [Fe/H]})
\label{eq:eq1}
\end{equation}

\noindent
where $P$ and [Fe/H] are the observed period and iron abundance, while $P_0$ is a pivoting period.  Therefore, in its most general form the $ABL$ function we want to use is: 

\begin{equation}
{\rm ABL}=\varpi 10^{0.2 m -2}=10^{0.2(\alpha+(\beta+\delta {\rm [Fe/H])}(\log P-\log P_0) +\gamma {\rm [Fe/H]})} \label{eq:abl}  
\end{equation}
\noindent
where P, [Fe/H], $\varpi$ and the apparent magnitude/Wesenheit are the observables, while the unknowns are the four parameters of Eq.~\ref{eq:eq1}.
Following ~\citet{Ripepi2020a}, we considered PL/PW relations with different number of parameters, i.e. with no metallicity dependence (two parameters), with metallicity dependence only on the intercept (three parameters) and with metallicity dependence on both the slope and intercept (four parameters). As mentioned above, we considered the following four relations: $PLK_s$, $PWJK_s$, $PWVK_s$ and $PWHVI$. For each relation we calculated the solution with two, three and four parameters separately for the case of Lit and L18 metallicities, respectively.       
We also performed the $ABL$ fitting to the data using two different approaches which are described in detail in the following sections.

\subsection{Nonlinear fitting approach}
A first approach to fit the eq.~\ref{eq:abl} has been achieved by using the \textit{nls} function from the R environment \citep{cranR}. It allows to perform a nonlinear least square regression (NL-LSQs hereafter) based on the Gauss--Newton minimization algorithm \citep[see e.g.][]{bat88}. 
The different uncertainties on the $ABL$ magnitudes have been taken into account by weighting the fit with standard $w_i=1/\sigma_i^2$ factors, where the $\sigma_i$ term contains the contributions by observed magnitudes errors, reddening errors and parallax errors.

To estimate parameter errors a bootstrap approach has been applied. In particular, 100 simulated data sets have been obtained by resampling the original one, allowing repeated points. Then the fitting technique has been applied on all simulated samples and the coefficients of the Eq.~\ref{eq:abl} were recalculated 100 times. Finally a statistical analysis of the obtained coefficient distributions allowed to estimate their errors, defined as the robust standard deviations ($std_{robust}=1.4826\cdot std$) of the quoted distributions.

Since the tested models have different number of parameters, we have also provided two estimators of the goodness of the models by calculating the Akaike information criterion \citep[AIC][]{Akaike2011} and the Bayesian information criterion \citep[BIC]{sch78}. They are defined respectively $AIC=2k - 2\ln{\mathcal{L}}$ and $BIC=k\cdot \ln N - 2\ln{\mathcal{L}}$, where $k$ is the number of parameters in the fit and $\ln{\mathcal{L}}$ is given by:
\begin{equation}
    \ln{\mathcal{L}} = -0.5\sum_{i=1}^N \left [ \left (\frac{res_i}{s_i}\right )^2 + \ln (2 \pi s_i^2)\right ]
\end{equation}
where $N$ is the number of fitted points, $res_i$ is the $i^{th}$ residual value, while the  term $s_i$ is the sum in quadrature of the $i^{th}$ uncertainty  and the intrinsic scatter around the fitted relation. We have set the intrinsic scatter of the fitted relations equal to half the rms of the residuals, but checked that this choice does not affect the results significantly.
The results of the fitting procedure are listed in Table~\ref{table:resNL}.

\subsection{Montecarlo Markov Chain (MCMC) approach.}

Complementary to the NL-LSQs approach, we performed a Markov Chain MonteCarlo (MCMC) method, which is a widely used technique that preserves a high efficiency even when the parameter space is large.
The idea is that, after a certain number of random steps (the length of the chain), moving towards a region of highest probability in the parameter space, the MCMC approximates the posterior distribution of a set of parameters.
In this work we took advantage of the {\sc EMCEE} package developed in Python\footnote{https://emcee.readthedocs.io/en/stable/}, by running 1000 chains each one with a size of 1000 steps (or walkers).
The posterior distribution in the logarithmic form is defined, except for a normalization constant term, as 
\begin{equation}
    \log P(\theta|X) = \log P(X|\theta) + \log P(\theta)
    \label{eq:posterior}
\end{equation}
\noindent
where $\theta$ represents the ensemble of parameters (i.e. $\alpha,~\beta~,\gamma,~\delta$) and X is the observable, i.e. the left-hand side of Eq.~\ref{eq:abl}.
$P(X|\theta)$ is the likelihood function, that we set, again in logarithmic form, as
\begin{equation}
    \ln{\mathcal{L}} = \log P(X|\theta) = -0.5 \cdot \sum (\frac{X - model}{\Delta X})^2
\end{equation}
\noindent
where $model$ is the right-hand side of Eq.~\ref{eq:abl}. Note that the minus sign indicates that the MCMC walks towards global minima.
The function $P(\theta)$ indicates our priors, that we set, in order to achieve a convergence in a finite number of walkers to $-10 < \alpha < 0$;  $-6 < \beta < 0$; $-2 < \gamma < 2$ and $-5 < \delta < 5$.\\
Our best estimation of each parameter is represented by the median of the posterior distribution defined in Eq.~\ref{eq:posterior} and sampled via the MCMC. We set as uncertainties the 16th and 84th percentile.
As discussed in the previous section the AIC and BIC are calculated also in this case to have an estimate of the goodness of each model.
The results of the fitting procedure with the MCMC method are listed in Table~\ref{table:resMCMC}. 
A comparison between the coefficient of the $PLZ$/$PWZ$ relations calculated with the NL-LSQs and the MCMC methods are shown in Fig.~\ref{fig:parComparison}. The agreement is excellent for all the coefficients, and this is   
particularly comforting, as the two approaches are completely different and independent of each other.

\section{Discussion}

To test the goodness of our $PLZ$/$PWZ$ relations, we apply them to the DCEPs in the LMC, derive its  distance modulus and compare it to that geometrically derived (from eclipsing binaries) by \citet{Pietrzynski2019}, i.e. $\mu_{LMC}=18.48\pm0.03$ mag (including systematic errors), which is currently considered one of the most accurate estimates in the literature.

To this aim, we collected literature photometry in the bands of interest of this work. More specifically, we considered about 4500 DCEPs in the LMC, having: $V,I$ photometry by \citet[OGLE IV survey][]{Udalski2018}; $J,K_s$ photometry \citep[][Ripepi et al. in preparation]{Ripepi2012,Ripepi2020b}; $H$ \citep{Inno2016}. As for the periods, they were taken from the OGLE IV survey quoted above. 

Then, we calculated the absolute $M_{K_{0,S}}$ and Wesenheit magnitudes $W(J,K_s)$, $W(V,K_s)$, $W(H,V,I)$  for each LMC DCEP adopting the coefficient of the $PLZ$/$PWZ$ relations reported in Tables~\ref{table:resNL} and ~\ref{table:resMCMC}, using the periods by the OGLE IV survey quoted above and  assuming [Fe/H]$_{LMC}$=$-$0.33 dex \citep[][]{Romaniello2008}.

Having apparent and absolute $K_{0,S}$ and Wesenheit magnitudes, it is easy to obtain individual DM values for each LMC DCEP\_F. The median of the distributions of these DMs is reported in Tables~\ref{table:resNL} and ~\ref{table:resMCMC}. The random error on the median is also provided and was calculated by summing in quadrature the standard deviation of the mean of the residuals obtained for the MW relations and the similar quantity for the LMC. The uncertainty is completely dominated by the MW relations, as the contribution of the LMC is negligible, given the large number of adopted pulsators in the LMC.

\begin{figure}
   \centering
   \includegraphics[width=8.5cm]{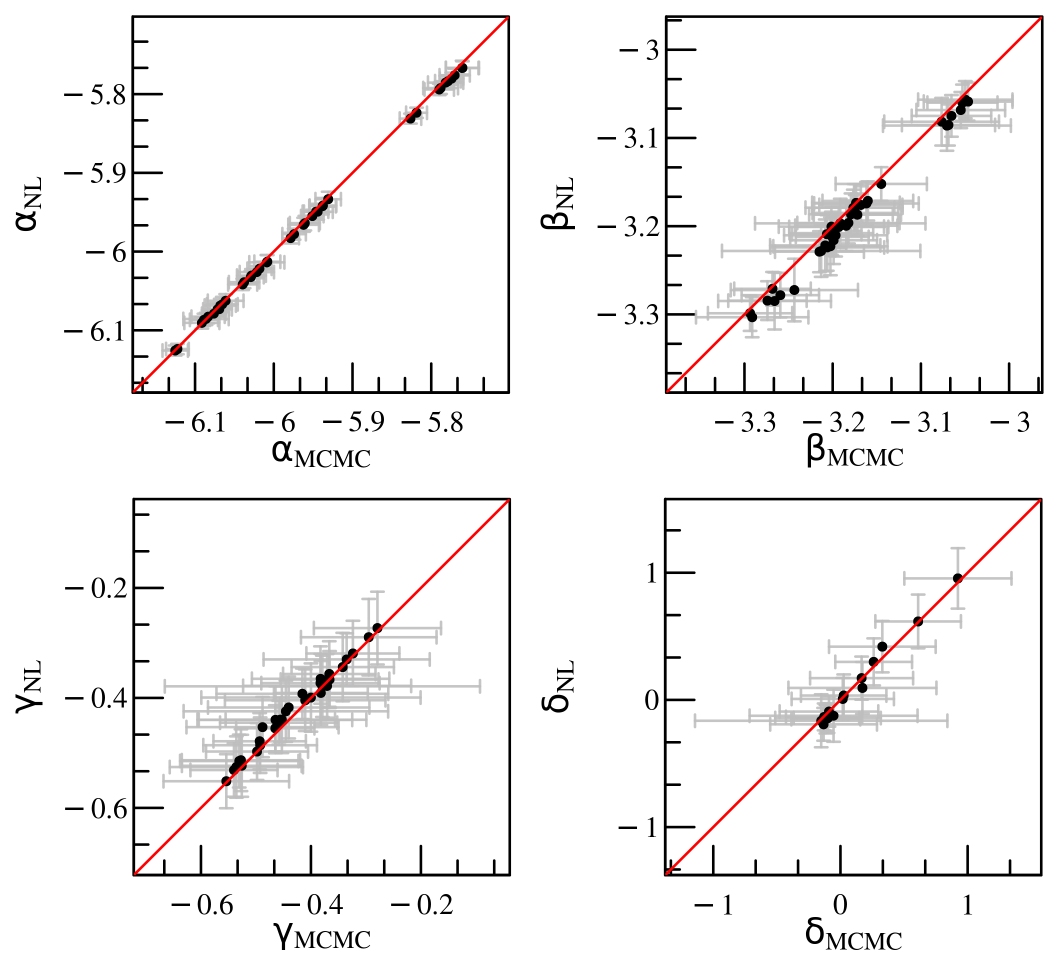}
      \caption{Comparison between the $PLZ$/$PWZ$ relations coefficients calculated with the NL-LSQ and MCMC methods reported in Table~\ref{table:resNL} and Table~\ref{table:resMCMC}. 
              }
         \label{fig:parComparison}
   \end{figure}

\begin{figure}
   \centering
   \includegraphics[width=8.5cm]{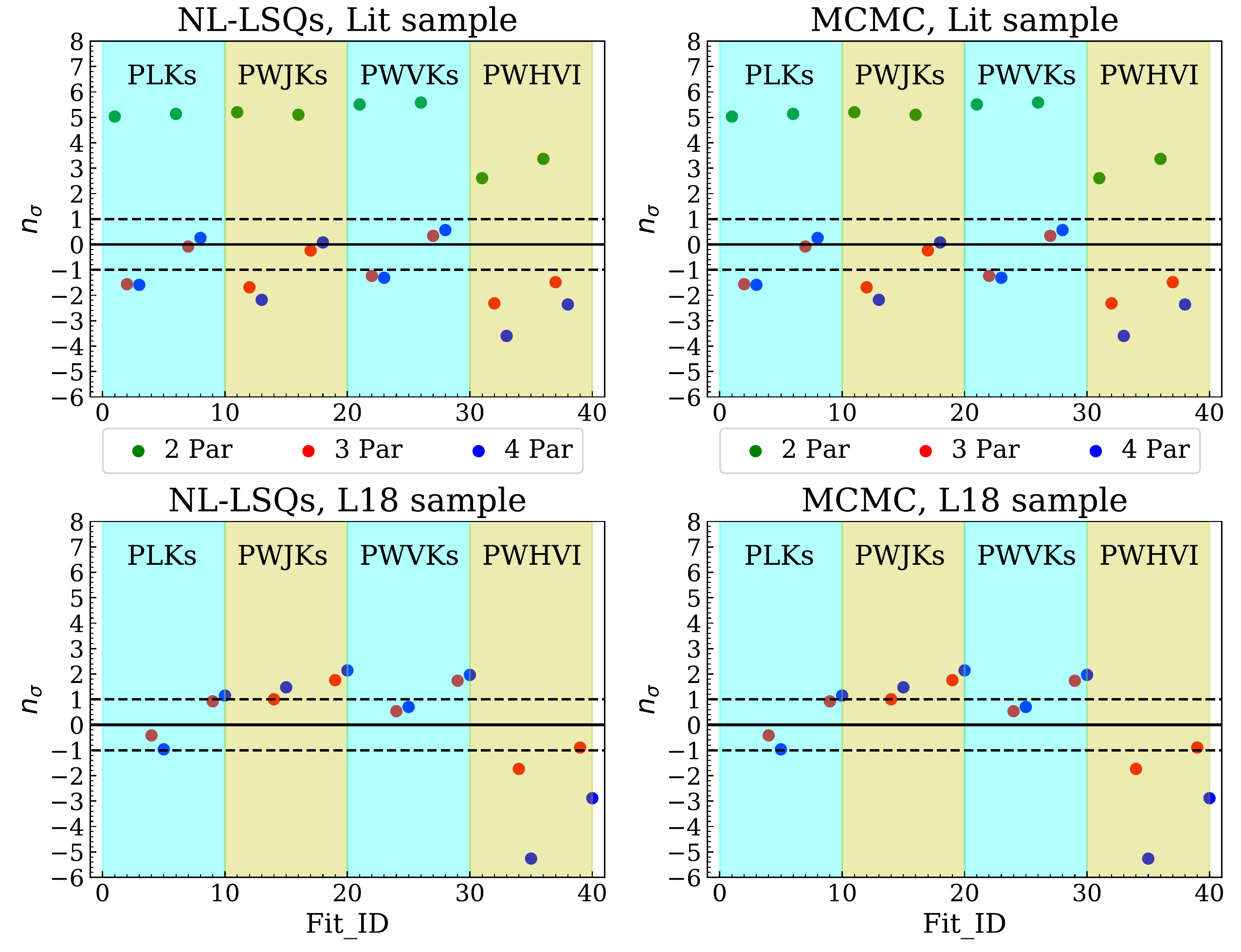}
      \caption{Number of $\sigma$ from the geometric distance of the LMC for the 40 fittings to the data shown in each of Table~\ref{table:resNL} and Table~\ref{table:resMCMC}.  
              }
         \label{fig:lmcDist}
   \end{figure}

To better visualize the results we decided to calculate the following quantity:

\begin{equation}
    n_\sigma = (\mu^{TW}_{LMC}-\mu^G_{LMC})/\sigma 
\end{equation}

\noindent 
where $\mu^G_{LMC}$ and $\mu^{TW}_{LMC}$ are the LMC DM as estimated by \citet{Pietrzynski2019} and 
 in this work, respectively; $\sigma$ is the uncertainty on the DM of the LMC obtained summing in quadrature the geometric uncertainty, equal to 0.03 mag with the random errors on $\mu^{TW}_{LMC}$ reported in the Tables~\ref{table:resNL} and ~\ref{table:resMCMC}. Given that the value of $\sigma$ is dominated by the uncertainty on $\mu^G_{LMC}$, we can approximately say that $n_\sigma$ is the deviation of our results with respect to the geometric distance of the LMC, in units of the uncertainty on this last quantity.
In Fig.~\ref{fig:lmcDist} left and right panels show the value of $n_\sigma$ for all the cases listed in Tables~\ref{table:resNL} and ~\ref{table:resMCMC}, respectively. The top and bottom panels report the cases adopting the Lit and L18 metallicity samples, respectively. An inspection of the figure allows us to draw some considerations: 

\begin{itemize}
    \item there is very little difference between the two approaches adopted for fitting the data. This is not surprising, as the coefficient of the $PLZ$/$PWZ$ calculated with NL-LSQand MCMC methods are very similar (see Fig.~\ref{fig:parComparison}). 
    \item the $PLK_s$, $PWJK_s$, $PWVK_s$ give always better distances to the LMC than the $PWHVI$ relation in the NL-LSQ case, while the 3 parameter case is as good as in the other bands for the L18 sample.
    \item the two parameter solution, i.e. without any metallicity term, is ruled out at more than the 5 $\sigma$ level in all the cases except for the PWHVI relation ($>3\sigma s$).
    \item the fits with the complete sample, i.e. including both DCEP\_F and DCEP\_1O (fit ids 7,8,17,18,27,28 in both Tables~\ref{table:resNL} and \ref{table:resMCMC}) provide better results in the case of the Lit sample, while they are equivalent in the case of the L18 sample.
\end{itemize}

\begin{figure}
   \centering
   \includegraphics[width=8.5cm]{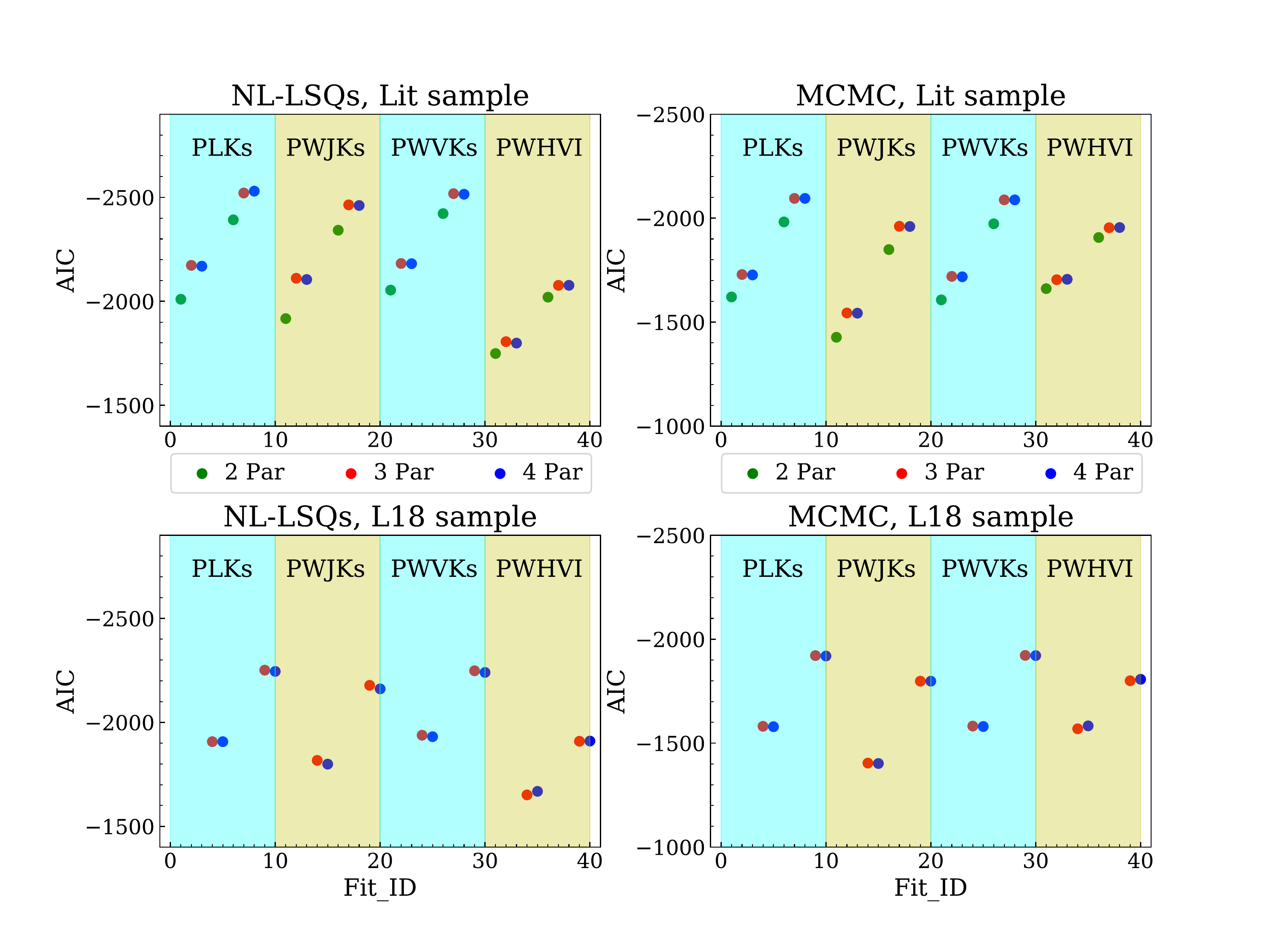}
      \caption{Significance of the fit (AIC) for the 40 fittings to the data shown in each of Table~\ref{table:resNL} and Table~\ref{table:resMCMC}.  
              }
         \label{fig:aic}
   \end{figure}

Figure~\ref{fig:aic} is very similar to Fig.~\ref{fig:lmcDist}, but this time on the y-axis we plotted the goodness of the fit AIC value. It is possible to see that the fit ids that provide the best LMC distances are also those with the lowest values of AIC, i.e. the best fits to the data. Also, the fits with four parameters, do not provide better AIC values with respect to the fits with three parameters.  

\begin{figure}
   \centering
   \includegraphics[width=8.5cm]{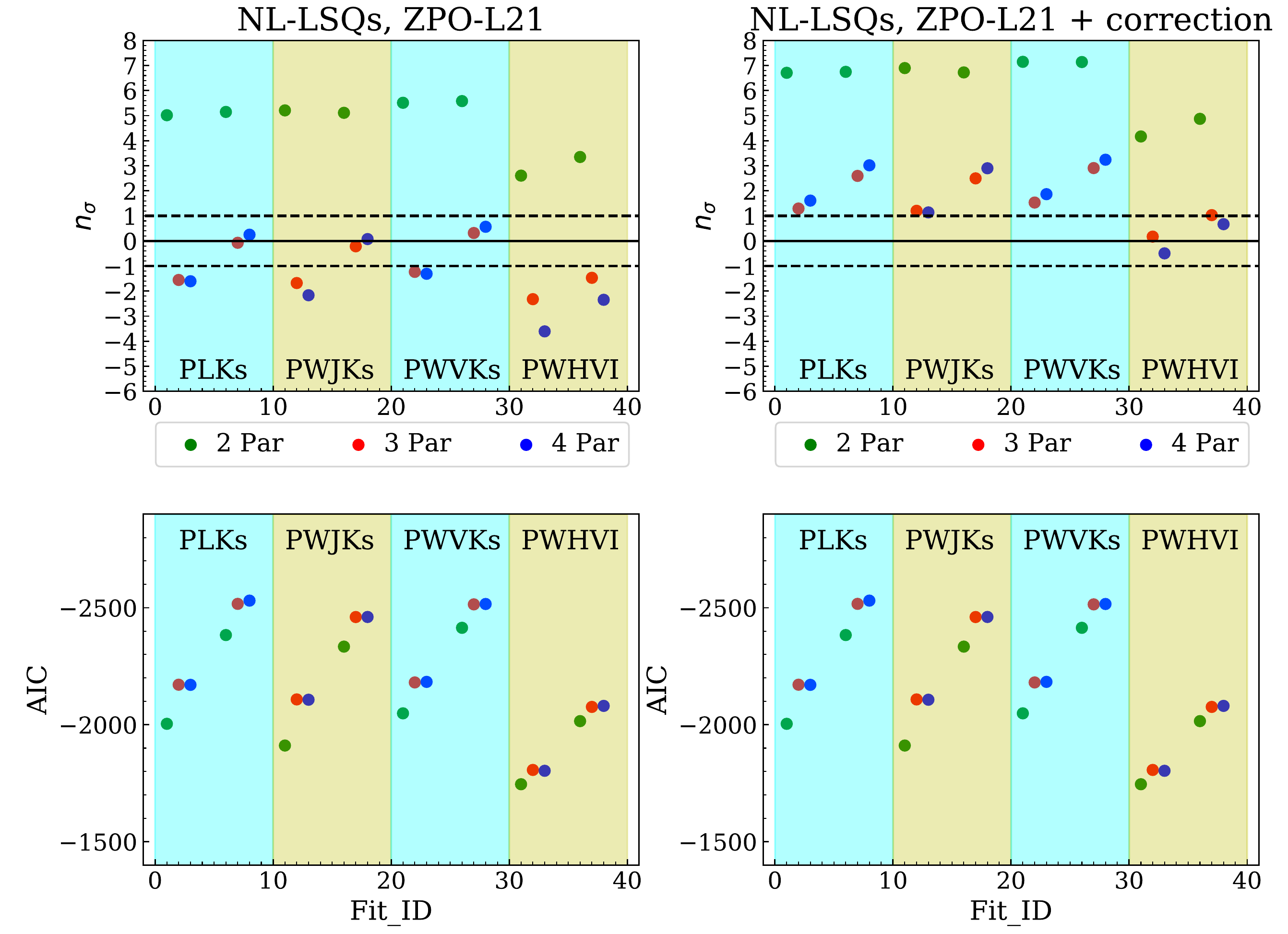}
      \caption{Comparison between the nonlinear case fitting solution without (left panels) and with the parallax ZPO by \citet{Riess2021} (right panels). The top panels show the number of $\sigma s$ from the geometric distance of the LMC, while the bottom panels display the AIC values relative to each case in the top panels.   
              }
         \label{fig:distance_riess}
   \end{figure}

\begin{figure}
   \centering
   \includegraphics[width=8.5cm]{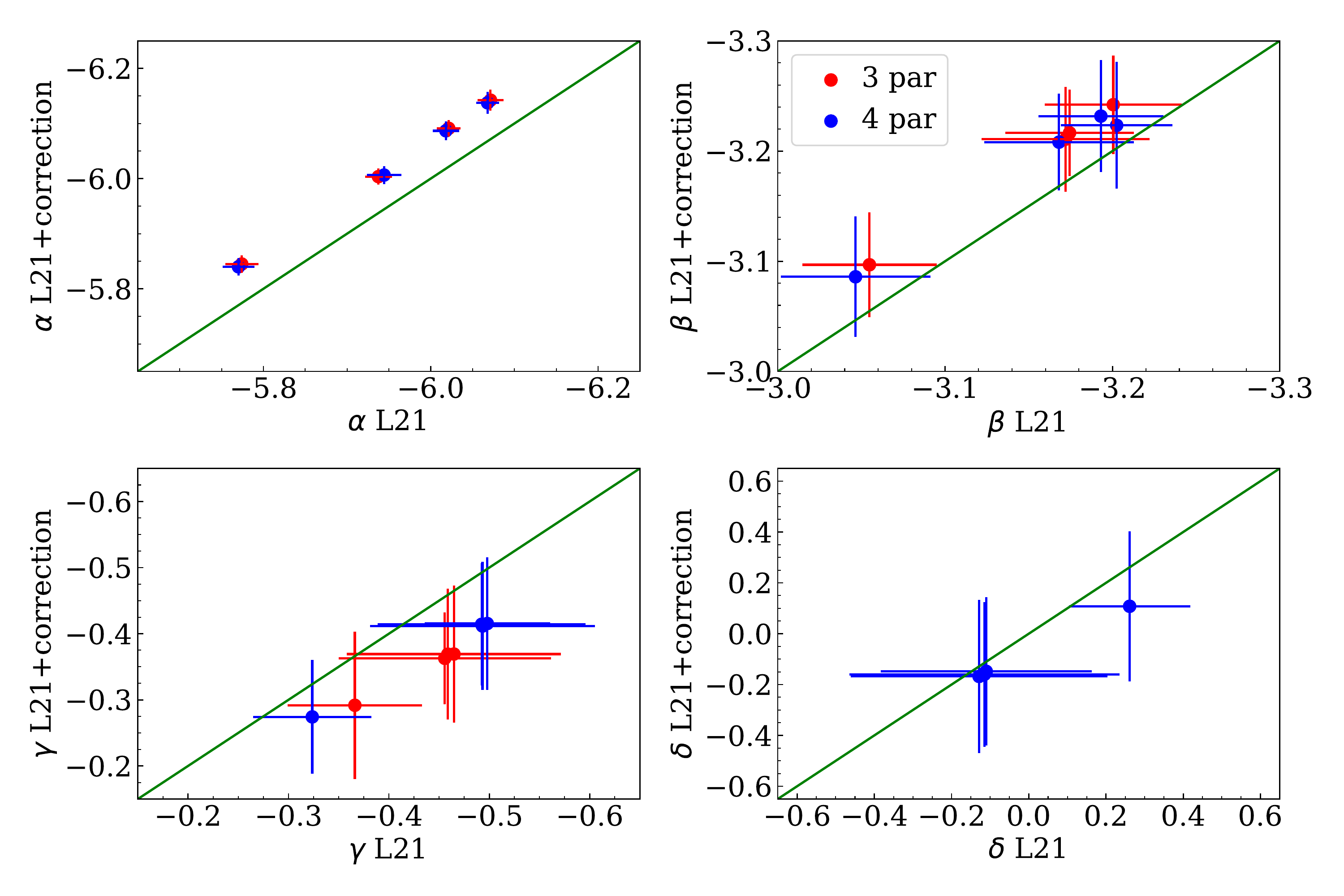}
      \caption{Parameters of the $PLZ$/$PWZ$ relations ($\alpha$, $\beta$ and $\gamma$, $\delta$) in the cases with 3 and 4 parameters (red and blue filled circles, respectively). The x- and y-axis show the results obtained in the NL-LSQs fitting solution without and with the parallax ZPO by \citet{Riess2021}, respectively.}
         \label{fig:parameters_riess}
   \end{figure}

As we have mentioned before, several authors have noticed that the  individual parallax ZPO by \citet{Lindegren2021b} are overestimated in a measure that goes between 10 and 15 $\mu$as. To verify the effect of this correction on our results, we decided to recalculate the distance of the LMC using the correction of 14 $\mu$as, according to \citet{Riess2021} who calculated it in their derivation of the $PWHVI$ relation for DCEPs. For simplicity, we considered only the case of the NL-LSQ fit with the Lit sample. The results are shown in Table~\ref{table:resNLRiess} which lists the same quantities of Table~\ref{table:resNL}\footnote{We did not recalculate the parameters in the MCMC case because, as shown in Fig.~\ref{fig:parComparison}, the NL and MCMC approaches provide almost identical results.} and in 
Fig.~\ref{fig:distance_riess}, where the top and bottom panels represent the same quantities of Fig.~\ref{fig:lmcDist} and ~\ref{fig:aic}, respectively.
An analysis of the figure reveals that the introduction of the correction worsens the results concerning the distance of the LMC for all the relations with the exception of the $PWHVI$, whose performance improves significantly, even if the corresponding AIC parameter is always worse than what obtained from the other $PLZ/PWZ$ relations. To better understand this behaviour it is instructive to check how the coefficients of the $PLZ/PWZ$ relations vary without/with the parallax ZPO corrections. This is shown in Fig.~\ref{fig:parameters_riess} for the same cases of Fig.~\ref{fig:distance_riess} except for the 2 parameter case which is scarcely informative. It can be seen that in all the considered cases, the introduction of the correction  increases the absolute values of $\alpha$ and $\beta$ by $\sim$1\%, while $\gamma$ is decreased (in absolute value) by $\sim$18-20\% and $\sim$13-16\% in the 3 and 4 parameter cases, respectively. The absolute value of $\delta$ is decreased by $\sim$30\%. Since the $\alpha$ and $\beta$ coefficients remain practically unaffected, the obvious conclusion of this experiment is that the parallax ZPO correction and the metallicity terms are degenerate. The introduction of the ZPO correction is compensated by lower (in absolute sense) metallicity terms, both in the cases of three and four parameters. In the case of four parameters, the $\gamma$ term decrease is smaller, since also the $\delta$ coefficient decreases. 

In all cases, without and with the introduction of the parallax ZPO correction, our results provide metallicity terms on the intercept that are larger than those found in the recent literature by \citet[][]{Riess2021} and \citet[][]{Breuval2021}, even if, the $\gamma$ coefficient of the $PWHVI$
relation, with the correction is consistent within the errors with that estimated by \citet[][]{Riess2021} that used the same $PWHVI$ relation and the same parallax ZPO correction. 

\begin{table*}
\caption{Results of the fitting of the data with the NL-LSQs method. The ID identifies each different fit to the data; $\alpha$, $\beta$, $\gamma$ and $\delta$ are the coefficient of the $PLZ$/$PWZ$ relations; n.MW is the number of MW DCEPs adopted in each fit; BIC and AIC are the goodness-of-fit parameters discussed in the text; Mode identifies the sample adopted in the fit: F means F-mode DCEPs only, F+1O means that both F- and 1O-mode pulsators are used; Source shows the origin of the metallicities; Relation specifies the relationship used in the fit; $\mu_{LMC}$ represents the distance modulus of the LMC obtained with the specific $PLZ$ or $PWZ$ relationship; n.LMC is the number of LMC DCEPs adopted to calculate the $\mu_{LMC}$ value. The lines highlithed in grey represent our best solution for each $PLZ$/$PWZ$ group.}
\label{table:resNL} 
\footnotesize\setlength{\tabcolsep}{3pt}
\centering          
\begin{tabular}{ccccccccccccc} 
\hline\hline             
ID & $\alpha$ & $\beta$ & $\gamma$ & $\delta$ & n.MW & BIC & AIC & Mode & Source & Relation & $\mu_{LMC}$ & n.LMC\\
\hline
1 &   $-5.819 \pm 0.013$ &  $-3.160 \pm 0.051$ &   &  & 381 &                                         $-$1996  &  $-$2003  & F & Lit & $PLK_s$   & 18.631$\pm$0.011  &     2432 \\         
2 &   $-5.760 \pm 0.021$ &  $-3.069 \pm 0.053$ &  $-0.530 \pm 0.108$ &  & 381 &                      $-$2159  &  $-$2171 & F & Lit & $PLK_s$     & 18.433$\pm$0.010  &     2432 \\
3 &   $-5.761 \pm 0.021$ &  $-3.070 \pm 0.072$ &  $-0.527 \pm 0.108$ & $0.018 \pm 0.402$ & 381 &     $-$2154  &  $-$2170 & F & Lit & $PLK_s$     & 18.432$\pm$0.010  &     2432 \\
4 &   $-5.779 \pm 0.016$ &  $-3.065 \pm 0.045$ &  $-0.488 \pm 0.138$ &  & 349 &                      $-$1894  &  $-$1906 & F & L18 & $PLK_s$     & 18.467$\pm$0.011  &     2432 \\
5 &   $-5.782 \pm 0.022$ &  $-3.077 \pm 0.065$ &  $-0.464 \pm 0.140$ & $0.173 \pm 0.582$ & 349 &      $-$1895  &  $-$1910 & F & L18 & $PLK_s$    & 18.451$\pm$0.010  &     2432 \\
6 &   $-5.826 \pm 0.014$ &  $-3.145 \pm 0.052$ &   &  & 448 &                                         $-$2375  &  $-$2383 & F+1O & Lit & $PLK_s$ & 18.634$\pm$0.010  &     4396 \\
\rowcolor{lightgray} 7 &   $-5.774 \pm 0.015$ &  $-3.055 \pm 0.050$ &  $-0.456 \pm 0.099$ &  & 448 &                     $-$2504  &  $-$2516 & F+1O & Lit & $PLK_s$   & 18.478$\pm$0.009  &     4396 \\
8 &   $-5.770 \pm 0.019$ &  $-3.046 \pm 0.050$ &  $-0.493 \pm 0.084$ & $-0.129 \pm 0.300$ & 448 &   $-$2514  &  $-$2530 & F+1O & Lit & $PLK_s$   & 18.488$\pm$0.009  &     4396 \\
9 &   $-5.790 \pm 0.019$ &  $-3.053 \pm 0.040$ &  $-0.416 \pm 0.156$ &  & 411 &                    $-$2236  &  $-$2248 & F+1O & L18 & $PLK_s$    & 18.508$\pm$0.009  &     4396 \\
10 &  $-5.788 \pm 0.022$ &  $-3.050 \pm 0.053$ &  $-0.440 \pm 0.147$ & $-0.090 \pm 0.388$ & 411 & $-$2229  &  $-$2245 & F+1O & L18 & $PLK_s$     & 18.514$\pm$0.009  &     4396 \\
11 &  $-6.122 \pm 0.014$ &  $-3.291 \pm 0.064$ &   &  & 381 &                                      $-$1903  &  $-$1911 & F & Lit & $PWJK_s$      & 18.636$\pm$0.011  &     2423 \\
12 &  $-6.061 \pm 0.023$ &  $-3.197 \pm 0.058$ &  $-0.554 \pm 0.114$ &  & 381 &                    $-$2096  &  $-$2108 & F & Lit & $PWJK_s$    &   18.429$\pm$0.010  &     2423 \\
13 &  $-6.065 \pm 0.020$ &  $-3.213 \pm 0.112$ &  $-0.526 \pm 0.109$ & $0.166 \pm 0.405$ & 381 &   $-$2091  &  $-$2106 & F & Lit & $PWJK_s$    &   18.415$\pm$0.010  &     2423 \\
14 &  $-6.086 \pm 0.029$ &  $-3.202 \pm 0.047$ &  $-0.379 \pm 0.287$ &  & 349 &                    $-$1804  &  $-$1815 & F & L18 & $PWJK_s$    &   18.510$\pm$0.011  &     2423 \\
15 &  $-6.083 \pm 0.021$ &  $-3.191 \pm 0.097$ &  $-0.400 \pm 0.199$ & $-0.151 \pm 0.992$ & 349 &  $-$1786  &  $-$1801 & F & L18 & $PWJK_s$    &   18.524$\pm$0.010  &     2423 \\
16 &  $-6.125 \pm 0.016$ &  $-3.294 \pm 0.048$ &   &  & 448 &                                      $-$2325  &  $-$2333 & F+1O & Lit & $PWJK_s$    &18.633$\pm$0.010  &     4360 \\
\rowcolor{lightgray} 17 &  $-6.071 \pm 0.018$ &  $-3.201 \pm 0.041$ &  $-0.465 \pm 0.071$ &  & 448 &                    $-$2448  &  $-$2460 & F+1O & Lit & $PWJK_s$    &18.473$\pm$0.009  &     4360 \\
18 &  $-6.068 \pm 0.020$ &  $-3.193 \pm 0.056$ &  $-0.498 \pm 0.093$ & $-0.115 \pm 0.271$ & 448 &  $-$2444  &  $-$2460 & F+1O & Lit & $PWJK_s$    &18.482$\pm$0.009  &     4360 \\
19 &  $-6.091 \pm 0.022$ &  $-3.207 \pm 0.039$ &  $-0.335 \pm 0.151$ &  & 411 &                    $-$2162  &  $-$2174 & F+1O & L18 & $PWJK_s$    &18.533$\pm$0.009  &     4360 \\
20 &  $-6.089 \pm 0.018$ &  $-3.201 \pm 0.060$ &  $-0.371 \pm 0.153$ & $-0.133 \pm 0.420$ & 411 &  $-$2145  &  $-$2161 & F+1O & L18 & $PWJK_s$    &18.544$\pm$0.009  &     4360 \\
21 &  $-6.069 \pm 0.012$ &  $-3.274 \pm 0.044$ &   &  & 369 &                                    $-$2040  &  $-$2048 & F & Lit & $PWVK_s$     &    18.645$\pm$0.012  &     2268 \\
22 &  $-6.008 \pm 0.017$ &  $-3.182 \pm 0.058$ &  $-0.540 \pm 0.079$ &  & 369 &                   $-$2169  &  $-$2180 & F & Lit & $PWVK_s$    &    18.443$\pm$0.011  &     2268 \\
23 &  $-6.009 \pm 0.022$ &  $-3.185 \pm 0.058$ &  $-0.536 \pm 0.121$ & $0.025 \pm 0.336$ & 369 &  $-$2167  &  $-$2183 & F & Lit & $PWVK_s$    &    18.441$\pm$0.010  &     2268 \\
24 &  $-6.029 \pm 0.017$ &  $-3.179 \pm 0.047$ &  $-0.446 \pm 0.186$ &  & 338 &                   $-$1925  &  $-$1936 & F & L18 & $PWVK_s$    &    18.496$\pm$0.011  &     2268 \\
25 &  $-6.028 \pm 0.018$ &  $-3.176 \pm 0.055$ &  $-0.453 \pm 0.154$ & $-0.054 \pm 0.660$ & 338 & $-$1918  &  $-$1933 & F & L18 & $PWVK_s$    &    18.501$\pm$0.011  &     2268 \\
26 &  $-6.076 \pm 0.015$ &  $-3.268 \pm 0.043$ &   &  & 434 &                                     $-$2406  &  $-$2414 & F+1O & Lit & $PWVK_s$    & 18.647$\pm$0.010  &     4108 \\
\rowcolor{lightgray} 27 &  $-6.022 \pm 0.022$ &  $-3.174 \pm 0.049$ &  $-0.459 \pm 0.107$ &  & 434 &                   $-$2502  &  $-$2514 & F+1O & Lit & $PWVK_s$    & 18.490$\pm$0.009  &     4108 \\
28 &  $-6.018 \pm 0.015$ &  $-3.168 \pm 0.047$ &  $-0.492 \pm 0.103$ & $-0.110 \pm 0.257$ & 434 & $-$2499  &  $-$2516 & F+1O & Lit & $PWVK_s$    & 18.497$\pm$0.009  &     4108 \\
29 &  $-6.040 \pm 0.018$ &  $-3.177 \pm 0.047$ &  $-0.382 \pm 0.125$ &  & 398 &                   $-$2233  &  $-$2245 & F+1O & L18 & $PWVK_s$    & 18.532$\pm$0.009  &     4108 \\
30 &  $-6.038 \pm 0.019$ &  $-3.174 \pm 0.050$ &  $-0.410 \pm 0.146$ & $-0.098 \pm 0.415$ & 398 &   $-$2225  &  $-$2241 & F+1O & L18 & $PWVK_s$  & 18.539$\pm$0.009  &     4108 \\
31 &  $-5.974 \pm 0.015$ &  $-3.259 \pm 0.044$ &   &  & 276 &                                     $-$1738  &  $-$1745 & F & Lit & $PWHVI$    &     18.558$\pm$0.011  &     2000 \\
32 &  $-5.931 \pm 0.016$ &  $-3.206 \pm 0.064$ &  $-0.382 \pm 0.085$ &  & 276 &                   $-$1795  &  $-$1806 & F & Lit & $PWHVI$    &     18.410$\pm$0.011  &     2000 \\
33 &  $-5.938 \pm 0.017$ &  $-3.243 \pm 0.072$ &  $-0.342 \pm 0.093$ & $0.329 \pm 0.419$ & 276 &  $-$1788  &  $-$1803 & F & Lit & $PWHVI$    &     18.372$\pm$0.010  &     2000 \\
34 &  $-5.946 \pm 0.015$ &  $-3.199 \pm 0.049$ &  $-0.383 \pm 0.108$ &  & 255 &                   $-$1641  &  $-$1651 & F & L18 & $PWHVI$    &     18.428$\pm$0.011  &     2000 \\
35 &  $-5.961 \pm 0.019$ &  $-3.266 \pm 0.064$ &  $-0.295 \pm 0.123$ & $0.924 \pm 0.421$ & 255 &  $-$1657  &  $-$1672 & F & L18 & $PWHVI$    &     18.322$\pm$0.010  &     2000 \\
36 &  $-5.979 \pm 0.011$ &  $-3.215 \pm 0.050$ &   &  & 317 &                                     $-$2007  &  $-$2015 & F+1O & Lit & $PWHVI$    &  18.581$\pm$0.010  &     2000 \\
\rowcolor{lightgray} 37 &  $-5.937 \pm 0.017$ &  $-3.172 \pm 0.049$ &  $-0.366 \pm 0.089$ &  & 317 &                   $-$2064  &  $-$2076 & F+1O & Lit & $PWHVI$    &  18.435$\pm$0.010  &     2000 \\
38 &  $-5.944 \pm 0.019$ &  $-3.203 \pm 0.065$ &  $-0.324 \pm 0.085$ & $0.261 \pm 0.302$ & 317 &  $-$2065  &  $-$2080 & F+1O & Lit & $PWHVI$    &  18.409$\pm$0.009  &     2000 \\
39 &  $-5.951 \pm 0.014$ &  $-3.162 \pm 0.059$ &  $-0.367 \pm 0.084$ &  & 292 &                   $-$1897  &  $-$1908 & F+1O & L18 & $PWHVI$    &  18.453$\pm$0.009  &     2000 \\
40 &  $-5.962 \pm 0.019$ &  $-3.208 \pm 0.063$ &  $-0.279 \pm 0.116$ & $0.611 \pm 0.336$ & 292 &  $-$1899  &  $-$1914 & F+1O & L18 & $PWHVI$    &  18.393$\pm$0.010  &     2000 \\
\hline                                   
\end{tabular}
\end{table*}

\begin{table*}
\caption{Fitting results from the MCMC approach. The meaning of the different columns is the same as in Table~\ref{table:resNL}.}
\label{table:resMCMC} 
\centering          
\begin{tabular}{ccccccccccccc} 
\hline\hline             
ID & $\alpha$ & $\beta$ & $\gamma$ & $\delta$ & n.MW & BIC & AIC & Mode & Source & Relation & $\mu_{LMC}$ & n.LMC\\   
\hline    
1   & $-5.824^{+0.006}_{-0.006}$  &$-3.171^{+0.022}_{-0.022}$  &                                &                                &  381  & -1613  &  -1621&        F     &    Lit   &  $PLK_s$     &     18.631$\pm$0.011  &     2432  \\
2   & $-5.767^{+0.008}_{-0.008}$  &$-3.085^{+0.023}_{-0.023}$  &   $   -0.515^{+0.048}_{-0.049}$ &                               &  381  & -1717  &  -1729&        F     &    Lit   &  $PLK_s$     &     18.438$\pm$0.010  &     2432  \\
3   & $-5.767^{+0.009}_{-0.009}$  &$-3.086^{+0.029}_{-0.029}$  &   $   -0.513^{+0.056}_{-0.056}$ & $   0.008^{+0.165}_{-0.164}$  &  381  & -1711  &  -1727&        F     &    Lit   &  $PLK_s$     &     18.438$\pm$0.010  &     2432  \\
4   & $-5.784^{+0.008}_{-0.007}$  &$-3.075^{+0.024}_{-0.023}$  &   $   -0.453^{+0.055}_{-0.055}$ &                               &  349  & -1569  &  -1581&        F     &    L18   &  $PLK_s$     &     18.480$\pm$0.011  &     2432  \\
5   & $-5.785^{+0.008}_{-0.008}$  &$-3.082^{+0.027}_{-0.027}$  &   $   -0.440^{+0.061}_{-0.061}$ & $   0.094^{+0.204}_{-0.203}$  &  349  & -1564  &  -1579&        F     &    L18   &  $PLK_s$     &     18.471$\pm$0.011  &     2432  \\
6   & $-5.831^{+0.006}_{-0.006}$  &$-3.152^{+0.019}_{-0.019}$  &                                     &                           &  448  & -1974  &  -1982&     F+1O     &    Lit   &  $PLK_s$     &     18.636$\pm$0.010  &     4396  \\
\rowcolor{lightgray} 7   & $-5.780^{+0.008}_{-0.008}$  &$-3.068^{+0.021}_{-0.020}$  &   $   -0.440^{+0.040}_{-0.041}$     &                           &  448  & -2083  &  -2095&     F+1O     &    Lit   &  $PLK_s$     &     18.482$\pm$0.009  &     4396  \\
8   & $-5.776^{+0.008}_{-0.008}$  &$-3.059^{+0.021}_{-0.021}$  &   $   -0.479^{+0.051}_{-0.050}$ & $  -0.134^{+0.106}_{-0.104}$  &  448  & -2078  &  -2095&     F+1O     &    Lit   &  $PLK_s$     &     18.493$\pm$0.009  &     4396  \\
9   & $-5.794^{+0.007}_{-0.007}$  &$-3.060^{+0.021}_{-0.021}$  &   $   -0.393^{+0.046}_{-0.046}$     &                           &  411  & -1908  &  -1921&     F+1O     &    L18   &  $PLK_s$     &     18.516$\pm$0.009  &     4396  \\
10  & $-5.792^{+0.007}_{-0.007}$  &$-3.057^{+0.021}_{-0.021}$  &   $   -0.417^{+0.056}_{-0.056}$ & $  -0.092^{+0.119}_{-0.118}$  &  411  & -1903  &  -1919&     F+1O     &    L18   &  $PLK_s$     &     18.523$\pm$0.009  &     4396  \\
11  & $-6.124^{+0.007}_{-0.007}$  &$-3.303^{+0.023}_{-0.023}$  &                                     &                           &  381  & -1419  &  -1427&        F     &    Lit   &  $PWJK_s$     &    18.633$\pm$0.011  &     2423  \\
12  & $-6.062^{+0.009}_{-0.008}$  &$-3.210^{+0.024}_{-0.024}$  &   $   -0.551^{+0.049}_{-0.050}$ &                               &  381  & -1532  &  -1544&        F     &    Lit   &  $PWJK_s$     &    18.426$\pm$0.010  &     2423  \\
13  & $-6.067^{+0.009}_{-0.009}$  &$-3.228^{+0.029}_{-0.029}$  &   $   -0.523^{+0.056}_{-0.056}$ & $   0.171^{+0.170}_{-0.167}$  &  381  & -1527  &  -1543&        F     &    Lit   &  $PWJK_s$     &    18.411$\pm$0.010  &     2423  \\
14  & $-6.086^{+0.008}_{-0.008}$  &$-3.208^{+0.024}_{-0.024}$  &   $   -0.379^{+0.056}_{-0.056}$ &                               &  349  & -1392  &  -1404&        F     &    L18   &  $PWJK_s$     &    18.507$\pm$0.011  &     2423  \\
15  & $-6.083^{+0.009}_{-0.008}$  &$-3.197^{+0.028}_{-0.028}$  &   $   -0.399^{+0.062}_{-0.062}$ & $  -0.164^{+0.208}_{-0.210}$  &  349  & -1387  &  -1402&        F     &    L18   &  $PWJK_s$     &    18.524$\pm$0.010  &     2423  \\
16  & $-6.126^{+0.006}_{-0.006}$  &$-3.299^{+0.020}_{-0.020}$  &                                &                                &  448  & -1841  &  -1849&     F+1O     &    Lit   &  $PWJK_s$     &    18.631$\pm$0.010  &     4360  \\
\rowcolor{lightgray} 17  & $-6.073^{+0.008}_{-0.008}$  &$-3.212^{+0.021}_{-0.021}$  &   $   -0.455^{+0.042}_{-0.042}$ &                               &  448  & -1948  &  -1961&     F+1O     &    Lit   &  $PWJK_s$     &    18.473$\pm$0.009  &     4360  \\
18  & $-6.069^{+0.009}_{-0.009}$  &$-3.201^{+0.023}_{-0.023}$  &   $   -0.497^{+0.052}_{-0.052}$ & $  -0.151^{+0.112}_{-0.111}$  &  448  & -1944  &  -1960&     F+1O     &    Lit   &  $PWJK_s$     &    18.485$\pm$0.009  &     4360  \\
19  & $-6.090^{+0.007}_{-0.007}$  &$-3.209^{+0.022}_{-0.022}$  &   $   -0.330^{+0.048}_{-0.048}$ &                               &  411  & -1786  &  -1798&     F+1O     &    L18   &  $PWJK_s$     &    18.533$\pm$0.009  &     4360  \\
20  & $-6.087^{+0.008}_{-0.008}$  &$-3.200^{+0.022}_{-0.022}$  &   $   -0.378^{+0.058}_{-0.058}$ & $  -0.192^{+0.129}_{-0.129}$  &  411  & -1782  &  -1798&     F+1O     &    L18   &  $PWJK_s$     &    18.550$\pm$0.009  &     4360  \\
21  & $-6.073^{+0.006}_{-0.006}$  &$-3.284^{+0.022}_{-0.022}$  &                                &                                &  369  & -1599  &  -1607&        F     &    Lit   &  $PWVK_s$     &    18.645$\pm$0.011  &     2268  \\
22  & $-6.013^{+0.008}_{-0.008}$  &$-3.196^{+0.023}_{-0.023}$  &   $   -0.531^{+0.049}_{-0.049}$ &                               &  369  & -1709  &  -1720&        F     &    Lit   &  $PWVK_s$     &    18.445$\pm$0.011  &     2268  \\
23  & $-6.014^{+0.009}_{-0.009}$  &$-3.200^{+0.029}_{-0.029}$  &   $   -0.526^{+0.056}_{-0.056}$ & $   0.034^{+0.166}_{-0.164}$  &  369  & -1703  &  -1718&        F     &    Lit   &  $PWVK_s$     &    18.442$\pm$0.011  &     2268  \\
24  & $-6.032^{+0.008}_{-0.008}$  &$-3.187^{+0.024}_{-0.024}$  &   $   -0.424^{+0.056}_{-0.056}$ &                               &  338  & -1570  &  -1582&        F     &    L18   &  $PWVK_s$     &    18.504$\pm$0.011  &     2268  \\
25  & $-6.030^{+0.008}_{-0.008}$  &$-3.179^{+0.027}_{-0.027}$  &   $   -0.440^{+0.061}_{-0.060}$ & $  -0.125^{+0.205}_{-0.205}$  &  338  & -1565  &  -1580&        F     &    L18   &  $PWVK_s$     &    18.516$\pm$0.011  &     2268  \\
26  & $-6.079^{+0.006}_{-0.006}$  &$-3.271^{+0.019}_{-0.019}$  &                                &                                &  434  & -1964  &  -1973&     F+1O     &    Lit   &  $PWVK_s$     &    18.649$\pm$0.010  &     4108  \\
\rowcolor{lightgray} 27  & $-6.026^{+0.008}_{-0.008}$  &$-3.185^{+0.020}_{-0.020}$  &   $   -0.445^{+0.040}_{-0.041}$ &                               &  434  & -2076  &  -2088&     F+1O     &    Lit   &  $PWVK_s$     &    18.493$\pm$0.009  &     4108  \\
28  & $-6.022^{+0.008}_{-0.008}$  &$-3.176^{+0.021}_{-0.022}$  &   $   -0.486^{+0.051}_{-0.051}$ & $  -0.141^{+0.101}_{-0.101}$  &  434  & -2072  &  -2088&     F+1O     &    Lit   &  $PWVK_s$     &    18.504$\pm$0.009  &     4108  \\
29  & $-6.042^{+0.007}_{-0.007}$  &$-3.179^{+0.021}_{-0.021}$  &   $   -0.365^{+0.045}_{-0.046}$ &                               &  398  & -1910  &  -1922&     F+1O     &    L18   &  $PWVK_s$     &    18.538$\pm$0.009  &     4108  \\
30  & $-6.039^{+0.008}_{-0.008}$  &$-3.174^{+0.021}_{-0.021}$  &   $   -0.404^{+0.056}_{-0.056}$ & $  -0.143^{+0.117}_{-0.117}$  &  398  & -1905  &  -1921&     F+1O     &    L18   &  $PWVK_s$     &    18.549$\pm$0.009  &     4108  \\
31  & $-5.978^{+0.007}_{-0.007}$  &$-3.278^{+0.025}_{-0.025}$  &                                &                                &  276  & -1654  &  -1661&        F     &    Lit   &   $PWHVI$     &    18.554$\pm$0.011  &     2000  \\
32  & $-5.934^{+0.009}_{-0.009}$  &$-3.224^{+0.026}_{-0.026}$  &   $   -0.391^{+0.058}_{-0.059}$ &                               &  276  & -1693  &  -1704&        F     &    Lit   &   $PWHVI$     &    18.403$\pm$0.011  &     2000  \\
33  & $-5.942^{+0.011}_{-0.010}$  &$-3.272^{+0.035}_{-0.035}$  &   $   -0.344^{+0.062}_{-0.062}$ & $   0.419^{+0.202}_{-0.199}$  &  276  & -1691  &  -1706&        F     &    Lit   &   $PWHVI$     &    18.352$\pm$0.011  &     2000  \\
34  & $-5.949^{+0.008}_{-0.008}$  &$-3.216^{+0.027}_{-0.027}$  &   $   -0.373^{+0.067}_{-0.067}$ &                               &  255  & -1558  &  -1569&        F     &    L18   &   $PWHVI$     &    18.428$\pm$0.011  &     2000  \\
35  & $-5.964^{+0.009}_{-0.009}$  &$-3.285^{+0.032}_{-0.032}$  &   $   -0.290^{+0.070}_{-0.070}$ & $   0.955^{+0.238}_{-0.235}$  &  255  & -1569  &  -1583&        F     &    L18   &   $PWHVI$     &    18.315$\pm$0.010  &     2000  \\
36  & $-5.983^{+0.007}_{-0.007}$  &$-3.229^{+0.023}_{-0.023}$  &                                &                                &  317  & -1899  &  -1907&     F+1O     &    Lit   &   $PWHVI$     &    18.579$\pm$0.010  &     2000  \\
\rowcolor{lightgray} 37  & $-5.942^{+0.009}_{-0.009}$  &$-3.187^{+0.024}_{-0.024}$  &   $   -0.367^{+0.052}_{-0.052}$ &                               &  317  & -1943  &  -1954&     F+1O     &    Lit   &   $PWHVI$     &    18.434$\pm$0.010  &     2000  \\
38  & $-5.949^{+0.010}_{-0.010}$  &$-3.223^{+0.033}_{-0.033}$  &   $   -0.319^{+0.059}_{-0.059}$ & $   0.299^{+0.185}_{-0.184}$  &  317  & -1940  &  -1955&     F+1O     &    Lit   &   $PWHVI$     &    18.403$\pm$0.009  &     2000  \\
39  & $-5.955^{+0.008}_{-0.008}$  &$-3.174^{+0.024}_{-0.024}$  &   $   -0.356^{+0.060}_{-0.060}$ &                               &  292  & -1789  &  -1800&     F+1O     &    L18   &   $PWHVI$     &    18.456$\pm$0.009  &     2000  \\
40  & $-5.966^{+0.009}_{-0.009}$  &$-3.222^{+0.029}_{-0.029}$  &   $   -0.273^{+0.066}_{-0.066}$ & $   0.617^{+0.211}_{-0.211}$  &  292  & -1792  &  -1807&     F+1O     &    L18   &   $PWHVI$     &    18.393$\pm$0.010  &     2000  \\
   \hline                                   
\end{tabular}
\end{table*}

\section{Summary and Conclusions}

In this paper we have presented new time-resolved HARPS-N@TNG high-resolution (R\,=\,115,000) spectroscopy (see Sect. 2) for a sample of 47 bright DCEPs plus 1 BLHER star\footnote{ASAS J162326-0941.0, previously classified as DCEP is actually a T2CEP of  BLHER type} with good {\it Gaia} DR3 parallaxes. This is the first paper of a project devoted to significantly enlarge the number of MW DCEPs with reliable abundance measurements, which we will then use to carefully  investigate the metallicity dependence of the PL/PW relations of these primary standard candles of the cosmic distance ladder. 

In Sect. 3 we have carried out a detailed analysis in order to derive, from each spectrum, main atmospheric parameters such as the effective temperature, surface gravity, microturbulence, and radial velocity. 
We have performed a complete characterization of the chemical abundances of 29 species for which we have detected and measured spectral lines. We found on average, abundance consistent with the solar values for the $\alpha$- and {\it iron}-peaks elements, as well as for aluminum, scandium, vanadium and copper. Among the $\alpha$-elements, sulfur is slightly under-abundant. Solar values have been inferred even for heavy elements such as strontium, yttrium and zirconium and, in general, even for rare earths. Barium is spread out over a wide range ($\approx$\,1~dex) of over-abundances. As expected for DCEPs, carbon and oxygen appear depleted with respect to the Sun and, on the contrary, sodium shows an average overabundance.
In general, our abundances are in agreement with the general trend of those by \citet{luck18}, both for what concern the correlation with iron and the Galactic gradient. Additionally, we confirmed the anti-correlation between oxygen and iron. 

In Sect. 4 we derived new, accurate estimates of the period and epoch of maximum light for all our program stars by combining $V$-band photometry from different sources available in the literature. The typical time baseline of the combined  time-series data is larger than 15 years for most of our targets. Periods and epochs inferred from the $V$-band data were used to fold the $B,I$ data, whenever available in the literature, and derive the corresponding average magnitudes.   
Intensity-averaged magnitudes in the $J,H,K_s$ bands were computed from the 2MASS single epoch photometry by means of a template fitting technique. The average precision of the 
intensity-averaged magnitudes we  obtained for our program stars is, for all bands,  on the order of a few hundredths of magnitude.
To estimate individual reddening values for all our program stars, we calculated and used new, accurate PC relations in the $(B-V)$ and $(V-I)$ colors, which showed to be in very good agreement with previous results in the literature.

In Sect. 5 our sample of 47 DCEPs  was complemented with  DCEPs for which iron abundances from high-resolution spectroscopy are available in the literature. We considered two separate samples, one comprising a compilation of literature metallicities and the other based exclusively on the work by L18. The comparison between the two metallicity scales reveals that the L18 results give on average slightly lower metallicities, apart from the metal-poor regime ([Fe/H]$\sim-0.3$-$0.4$ dex) where the trend is inverted.


In Sect. 6 we discussed the astrometric properties of our and literature samples, eliminating from the following analysis the stars with ruwe parameter larger than 1.4 and then correcting the Gaia parallax ZPO individually using the recipe by \citet[][]{Lindegren2021b}. The comparison between the ZPO-corrected Gaia and the HST parallaxes for the 7 stars having such HST measures, reveals agreement at 1$\sigma$, albeit with a rather large scatter. 
After removing the stars with poor astrometry, the total DCEPs sample adopted in this work, including the 47 stars presented here is of 448 and 411 stars for the Lit and L18 samples, respectively.

In Sect. 7 we faced the problem of deriving the $PLZ/PWZ$ relations using the data presented in the previous sections. To this aim, we adopted the ABL formulation that allows to treat linearly the parallax. We calculated the $PLZ/PWZ$ considering the following three cases: i) no metallicity term (2 parameters); ii)  with metallicity term on the intercept (3 parameters); iii)  with the metallicity term on both the slope and the intercept (4 parameters). We considered four different types of relations, involving NIR magnitudes: $PLK_s$, $PWJK_s$, $PWVK_s$ and $PWHVI$. The fitting procedure was carried out in two completely different ways: a standard NL-LSQs technique with a bootstrap routine to robustly calculate the errors and a Bayesian approach adopting a MCMC method to best estimate coefficients and errors. The fitting procedure has been carried out in a variety of cases: with 2,3,4 parameters, using the Lit or L18 metallicity samples and adopting separate DCEP\_F only or DCEP\_F+DCEP\_1O samples. In the end we have 40 cases that have been investigated with both the NL-LSQ and MCMC methods.  

In Sect. 8 we tested the $PLZ/PWZ$ relations calculated in this work against  the geometric distance to the LMC  estimated by \citet{Pietrzynski2019}, and using the AIC/BIC parameters as goodness-of-fit estimators. To estimate the LMC distance with our $PLZ/PWZ$ relations, we used literature data for $\sim$4500 DCEPs in the LMC and averaged out the single distance moduli obtained for each DCEP. 
We find that: 1) the two fitting methods adopted to derive the $PLZ/PWZ$ relations provide results in excellent agreement each other; 2) the adoption of the Lit or the L18 metallicities gives consistent results; 3) a significant metallicity term is needed to derive LMC distances $<1 \sigma$ from the geometric one. The best results are provided by the $PLK_s$, $PWJK_s$ and $PWVK_s$ relations with 3 parameters, the Lit metallicities and the full sample (F+1O-mode pulsators), i.e. the solutions 7, 17, 27 of Table~\ref{table:resNL}. In these cases we find a metalicity dependence of  $-0.456\pm$0.099, $-0.465\pm$0.071 and  $-0.459\pm$0.107 mag/dex, respectively. These fits are also those with the best AIC/BIC values. The metal dependence for the $PWHVI$ case is significantly lower: $-0.366\pm$0.089 mag/dex, with worser AIC/BIC values than the previous cases.   
 4) the fits with 4 parameters do not improve the results (AIC/BIC values equal to the 3 parameters case); 5) the inclusion of a correction by 14 $\mu$as by \citet[][]{Riess2021} to the Gaia parallaxes ZPO, improves significantly the results with the $PWHVI$ relations, but provide $> 1 \sigma$ LMC distances for the other relationships; 6) the inclusion of the correction to the Gaia parallaxes ZPO causes a reduction by $\sim$22\% (in absolute terms) of the metallicity terms both in the 3 and 4 parameters cases. Since the resulting LMC distances is still compatible with the geometric one, this occurrence suggests that the inclusion of the parallax ZPO correction and the metallicity terms are largely degenerate. 7) In all the cases considered here, except the $PWHVI$ relations with the ZPO correction, the metallicity terms we found are larger than those provided in the literature. 

The fact that the ZPO correction and the metallicity terms appear degenerate means that it is important to have independent estimates of the parallax ZPO. In this sense we will test the recipe to calculate individual parallax ZPOs recently published in pre-print form by \citet[][]{Groenewegen2021}. In any case, it seems that we are still not close to have a precision of 3-4 $\mu$as on the parallax ZPO that is mandatory to derive the H$_0$ constant with 1\% precision.


\section*{Acknowledgements}

We thank our anonymous Referee for their comments that helped us to improve the manuscript.
This work has made use of data from the European Space Agency (ESA) mission
{\it Gaia} (\url{https://www.cosmos.esa.int/gaia}), processed by the {\it Gaia} Data Processing and Analysis Consortium (DPAC,
\url{https://www.cosmos.esa.int/web/gaia/dpac/consortium}). Funding for the DPAC has been provided by national institutions, in particular the institutions participating in the {\it Gaia} Multilateral Agreement.
In particular, the Italian participation
in DPAC has been supported by Istituto Nazionale di Astrofisica
(INAF) and the Agenzia Spaziale Italiana (ASI) through grants I/037/08/0,
I/058/10/0, 2014-025-R.0, and 2014-025-R.1.2015 to INAF (PI M.G. Lattanzi).
V.R., M.M. and G.C. acknowledge partial support from the project "MITiC: MIning The Cosmos Big Data and Innovative Italian Technology for Frontier Astrophysics and Cosmology”  (PI B. Garilli).
This publication makes use of VOSA, developed under the Spanish Virtual Observatory project supported by the Spanish MINECO through grant AyA2017-84089.
VOSA has been partially updated by using funding from the European Union's Horizon 2020 Research and Innovation Programme, under Grant Agreement nº 776403 (EXOPLANETS-A).
This research has made use of the SIMBAD database,
operated at CDS, Strasbourg, France.
Based on data obtained from the ESO Science Archive Facility under request number 534967.

\section*{Data Availability}

The only proprietary data used in this paper are represented by the HARPS-N spectra. The reduced spectra are available upon request to the corresponding author (V. Ripepi)

%
%

\appendix
\section{Result of the fit}

\begin{table*}
\caption{Coefficients of the linear fit of the form A(X)=a+bA(Fe), used to reproduce the abundances of the elements as a function of iron, see Fig.~\ref{fig:patt_luck}.}
\label{fitgrad} 
\footnotesize\setlength{\tabcolsep}{3pt}
\centering          
\begin{tabular}{lrrr} 
\hline\hline             
El &         a~~~~~~~~~~~&        b~~~~~~~~~ &  scatter       \\
\hline
C  & -0.243 $\pm$ 0.008  &  0.03 $\pm$ 0.04 &   0.17  \\
O  & -0.008 $\pm$ 0.005  & -0.37 $\pm$ 0.03 &   0.11  \\
Na &  0.353 $\pm$ 0.006  & -0.01 $\pm$ 0.03 &   0.13  \\
Mg &  0.087 $\pm$ 0.005  &  0.04 $\pm$ 0.03 &   0.12  \\
Al &  0.107 $\pm$ 0.006  & -0.09 $\pm$ 0.03 &   0.13  \\
Si &  0.154 $\pm$ 0.002  & -0.17 $\pm$ 0.01 &   0.06  \\
S  &  0.040 $\pm$ 0.006  &  0.03 $\pm$ 0.03 &   0.13  \\
Ca &  0.044 $\pm$ 0.003  & -0.04 $\pm$ 0.01 &   0.07  \\
Sc &  0.338 $\pm$ 0.006  &  0.00 $\pm$ 0.03 &   0.13  \\
Ti &  0.136 $\pm$ 0.003  & -0.20 $\pm$ 0.01 &   0.07  \\
V  &  0.124 $\pm$ 0.006  & -0.42 $\pm$ 0.03 &   0.13  \\
Cr &  0.153 $\pm$ 0.003  & -0.07 $\pm$ 0.01 &   0.07  \\
Mn & -0.121 $\pm$ 0.005  &  0.14 $\pm$ 0.02 &   0.11  \\
Co &  0.212 $\pm$ 0.007  & -0.55 $\pm$ 0.04 &   0.15  \\
Ni & -0.017 $\pm$ 0.002  & -0.04 $\pm$ 0.01 &   0.05  \\
Cu &  0.136 $\pm$ 0.010  & -0.21 $\pm$ 0.06 &   0.22  \\
Zn & -0.117 $\pm$ 0.009  &  0.26 $\pm$ 0.05 &   0.20  \\
Sr &  0.495 $\pm$ 0.023  & -0.68 $\pm$ 0.13 &   0.41  \\
Y  &  0.205 $\pm$ 0.005  & -0.25 $\pm$ 0.02 &   0.11  \\
Zr &  0.427 $\pm$ 0.010  & -0.67 $\pm$ 0.06 &   0.23  \\
Ba &  0.223 $\pm$ 0.032  & -0.25 $\pm$ 0.13 &   0.33  \\
La &  0.175 $\pm$ 0.006  & -0.48 $\pm$ 0.03 &   0.13  \\
Ce &  0.087 $\pm$ 0.007  & -0.47 $\pm$ 0.04 &   0.16  \\
Pr & -0.438 $\pm$ 0.030  & -1.25 $\pm$ 0.14 &   0.15  \\
Nd &  0.014 $\pm$ 0.005  & -0.40 $\pm$ 0.03 &   0.12  \\
Sm & -0.031 $\pm$ 0.007  & -0.37 $\pm$ 0.04 &   0.15  \\
Eu & -0.236 $\pm$ 0.006  & -0.45 $\pm$ 0.03 &   0.13  \\
Gd &  0.137 $\pm$ 0.026  & -0.89 $\pm$ 0.11 &   0.18  \\

\hline                                   
\end{tabular}
\end{table*}

\begin{table*}
\caption{Coefficients of the linear fit of the form [X/H]=a+bR$_{GC}$, used to reproduce the abundances of the elements as a function of galactocentric distance, see Fig.~\ref{fig:grad_luck}.}
\label{fitgraddist} 
\footnotesize\setlength{\tabcolsep}{3pt}
\centering          
\begin{tabular}{lrrr} 
\hline\hline             
El &         a~~~~~~~~~~&        b~~~~~~~~~~ &  scatter       \\
\hline
C  &  0.23 $\pm$ 0.04 & -0.050 $\pm$ 0.004  &   0.58   \\
O  &  0.34 $\pm$ 0.02 & -0.037 $\pm$ 0.002  &   0.35   \\
Na &  0.77 $\pm$ 0.03 & -0.045 $\pm$ 0.003  &   0.56   \\
Mg &  0.52 $\pm$ 0.03 & -0.047 $\pm$ 0.003  &   0.56   \\
Al &  0.54 $\pm$ 0.03 & -0.047 $\pm$ 0.003  &   0.50   \\
Si &  0.53 $\pm$ 0.02 & -0.040 $\pm$ 0.002  &   0.36   \\
S  &  0.58 $\pm$ 0.03 & -0.058 $\pm$ 0.003  &   0.51   \\
Ca &  0.46 $\pm$ 0.03 & -0.045 $\pm$ 0.002  &   0.42   \\
Sc &  0.72 $\pm$ 0.03 & -0.041 $\pm$ 0.003  &   0.58   \\
Ti &  0.42 $\pm$ 0.02 & -0.031 $\pm$ 0.002  &   0.41   \\
V  &  0.30 $\pm$ 0.03 & -0.019 $\pm$ 0.003  &   0.48   \\
Cr &  0.50 $\pm$ 0.03 & -0.037 $\pm$ 0.002  &   0.45   \\
Mn &  0.28 $\pm$ 0.04 & -0.043 $\pm$ 0.003  &   0.60   \\
Fe &  0.41 $\pm$ 0.03 & -0.044 $\pm$ 0.002  &   0.41   \\
Co &  0.32 $\pm$ 0.03 & -0.012 $\pm$ 0.003  &   0.51   \\
Ni &  0.39 $\pm$ 0.03 & -0.043 $\pm$ 0.002  &   0.42   \\
Cu &  0.48 $\pm$ 0.05 & -0.037 $\pm$ 0.004  &   0.69   \\
Zn &  0.47 $\pm$ 0.05 & -0.063 $\pm$ 0.004  &   0.75   \\
Sr &  0.50 $\pm$ 0.09 & -0.001 $\pm$ 0.008  &   1.02   \\
Y  &  0.45 $\pm$ 0.03 & -0.026 $\pm$ 0.002  &   0.48   \\
Zr &  0.42 $\pm$ 0.04 &  0.000 $\pm$ 0.004  &   0.70   \\
Ba &  0.44 $\pm$ 0.12 & -0.027 $\pm$ 0.011  &   0.55   \\
La &  0.35 $\pm$ 0.03 & -0.019 $\pm$ 0.002  &   0.46   \\
Ce &  0.19 $\pm$ 0.03 & -0.011 $\pm$ 0.003  &   0.56   \\
Pr & -0.66 $\pm$ 0.12 &  0.025 $\pm$ 0.012  &   0.15   \\
Nd &  0.15 $\pm$ 0.03 & -0.015 $\pm$ 0.002  &   0.47   \\
Sm &  0.11 $\pm$ 0.03 & -0.015 $\pm$ 0.003  &   0.55   \\
Eu & -0.06 $\pm$ 0.03 & -0.019 $\pm$ 0.002  &   0.47   \\
Gd &  0.34 $\pm$ 0.10 & -0.023 $\pm$ 0.010  &   0.17   \\ 
\hline                                   
\end{tabular}
\end{table*}

\section{Light curves in $V$-band for the target stars}
\label{appendix:appendix_a}

\begin{figure*}
 \centering
 \includegraphics[width=\textwidth]{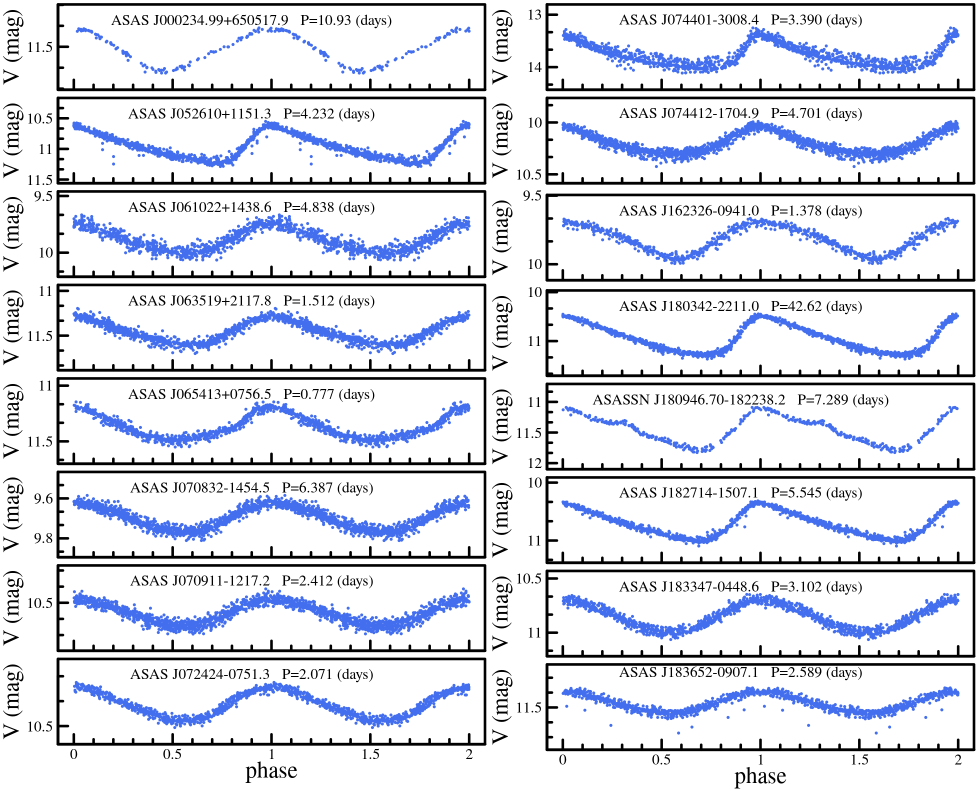}
\caption{$V$-band light curves of the program stars (see Table~\ref{Tab:log} for details)}
  \label{fig:LC}
\end{figure*}

\renewcommand{\thefigure}{\arabic{figure} (Cont.)}
\addtocounter{figure}{-1}

\begin{figure*}
 \centering
 \includegraphics[width=\textwidth]{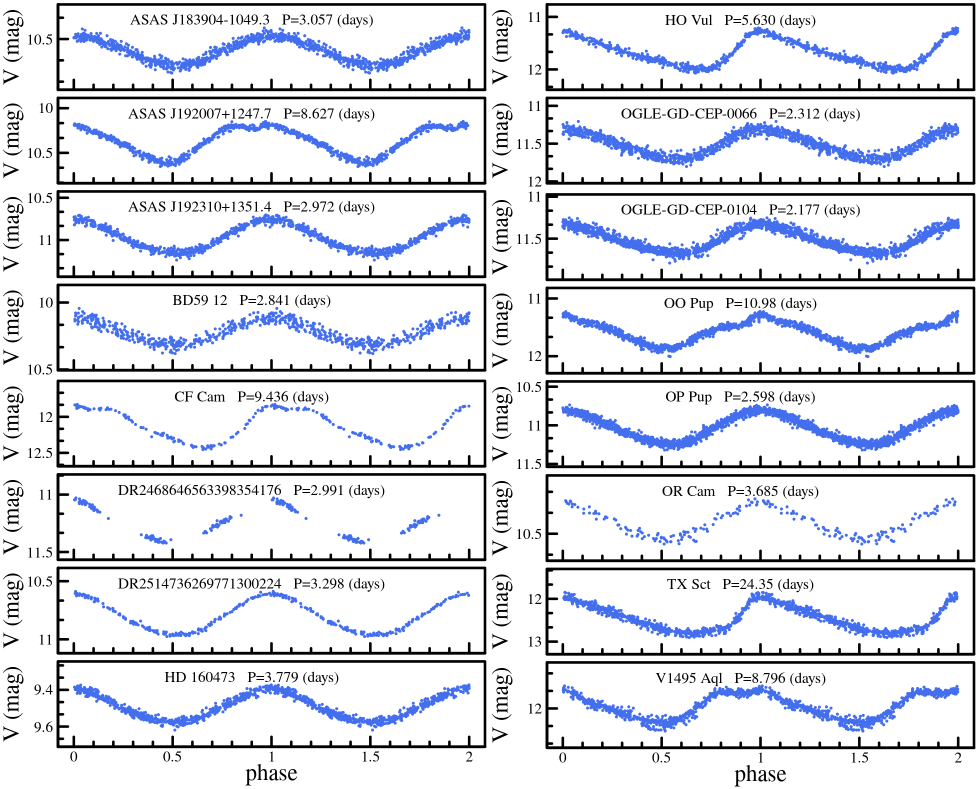}
\caption{}
\end{figure*}

\renewcommand{\thefigure}{\arabic{figure}}
\addtocounter{figure}{-1}

\begin{figure*}
 \centering
 \includegraphics[width=\textwidth]{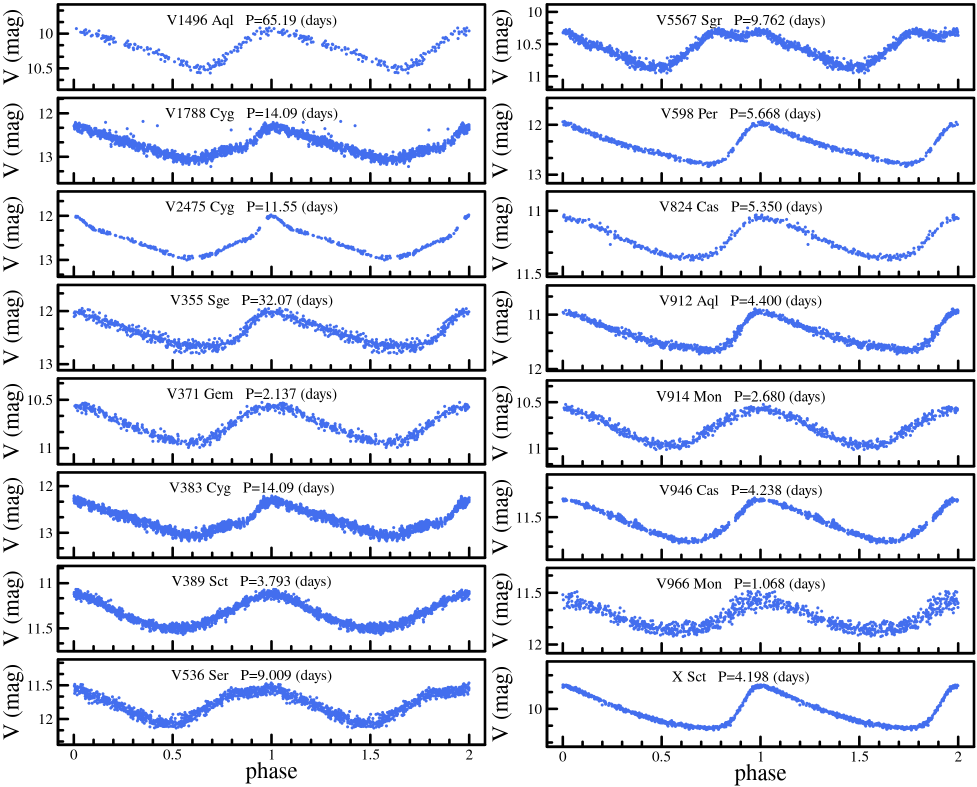}
\caption{}
\end{figure*}

\begin{table*}
\caption{Same as in Table~\ref{table:resNL}, but adopting the parallax ZPO correction by \citet{Riess2021}}
\label{table:resNLRiess} 
\footnotesize\setlength{\tabcolsep}{3pt}
\centering          
\begin{tabular}{ccccccccccccc} 
\hline\hline             
ID & $\alpha$ & $\beta$ & $\gamma$ & $\delta$ & n.MW & BIC & AIC & Mode & Source & Relation & $\mu_{\rm LMC}$ & n.LMC\\
\hline
   1    & $-$5.879$\pm$0.014  & $-$3.184$\pm$0.054  &                   &                     &  381 &  $-$2125 &  $-$2133 &      F  &   Lit  &   PLK  &  18.681$\pm$0.011  &     2432 \\
   2    & $-$5.831$\pm$0.019  & $-$3.108$\pm$0.054  &  -0.437$\pm$0.108 &                     &  381 &  $-$2208 &  $-$2220 &      F  &   Lit  &   PLK  &  18.519$\pm$0.010  &     2432 \\
   3    & $-$5.828$\pm$0.019  & $-$3.098$\pm$0.072  &  -0.456$\pm$0.111 & $-$0.110$\pm$0.341  &  381 &  $-$2199 &  $-$2215 &      F  &   Lit  &   PLK  &  18.528$\pm$0.011  &     2432 \\
   4    & $-$5.843$\pm$0.018  & $-$3.099$\pm$0.048  &  -0.410$\pm$0.181 &                     &  349 &  $-$1957 &  $-$1968 &      F  &   L18  &   PLK  &  18.543$\pm$0.011  &     2432 \\
   5    & $-$5.844$\pm$0.021  & $-$3.102$\pm$0.063  &  -0.405$\pm$0.139 &    0.039$\pm$0.629  &  349 &  $-$1951 &  $-$1967 &      F  &   L18  &   PLK  &  18.540$\pm$0.011  &     2432 \\
   6    & $-$5.887$\pm$0.013  & $-$3.169$\pm$0.037  &                   &                     &  448 &  $-$2476 &  $-$2485 &   F+1O  &   Lit  &   PLK  &  18.682$\pm$0.010  &     4396 \\
\rowcolor{lightgray}   7    & $-$5.845$\pm$0.016  & $-$3.097$\pm$0.048  &  -0.363$\pm$0.069 &                     &  448 &  $-$2570 &  $-$2582 &   F+1O  &   Lit  &   PLK  &  18.558$\pm$0.009  &     4396 \\
   8    & $-$5.840$\pm$0.016  & $-$3.086$\pm$0.055  &  -0.412$\pm$0.097 & $-$0.168$\pm$0.302  &  448 &  $-$2566 &  $-$2582 &   F+1O  &   Lit  &   PLK  &  18.571$\pm$0.010  &     4396 \\
   9    & $-$5.854$\pm$0.022  & $-$3.089$\pm$0.042  &  -0.337$\pm$0.117 &                     &  411 &  $-$2297 &  $-$2309 &   F+1O  &   L18  &   PLK  &  18.580$\pm$0.009  &     4396 \\
  10    & $-$5.852$\pm$0.017  & $-$3.084$\pm$0.043  &  -0.372$\pm$0.126 & $-$0.127$\pm$0.366  &  411 &  $-$2297 &  $-$2313 &   F+1O  &   L18  &   PLK  &  18.590$\pm$0.009  &     4396 \\
  11    & $-$6.182$\pm$0.015  & $-$3.315$\pm$0.067  &                   &                     &  381 &  $-$2057 &  $-$2065 &      F  &   Lit  &  PWJK  &  18.687$\pm$0.011  &     2423 \\
  12    & $-$6.132$\pm$0.020  & $-$3.236$\pm$0.058  &  -0.458$\pm$0.099 &                     &  381 &  $-$2109 &  $-$2121 &      F  &   Lit  &  PWJK  &  18.516$\pm$0.010  &     2423 \\
  13    & $-$6.132$\pm$0.017  & $-$3.238$\pm$0.076  &  -0.454$\pm$0.093 &    0.022$\pm$0.382  &  381 &  $-$2105 &  $-$2121 &      F  &   Lit  &  PWJK  &  18.514$\pm$0.011  &     2423 \\
  14    & $-$6.150$\pm$0.024  & $-$3.236$\pm$0.053  &  -0.297$\pm$0.285 &                     &  349 &  $-$1877 &  $-$1889 &      F  &   L18  &  PWJK  &  18.588$\pm$0.010  &     2423 \\
  15    & $-$6.145$\pm$0.025  & $-$3.216$\pm$0.074  &  -0.338$\pm$0.213 & $-$0.299$\pm$0.867  &  349 &  $-$1864 &  $-$1879 &      F  &   L18  &  PWJK  &  18.616$\pm$0.010  &     2423 \\
  16    & $-$6.185$\pm$0.015  & $-$3.317$\pm$0.048  &                   &                     &  448 &  $-$2463 &  $-$2472 &   F+1O  &   Lit  &  PWJK  &  18.682$\pm$0.009  &     4360 \\
\rowcolor{lightgray}  17    & $-$6.142$\pm$0.019  & $-$3.242$\pm$0.045  &  -0.369$\pm$0.104 &                     &  448 &  $-$2499 &  $-$2511 &   F+1O  &   Lit  &  PWJK  &  18.555$\pm$0.009  &     4360 \\
  18    & $-$6.138$\pm$0.020  & $-$3.232$\pm$0.051  &  -0.415$\pm$0.100 & $-$0.160$\pm$0.284  &  448 &  $-$2491 &  $-$2508 &   F+1O  &   Lit  &  PWJK  &  18.567$\pm$0.009  &     4360 \\
  19    & $-$6.156$\pm$0.019  & $-$3.242$\pm$0.052  &  -0.253$\pm$0.178 &                     &  411 &  $-$2255 &  $-$2267 &   F+1O  &   L18  &  PWJK  &  18.607$\pm$0.009  &     4360 \\
  20    & $-$6.153$\pm$0.027  & $-$3.235$\pm$0.050  &  -0.300$\pm$0.174 & $-$0.175$\pm$0.307  &  411 &  $-$2244 &  $-$2260 &   F+1O  &   L18  &  PWJK  &  18.621$\pm$0.009  &     4360 \\
  21    & $-$6.128$\pm$0.016  & $-$3.299$\pm$0.058  &                   &                     &  369 &  $-$2146 &  $-$2154 &      F  &   Lit  &  PWVK  &  18.694$\pm$0.012  &     2268 \\
  22    & $-$6.077$\pm$0.020  & $-$3.222$\pm$0.053  &  -0.449$\pm$0.109 &                     &  369 &  $-$2199 &  $-$2211 &      F  &   Lit  &  PWVK  &  18.526$\pm$0.011  &     2268 \\
  23    & $-$6.074$\pm$0.024  & $-$3.211$\pm$0.056  &  -0.469$\pm$0.121 & $-$0.112$\pm$0.383  &  369 &  $-$2191 &  $-$2207 &      F  &   Lit  &  PWVK  &  18.536$\pm$0.011  &     2268 \\
  24    & $-$6.093$\pm$0.020  & $-$3.215$\pm$0.048  &  -0.367$\pm$0.198 &                     &  338 &  $-$1965 &  $-$1976 &      F  &   L18  &  PWVK  &  18.571$\pm$0.011  &     2268 \\
  25    & $-$6.089$\pm$0.022  & $-$3.202$\pm$0.077  &  -0.394$\pm$0.169 & $-$0.200$\pm$0.657  &  338 &  $-$1962 &  $-$1978 &      F  &   L18  &  PWVK  &  18.591$\pm$0.011  &     2268 \\
  26    & $-$6.135$\pm$0.014  & $-$3.293$\pm$0.039  &                   &                     &  434 &  $-$2523 &  $-$2531 &   F+1O  &   Lit  &  PWVK  &  18.694$\pm$0.010  &     4108 \\
\rowcolor{lightgray}  27    & $-$6.091$\pm$0.015  & $-$3.217$\pm$0.039  &  -0.369$\pm$0.099 &                     &  434 &  $-$2587 &  $-$2600 &   F+1O  &   Lit  &  PWVK  &  18.567$\pm$0.009  &     4108 \\
  28    & $-$6.087$\pm$0.017  & $-$3.208$\pm$0.044  &  -0.414$\pm$0.093 & $-$0.148$\pm$0.292  &  434 &  $-$2585 &  $-$2601 &   F+1O  &   Lit  &  PWVK  &  18.577$\pm$0.009  &     4108 \\
  29    & $-$6.103$\pm$0.016  & $-$3.213$\pm$0.046  &  -0.305$\pm$0.155 &                     &  398 &  $-$2307 &  $-$2319 &   F+1O  &   L18  &  PWVK  &  18.602$\pm$0.010  &     4108 \\
  30    & $-$6.101$\pm$0.019  & $-$3.209$\pm$0.042  &  -0.343$\pm$0.140 & $-$0.132$\pm$0.387  &  398 &  $-$2288 &  $-$2304 &   F+1O  &   L18  &  PWVK  &  18.612$\pm$0.010  &     4108 \\
  31    & $-$6.032$\pm$0.012  & $-$3.286$\pm$0.047  &                   &                     &  276 &  $-$1778 &  $-$1785 &      F  &   Lit  & PWHVI  &  18.605$\pm$0.011  &     2000 \\
  32    & $-$5.997$\pm$0.016  & $-$3.243$\pm$0.046  &  -0.310$\pm$0.087 &                     &  276 &  $-$1804 &  $-$1815 &      F  &   Lit  & PWHVI  &  18.485$\pm$0.011  &     2000 \\
  33    & $-$6.000$\pm$0.020  & $-$3.262$\pm$0.068  &  -0.289$\pm$0.103 &    0.171$\pm$0.436  &  276 &  $-$1806 &  $-$1820 &      F  &   Lit  & PWHVI  &  18.465$\pm$0.011  &     2000 \\
  34    & $-$6.006$\pm$0.018  & $-$3.232$\pm$0.057  &  -0.328$\pm$0.153 &                     &  255 &  $-$1660 &  $-$1671 &      F  &   L18  & PWHVI  &  18.493$\pm$0.010  &     2000 \\
  35    & $-$6.019$\pm$0.019  & $-$3.289$\pm$0.066  &  -0.252$\pm$0.134 &    0.791$\pm$0.370  &  255 &  $-$1669 &  $-$1683 &      F  &   L18  & PWHVI  &  18.402$\pm$0.010  &     2000 \\
  36    & $-$6.036$\pm$0.016  & $-$3.245$\pm$0.046  &                   &                     &  317 &  $-$2057 &  $-$2064 &   F+1O  &   Lit  & PWHVI  &  18.626$\pm$0.010  &     2000 \\
\rowcolor{lightgray}  37    & $-$6.004$\pm$0.015  & $-$3.211$\pm$0.048  &  -0.292$\pm$0.112 &                     &  317 &  $-$2087 &  $-$2098 &   F+1O  &   Lit  & PWHVI  &  18.511$\pm$0.009  &     2000 \\
  38    & $-$6.006$\pm$0.016  & $-$3.223$\pm$0.058  &  -0.274$\pm$0.086 &    0.108$\pm$0.296  &  317 &  $-$2083 &  $-$2098 &   F+1O  &   Lit  & PWHVI  &  18.500$\pm$0.009  &     2000 \\
  39    & $-$6.011$\pm$0.016  & $-$3.197$\pm$0.047  &  -0.311$\pm$0.092 &                     &  292 &  $-$1905 &  $-$1916 &   F+1O  &   L18  & PWHVI  &  18.518$\pm$0.009  &     2000 \\
  40    & $-$6.020$\pm$0.017  & $-$3.235$\pm$0.062  &  -0.239$\pm$0.118 &    0.492$\pm$0.241  &  292 &  $-$1910 &  $-$1925 &   F+1O  &   L18  & PWHVI  &  18.470$\pm$0.009  &     2000 \\
\hline                                   
\end{tabular}
\end{table*}

\end{document}